\documentclass{aa}

\usepackage{graphics,graphicx,subfigure}
\usepackage{natbib}
\usepackage{txfonts,bm}
\usepackage{color}
\begin{document}

\title{Opacity of fluffy dust aggregates}

\author{Akimasa Kataoka\inst{1,2} \and Satoshi Okuzumi \inst{3} \and Hidekazu Tanaka \inst{4}\and Hideko Nomura \inst{3}}
\institute{
{Department of Astronomical Science, School of Physical Sciences, Graduate University for Advanced Studies (SOKENDAI), Mitaka, Tokyo 181-8588, Japan\\
\email{akimasa.kataoka@nao.ac.jp}
\and
National Astronomical Observatory of Japan, Mitaka, Tokyo 181-8588, Japan
\and
Department of Earth and Planetary Sciences, Tokyo Institute of Technology, Meguro, Tokyo, 152-8551, Japan
\and
Institute of Low Temperature Science, Hokkaido University, Kita, Sapporo 060-0819, Japan}
}
\abstract
 {
 Dust grains coagulate to form dust aggregates in protoplanetary disks.
 Their porosity can be extremely high in the disks.
 Although disk emission may come from fluffy dust aggregates, the emission has been modeled with compact grains.
}
{
We aim to reveal the mass opacity of fluffy aggregates from infrared to millimeter wavelengths with the filling factor ranging from 1 down to $10^{-4}$.
}
{
We use Mie calculations with an effective medium theory.
The monomers are assumed to be 0.1 ${\rm \mu m}$ sized grains, which is much shorter than the wavelengths that we focus on.
}
{
We find that the absorption mass opacity of fluffy aggregates are characterized by the product $a\times f$, where $a$ is the dust radius and $f$ is the filling factor, except for the interference structure.
The scattering mass opacity is also characterized by $af$ at short wavelengths while it is higher in more fluffy aggregates at long wavelengths.
We also derive the analytic formula of the mass opacity and find that it reproduces the Mie calculations.
We also calculate the expected difference of the emission between compact and fluffy aggregates in protoplanetary disks with a simple dust growth and drift model.
We find that compact grains and fluffy aggregates can be distinguished by the radial distribution of the opacity index $\beta$.
The previous observation of the radial distribution of $\beta$ is consistent with the fluffy case, but more observations are required to distinguish between fluffy or compact.
In addition, we find that the scattered light would be another way to distinguish between compact grains and fluffy aggregates.
}
{}

\keywords{planets and satellites: formation -- methods: numerical and analytical -- protoplanetary disks}
\maketitle

\section{introduction}
Optical properties of dust grains have been investigated by many authors to understand the emission from various kinds of astronomical objects.
In protoplanetary disks, dust grains are important not only as the emitter of radiation, but also as the seeds of planets.
The size of dust grains increases by coagulation from submicron size to millimeter size or larger.
A number of radio observations suggest that dust grains have been grown to millimeter-sized grains in protoplanetary disks \citep{AndrewsWilliams05, Isella09, Ricci10a, Ricci10b,Guilloteau11, vanderMarel13}.

The silicate feature at 10 ${\rm \mu m}$ is evidence of grain growth \citep[e.g.,][]{vanBoekel05}.
The infrared observations suggest that the size of silicate dust grains is spreading from $0.1 {\rm ~\mu m}$ to a few ${\rm \mu m}$.
The infrared emission is expected to come from the surface region of protoplanetary disks.
Tiny grains are kinematically well coupled to the disk gas and thus stirred up to the disk surface.
Thus, we cannot carry out information of dust grains larger than the micron size from infrared observations.
In addition, infrared scattered light images of protoplanetary disks are less luminous than expected from other observations. 
This may infer the presence of large compact grains or porous aggregates at the disk surface \citep{Mulders13}.

The opacity index at submillimeter wavelengths is used as another clue of grain growth \citep{Beckwith90, BeckwithSargent91,MiyakeNakagawa93}.
The most striking evidence of dust growth is the opacity index $\beta$, where $\kappa_{\nu}\propto \nu^{\beta}$; $\beta$ is estimated from observed flux slope $\alpha$, where $F_{\nu}\propto \nu^{\alpha}$.
If the dust emission is optically thin, the dust slope has a relation of $\beta=\alpha-2$.
The index $\beta$ is typically from 1 to 0 in protoplanetary disks, which means grain growth in protoplanetary disks \citep{AndrewsWilliams05, Lommen10, Perez12}.
The recent observations using radio interferometers have revealed the radial profile of $\beta$.
\citet{Perez12} made a model fit of $\beta$ and suggested that $\beta$ is different between in the inner and outer part of the disk.
Thus, the dust grains in the inner part of the disk are expected to grow to a larger size.

Although the protoplanetary disk emissions are usually modeled with compact dust grains, recent numerical simulations have shown that dust grains coagulate to form fluffy structure, especially in the case of icy dust aggregates.
With low speed collisions, dust grains form fluffy aggregates.
However, it has been shown that aggregate are not effective in compressing the fluffy dust aggregates.
\citet{Wada08} and \citet{Suyama08, Suyama12} investigated collisional compression of icy dust aggregates, and \citet{Okuzumi12} performed coagulation simulations including the collisional compression.
They revealed that the initial fractal growth stops when the collisional energy exceeds the rolling energy.
They derived that the achievable lowest filling factor is $\sim 10^{-5} (m_{\rm roll}/10^{-4} {\rm~g})$, where $m_{\rm roll}$ is the aggregate mass when the impact energy is equal to the rolling energy.
Moreover, \citet{Kataoka13a,Kataoka13b} introduced the static compression of dust aggregates.
They showed that the filling factor decreases to as low as $10^{-4}$ even when considering the effects of the static compression. 
However, the porosity evolution of icy dust aggregates has not been confirmed by laboratory experiments yet.

The icy and fluffy aggregates are expected to overcome theoretical problems in planetesimal formation.
Fluffy aggregates are expected to overcome the radial drift barrier \citep{Okuzumi12, Kataoka13b} and the bouncing barrier \citep{Wada11}.
Moreover, if particles are composed of ice, the dust aggregates overcome the fragmentation barrier because they are sticky \citep{Wada09,Wada13}. 

Dust coagulation has also been investigated in laboratory experiments.
As an analog to silicate dust grains, which are expected to be inside the snowline in protoplanetary disks, silica particles have been used in laboratory experiments.
Conditions for bouncing and fragmentation have been studied in laboratory experiments \citep{BlumWurm08, Zsom10} and some scenarios for planetesimal formation breaking through the bouncing barrier have been proposed \citep{Windmark12, Drazkowska13}.
From the viewpoint of porosity evolution, silicate dust aggregates are expected to be less fluffy than icy dust aggregates because the surface energy of silicate is lower than ice.
Microgravity experiments have confirmed the hit-and-stick process of forming fluffy dust aggregates \citep{Kothe13}.
However, further growth concerning compression is still uncertain in laboratory experiments.
\citet{Zsom11b} performed numerical simulations of dust coagulation of silicate particles, using the hit-and-stick model proposed by \citet{Okuzumi+09}.
They showed that the filling factor of dust aggregates can reach $10^{-3}$ before the onset of compaction.
\footnote{
\citet{Zsom10} obtained less fluffy aggregates than \citet{Zsom11b} because \citet{Zsom10} adopted the porosity model proposed by \citet{Ormel07}, which is not as accurate as the model of \citet{Okuzumi+09}.
}


Observational constraints of porosity of dust aggregates in protoplanetary disks are important.
However, studies of interpreting disk observations have assumed $f\ge 0.1$ \citep[e.g.,][]{Birnstiel10b}, which is relatively compact compared with the extremely porous aggregates, whose filling factor is $10^{-4}$, as discussed above.
In this paper, as a first step to constrain the porosity of dust aggregates in protoplanetary disks, we investigate optical properties of dust aggregates including the extremely porous aggregates.

Opacity of porous aggregates has been investigated by several theoretical methods.
In the context of explaining cometary dust, scattering properties of BPCA and BCCA aggregates have been studied \citep{Kimura03, Kimura06, Kolokolova07}.
The number of constituent particles was limited to $\sim 60000$ ($\sim 10^{-10}$ g in mass if the particle size is 0.1 ${\rm \mu m}$), and the opacity was only studied at infrared wavelengths.
In the context of explaining the interstellar silicate feature, in addition, the effects of monomer shapes on optical properties at infrared wavelength have been also studied \citep{Min03,Min05,Min07}.
In this paper, we examine the absorption and scattering mass opacities of dust aggregates at wavelengths ranging from $1{\rm~\mu m}$ to 10 cm.
The aggregates have a size ranging from micron to kilometer and a filling factor ranging from 1 to $10^{-4}$. 

One of the popular methods for calculating the mass opacity of porous aggregates is the discrete dipole approximation (DDA) \citep{DraineFlatau94,Min06}.
This calculation takes a huge computational time for large aggregates.
To investigate the opacities of highly porous aggregates for a wide size range, the method would not be suitable.
In this paper, we aim to reveal the mass opacity of fluffy aggregates from infrared to millimeter wavelengths with the filling factor ranging from 1 down to $10^{-4}$.
Thus, we use the effective medium theory (EMT).
This method is fast in calculation but inaccurate in some parameters.
\citet{Kozasa92} have shown that EMT reproduces the absorption opacity of BCCA and BPCA clusters, whose constituent monomers are up to 1024, within a error of a factor of two.
The EMT is also known to be accurate for porous aggregates whose constituent particles are small compared with the wavelength of incident radiation \citep{Voshchinnikov05, Shen08}.
Because the dust aggregates considered in this paper are highly porous aggregates consisting of submicron-sized monomers, EMT would be a good approximation for calculations in this paper.
We note that the scattering opacity derived with EMT largely deviates from the actual value in some parameter space \citep{Shen09}.
The accuracy of EMT in a large parameter space should be tested in the future work.

This paper is organized as follows.
We describe the composition of dust grains and the calculating method of mass opacities in Section \ref{sec:method}.
We show the results of the absorption and scattering mass opacities of highly porous aggregates by using Mie theory with EMT in Section \ref{sec:result}.
We derive analytic formulae to reproduce the results in Section \ref{sec:piecewise}.
Then, we construct a simple dust growth and drift model in protoplanetary disks and propose a method to distinguish compact and fluffy aggregates in radio observations by using the slope at millimeter wavelengths, the so-called dust $\beta$, in Section \ref{sec:disk}.
Finally, we summarize and discuss the previous observations with porous aggregates in Section \ref{sec:summary}.

\section{Method}
\label{sec:method}
Here, we briefly summarize the definitions of optical properties following \citet[hereafter BH83]{BohrenHuffman83} and \citet{MiyakeNakagawa93}.
We consider a particle or an aggregate with radius $a$ and internal mass density $\rho$.
The radius of an aggregate represents the characteristic radius, defined as $a=\sqrt{5/3}a_{\rm g}$, where $a_{\rm g}$ is the gyration radius \citep{Mukai92}. 
We define the size parameter $x$ as
\begin{equation}
x=\frac{2\pi a}{\lambda},
\end{equation}
where $\lambda$ is the wavelength.
We also define the relative refractive index $m$ as
\begin{equation}
m=n+ik,
\end{equation}
where $n$ and $k$ are the real and imaginary parts of the refractive index, respectively.

\subsection{Dust grains: monomers}
We consider a dust aggregate, which consists of a number of monomers.
The monomers are assumed to be composed of silicate, organics, and water ice without any void structure.
The mass fractional abundance is set to be consistent with \citet{Pollack94}, where $\zeta_{\rm silicate}=2.64 \times 10^{-3}$, $\zeta_{\rm organics}=3.53 \times 10^{-3}$, and $\zeta_{\rm ice}=5.55 \times 10^{-3}$.
The internal densities of silicate, organics, and ice are taken to be $3.5 {\rm ~g~cm^{-3}}$, $1.5 {\rm ~g~cm^{-3}}$, and $0.92 {\rm ~g~cm^{-3}}$, respectively.
The mean internal density is therefore $1.68 {\rm ~g~cm^{-3}}$.
The resultant volume fractions are 8 \% of silicate, 26 \% of organics, and 66 \% of water ice.
We use the refractive index of astronomical silicate from \citet{WeingartnerDraine01}, organics from \citet{Pollack94}, and water ice from \citet{Warren84}.

Here, we discuss the structure of a monomer composed of various materials.
We assume that the monomer has a core-mantle structure, where silicate components are inside and ice and organics cover the silicate core.
This assumption is reasonable because the condensation temperature of silicates is much higher than those of ices and organics.
\footnote{
The material properties of organics, such as surface energy and Young's modulus, are still uncertain, but it is considered to be similar to those of ice \citep[e.g.,][]{Kudo02}.
}
The collisional and static compression and fragmentation velocity are determined by the surface material of monomers, which is expected to be ice or organics.
Therefore, the structure and fluffiness of the aggregates are expected to be similar to icy aggregates \citep{Okuzumi12, Kataoka13b}.

The effective refractive index of the mixture can be derived from dielectric functions.
By using Maxwell-Garnett rule without voids, the effective dielectric function is obtained as
\begin{equation}
\epsilon_{\rm mix}=\frac{\Sigma f_{j}\gamma_{j} \epsilon_{j}}{\Sigma f_{j}\gamma_{j}},
\end{equation}
where
\begin{equation}
\gamma_{j}=\frac{3}{\epsilon_{j}+2},
\end{equation}
and $f_{j}$ and $\epsilon_{j}$ represent the volume filling factor and the dielectric function of each species.
The dielectric function is related to the refractive index as $\epsilon=m^{2}$.

\begin{figure}[htbp]
 \begin{center}
  \includegraphics[width=80mm]{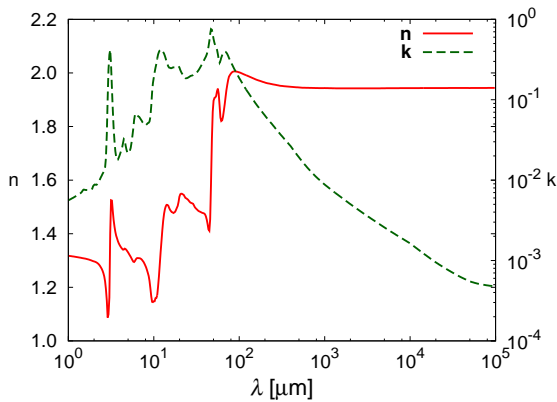}
 \end{center}
 \caption{
 The complex refractive index of the mixture of silicate, organics, and water ice.
 }
 \label{fig:check_ref}
\end{figure}

Figure \ref{fig:check_ref} shows the real and imaginary part of the effective refractive index of the mixture.
We use this effective refractive index as the material refractive index in the following discussion. 

The simplified fractional abundance based on \citet{Pollack94} has been widely used in several papers \citep[e.g.,][]{DAlessio01, Tanaka05, Isella09, Ricci10a, Ricci10b, Perez12}.
Some recent studies use the dielectric functions of carbonaceous material \citep{Zubko96} instead of organics \citep{Pollack94,LiGreenberg97}.
In protoplanetary disks, the carbonaceous materials would interact with other species to produce organics.
Thus, we use the dielectric function of organics based on \citet{Pollack94} in this paper.
However, the optical properties of organics in protoplanetary disks have large uncertainties because astronomical organics may be different from laboratory data.

\subsection{Aggregates of monomers}
To calculate the opacity of fluffy aggregates, we use the effective medium theory again.
In the case of the mixture of monomers and voids, Maxwell-Garnett theory is applicable to obtain the effective dielectric function as
\begin{equation}
\epsilon_{\rm eff}=\frac{1+2fF}{1-fF},
\label{eq:eps_eff}
\end{equation}
where
\begin{equation}
F=\frac{\epsilon_{\rm mix}-1}{\epsilon_{\rm mix}+2},
\end{equation}
$\epsilon_{\rm mix}$ is the effective dielectric function of the mixture, and $f$ is the volume filling factor of the aggregate.

We will investigate the mass opacity of dust aggregates for a wide range of the dust radius $a$ and the filling factor $f$.
We adopt the Mie calculation with the effective medium theory described above.
\citet{Voshchinnikov05} show that the EMT is a good approximation when the inclusions are smaller than the wavelengths of radiation.
Here, the monomer size is $0.1{\rm ~\mu m}$ while the wavelengths are larger than 1 ${\rm \mu m}$.
Thus, the EMT would be a good approximation in the calculations in this paper.
The filling factor is expected to decrease to $f\sim 10^{-4}$ and the dust radius grows from micron to kilometer \citep{Kataoka13b}.
Therefore, we will investigate the mass opacity in such parameter space.

We note that we do not choose a set of $a$ and $f$ where both $a$ and $f$ are too small.
In the porosity evolution scenario proposed by \citet{Kataoka13b}, the dust aggregates grow as fractals in the very early stage of the coagulation.
This stage corresponds to the lower limit of $a$ and $f$.
In this paper, we consider a set of  $a$ and $f$ where $af \geq 0.1 {\rm \mu m}$.

\subsection{Mass opacity}
We use the Mie calculation with the effective medium theory to calculate dimensionless absorption and scattering coefficients $Q_{\rm abs}$ and $Q_{\rm sca}$.
Then, we obtain absorption and scattering mass opacities defined as
\begin{equation}
\kappa_{\rm abs}=\frac{\pi a^2}{m}Q_{\rm abs}, 
\end{equation}
\begin{equation}
\kappa_{\rm sca}=\frac{\pi a^2}{m}Q_{\rm sca}.
\end{equation}
We note that the mass opacities are given per gram of dust.
To obtain the mass opacities per gram of gas, one should divide the mass opacities by the dust-to-gas mass ratio.

\section{Results}
\label{sec:result}
\subsection{Absorption mass opacity}
The absorption mass opacity of porous dust aggregates strongly depends on their size and filling factor.
In protoplanetary disks, radio emission at millimeter wavelengths provide optically thin emission, in other words, directly reflects the opacity.
Therefore, we aim to reveal what properties of dust aggregates determine the mass opacity.

\begin{figure*}
 \begin{center}
 	\subfigure{
 	 \includegraphics[width=80mm]{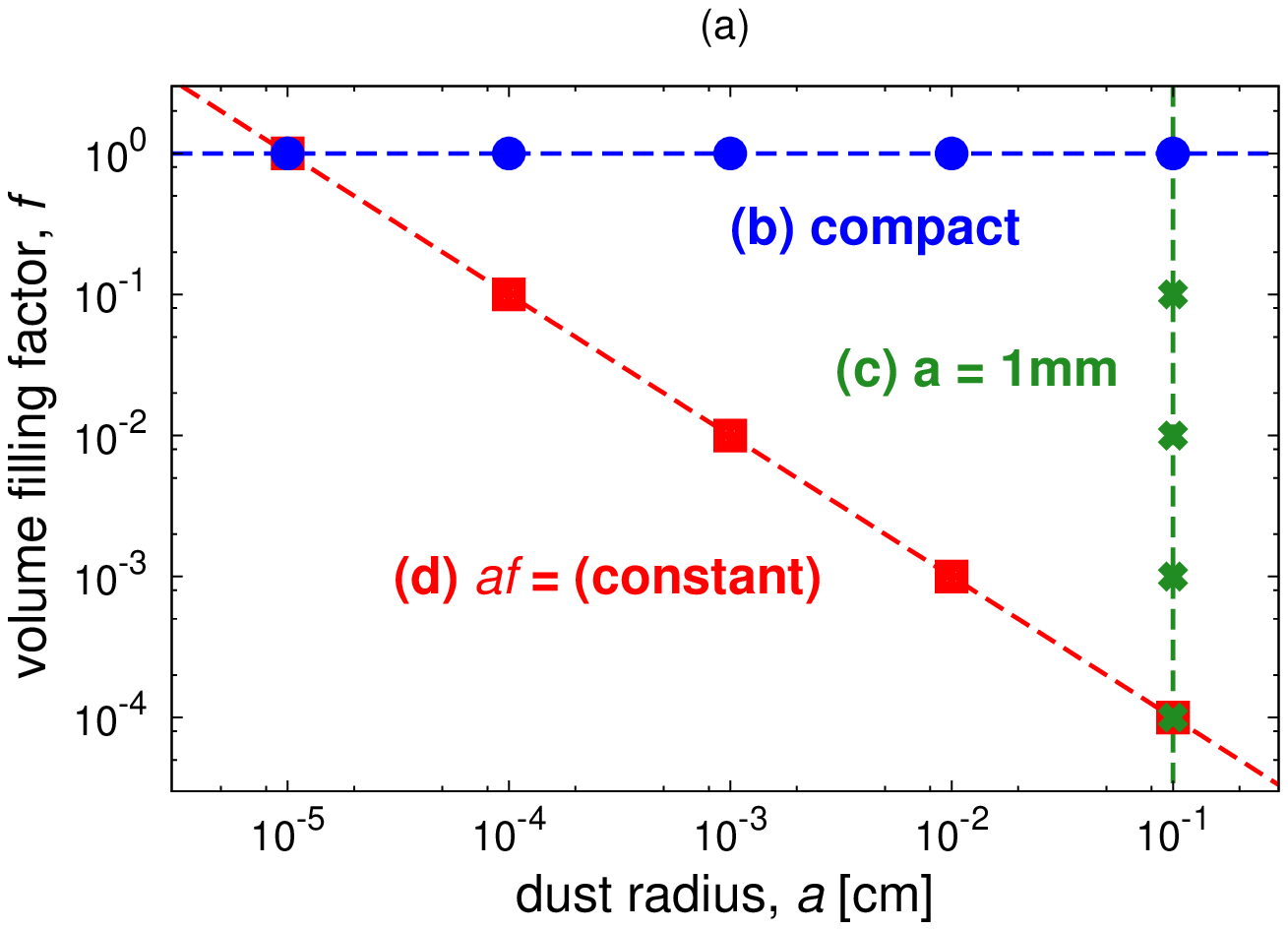}
	}
 	\subfigure{
 	 \includegraphics[width=80mm]{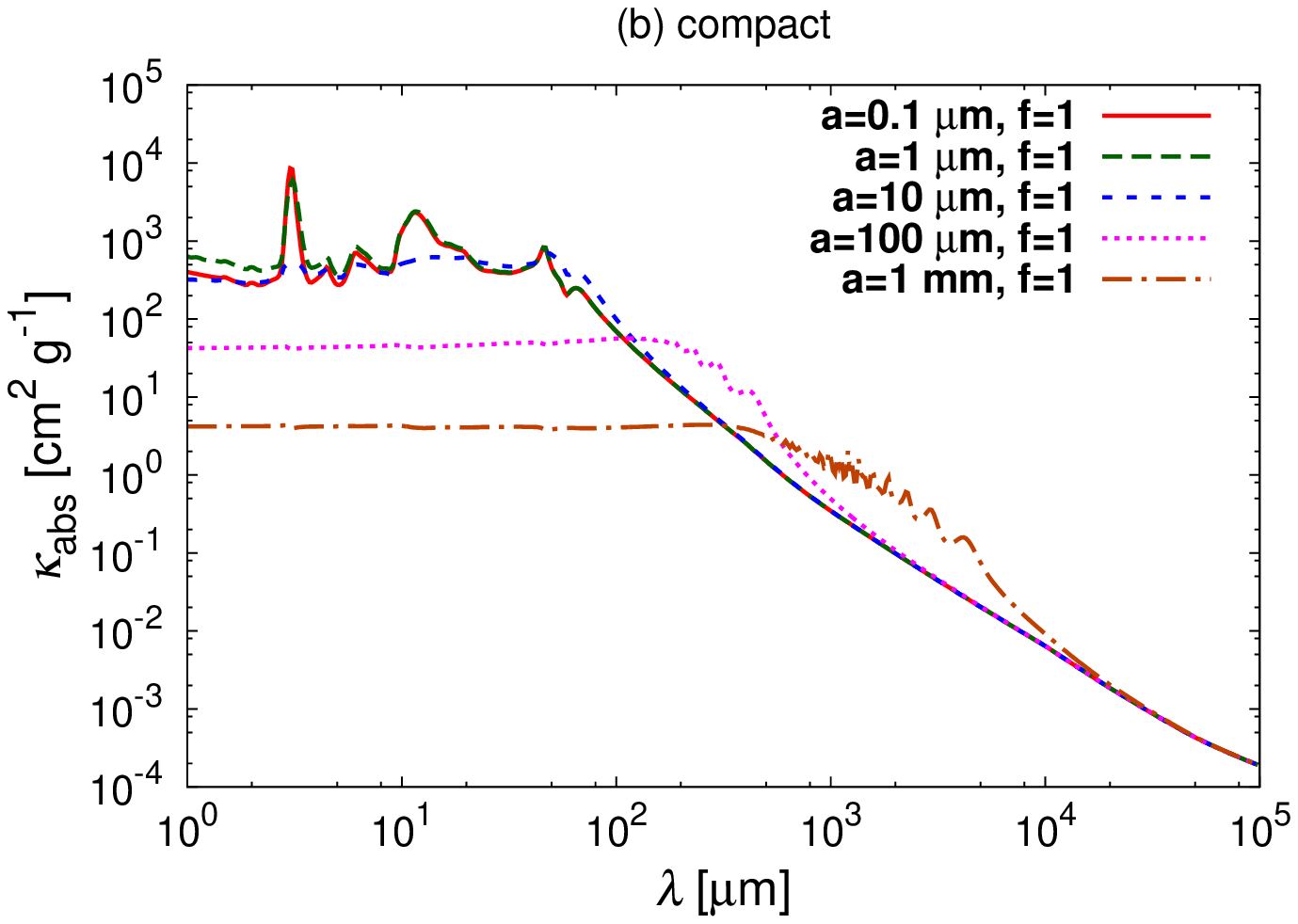}
	}\\
 	\subfigure{
 	 \includegraphics[width=80mm]{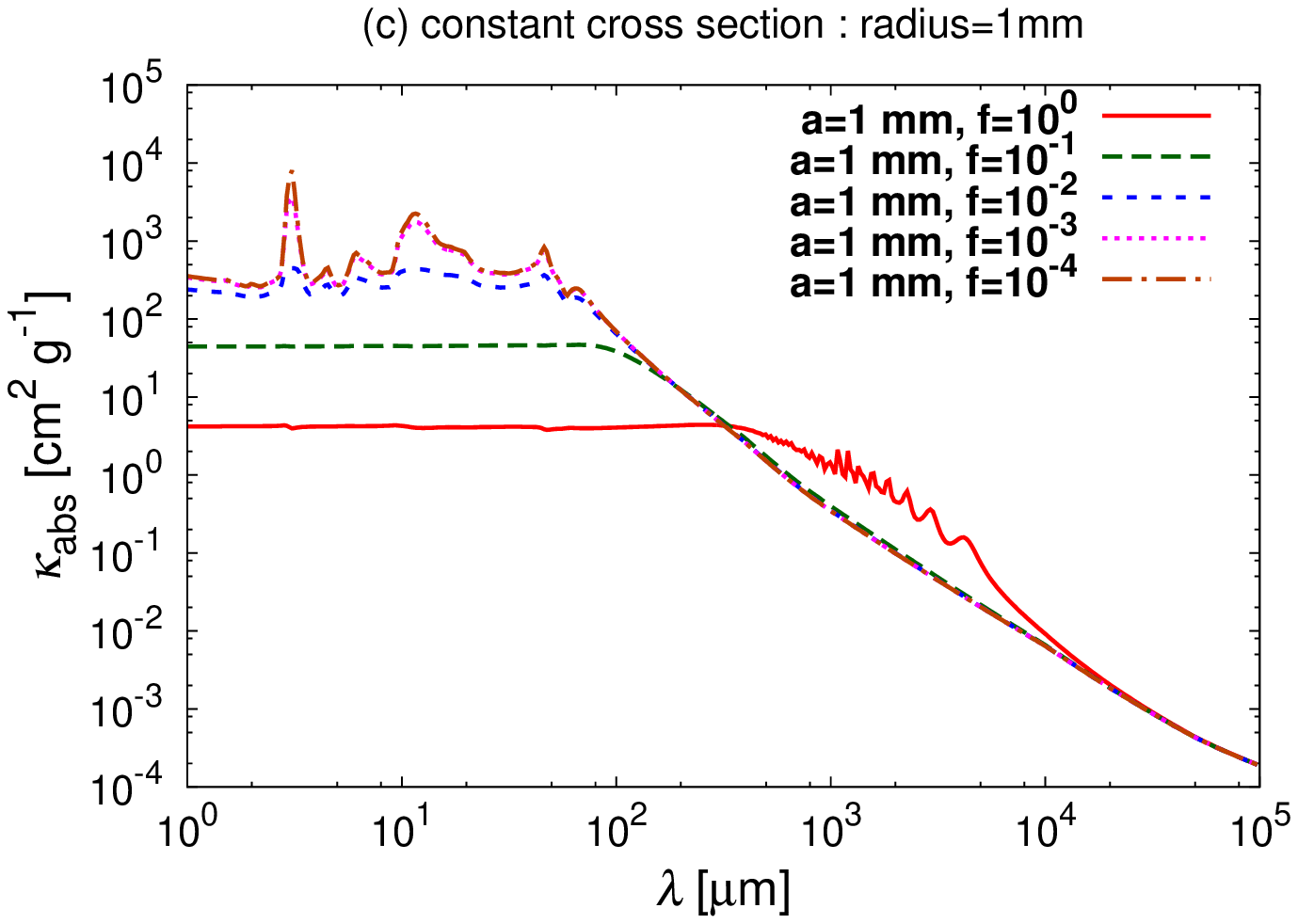}
	}
 	\subfigure{
 	 \includegraphics[width=80mm]{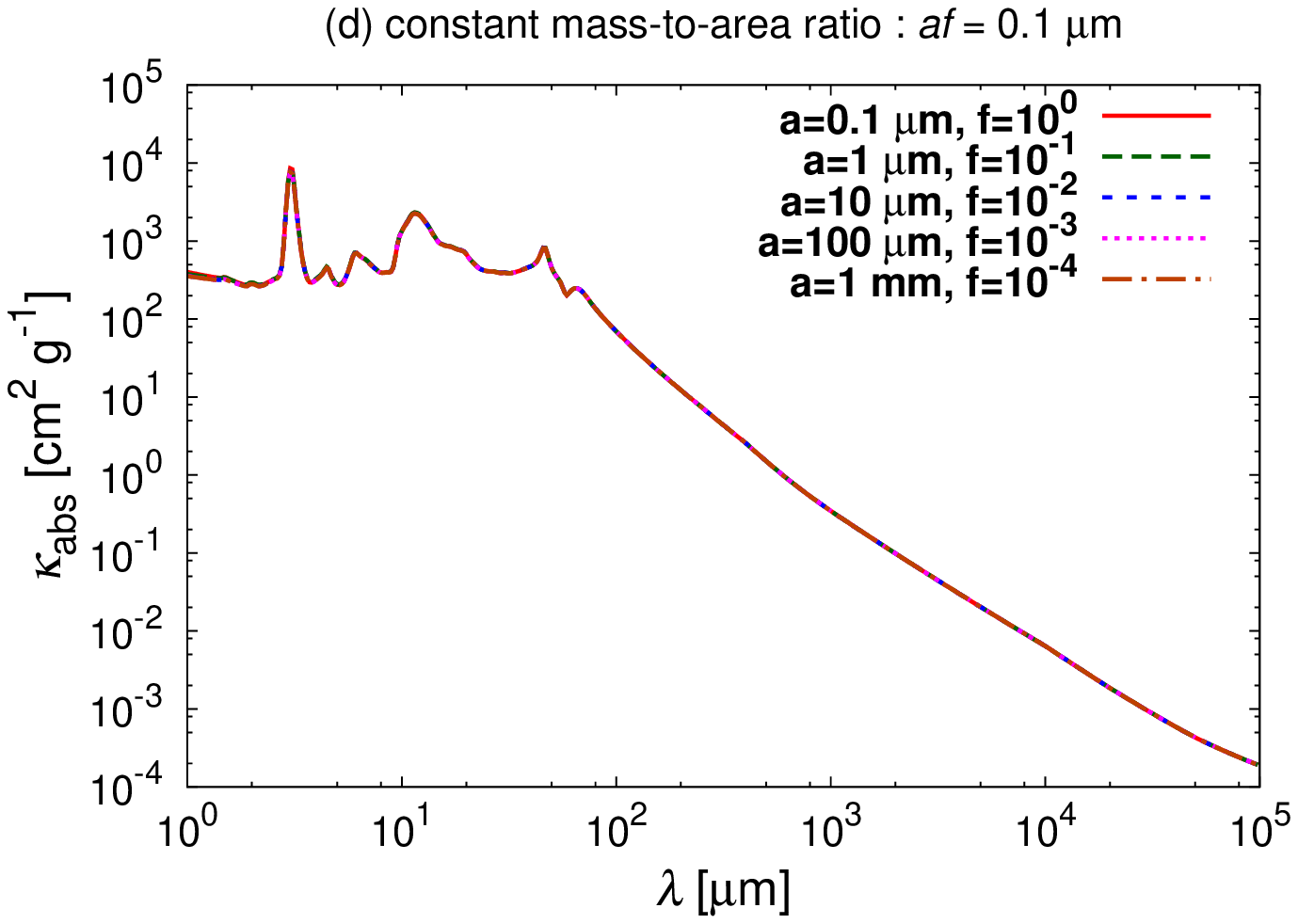}
	}
 \end{center}
 \caption{
 Absorption mass opacities for each dust radius and filling factor.
 (a) The parameter space in the volume filling factor $f$ against the dust radius $a$ to investigate the mass opacity.
 The dotted lines correspond to (b) (c) and (d).
 (b) The absorption mass opacity when the dust radius $a$ changes while $f=1$.
 (c) When the dust filling factor changes while $a=1$ mm.
 (d) When both $a$ and $f$ change while $af$ keeps constant.
 The constant $af$ corresponds to the same mass-to-area ratio of the dust aggregates.
 }
 \label{fig:abs_general}
\end{figure*}
Figure \ref{fig:abs_general} shows the dependency of the mass opacities of dust aggregates on the dust radius $a$ and the filling factor $f$.
Figure  \ref{fig:abs_general} (b) shows the mass opacity of different dust radius while the filling factor is fixed at unity (i.e., compact growth).
The absorption mass opacity from optical to infrared wavelengths decreases as the aggregate size increases, and it has an enhancement because of the interference at the millimeter wavelengths depending on the aggregate size.
This trend is well-known as the grain-growth effects on the dust opacity.
Figure  \ref{fig:abs_general} (c) shows the mass opacity of different filling factors while the dust radius is fixed at $a=1$ mm.
The mass opacity strongly depends on the filling factor but cannot be characterized by one parameter.
Figure \ref{fig:abs_general} (d) shows that the mass opacity when both $a$ and $f$ change but $af$ remains constant.
The mass opacity is almost the same in this case.
This result suggests that the optical properties of fluffy dust aggregates are characterized by $af$.

Here, we introduce a new parameter $af$, where $a$ is the dust radius and $f$ is the filling factor.
Keeping $af$ constant corresponds to the constant mass-to-area ratio of the dust aggregates because (mass-to-area ratio) $\sim (a^{3}f) / (a^2) \sim af$.
We already find that the mass opacity is characterized by $af$ in the case of $af=0.1{\rm~ \mu m}$ in Fig. \ref{fig:abs_general} (d).
Next, we investigate whether the mass opacity is characterized by one parameter $af$ in the wide range of $af$.

\begin{figure*}
 \begin{center}
 	\subfigure{
 	 \includegraphics[width=80mm]{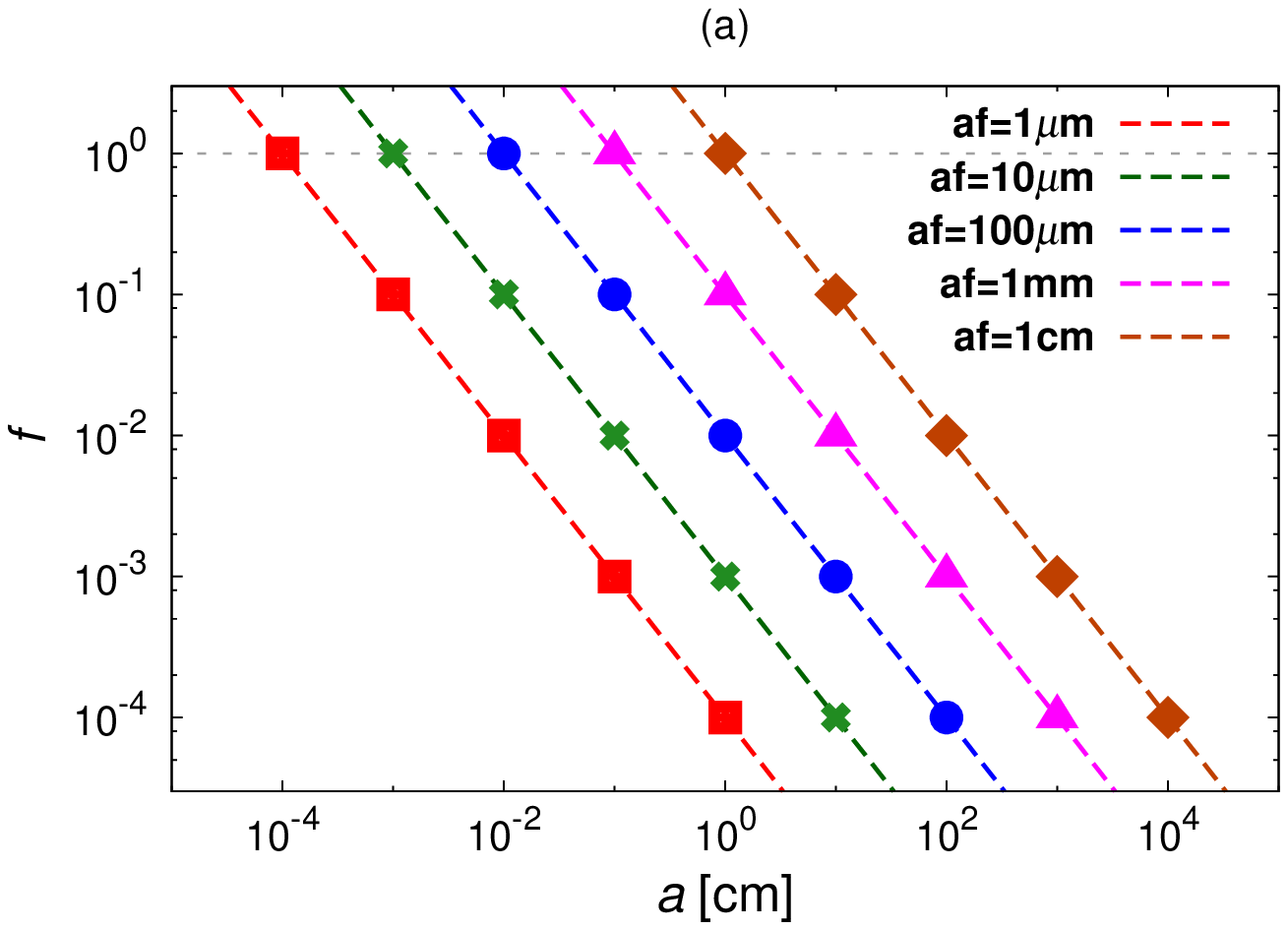}
	}
 	\subfigure{
 	 \includegraphics[width=80mm]{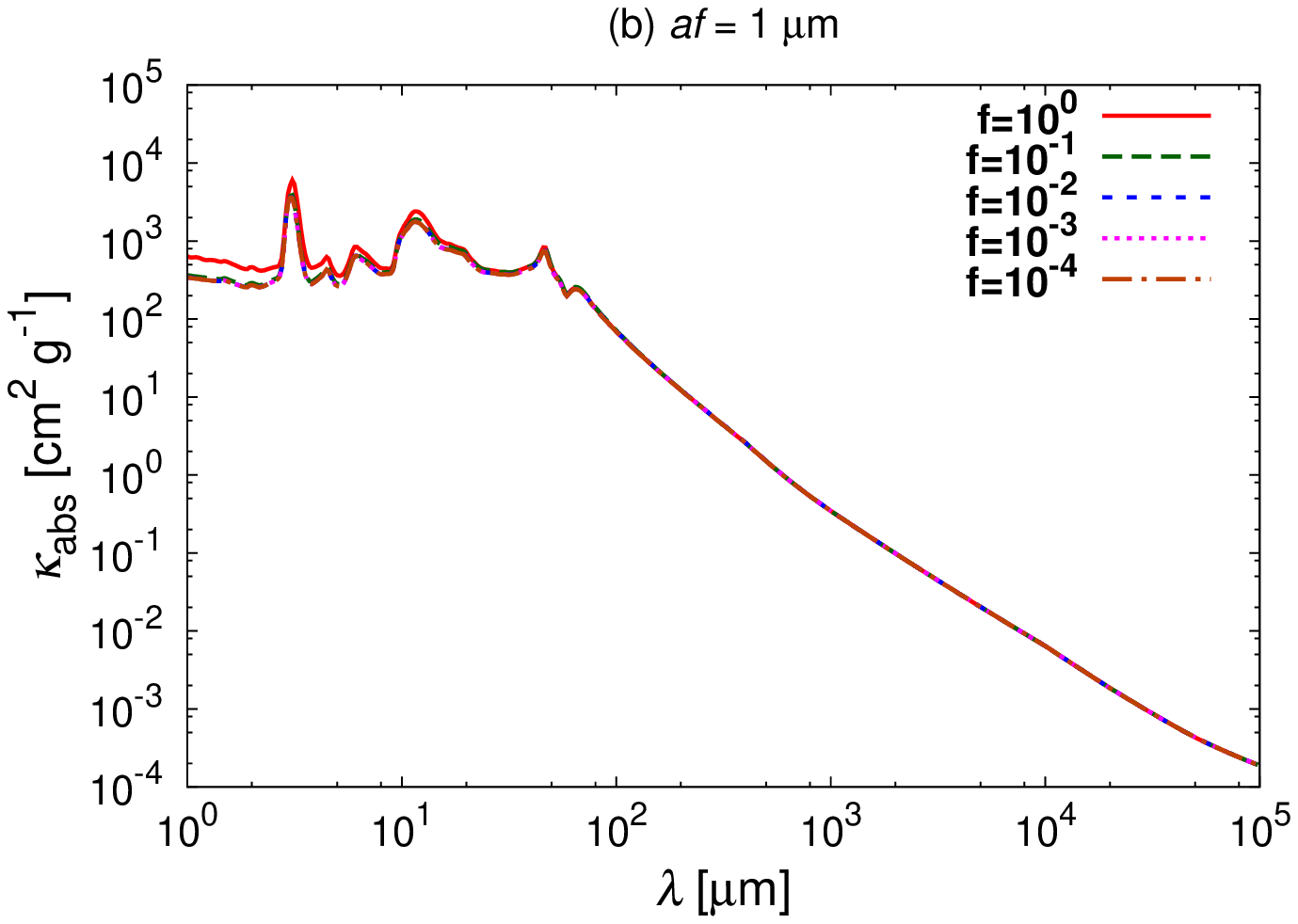}
	}\\
 	\subfigure{
 	 \includegraphics[width=80mm]{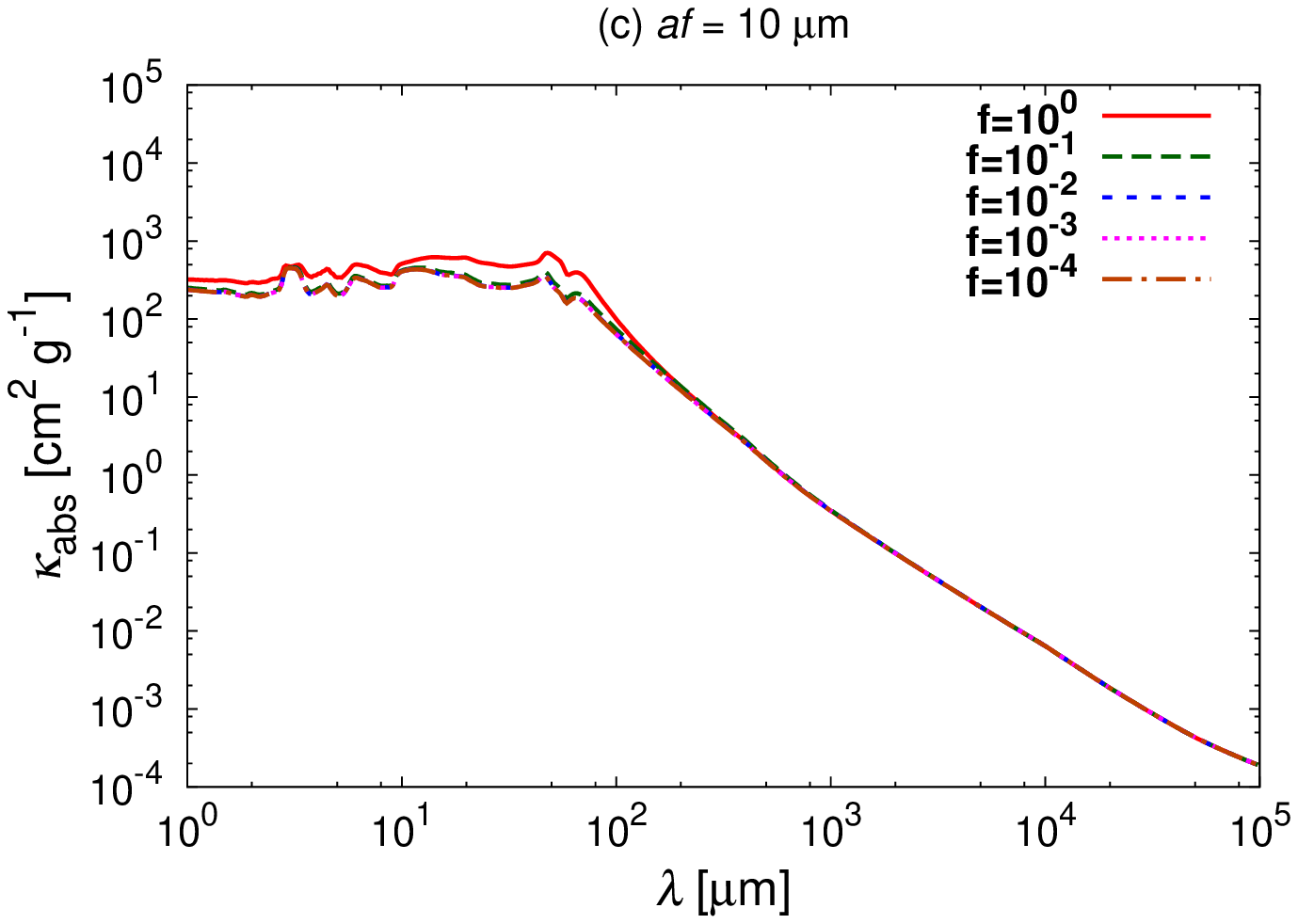}
	}
 	\subfigure{
 	 \includegraphics[width=80mm]{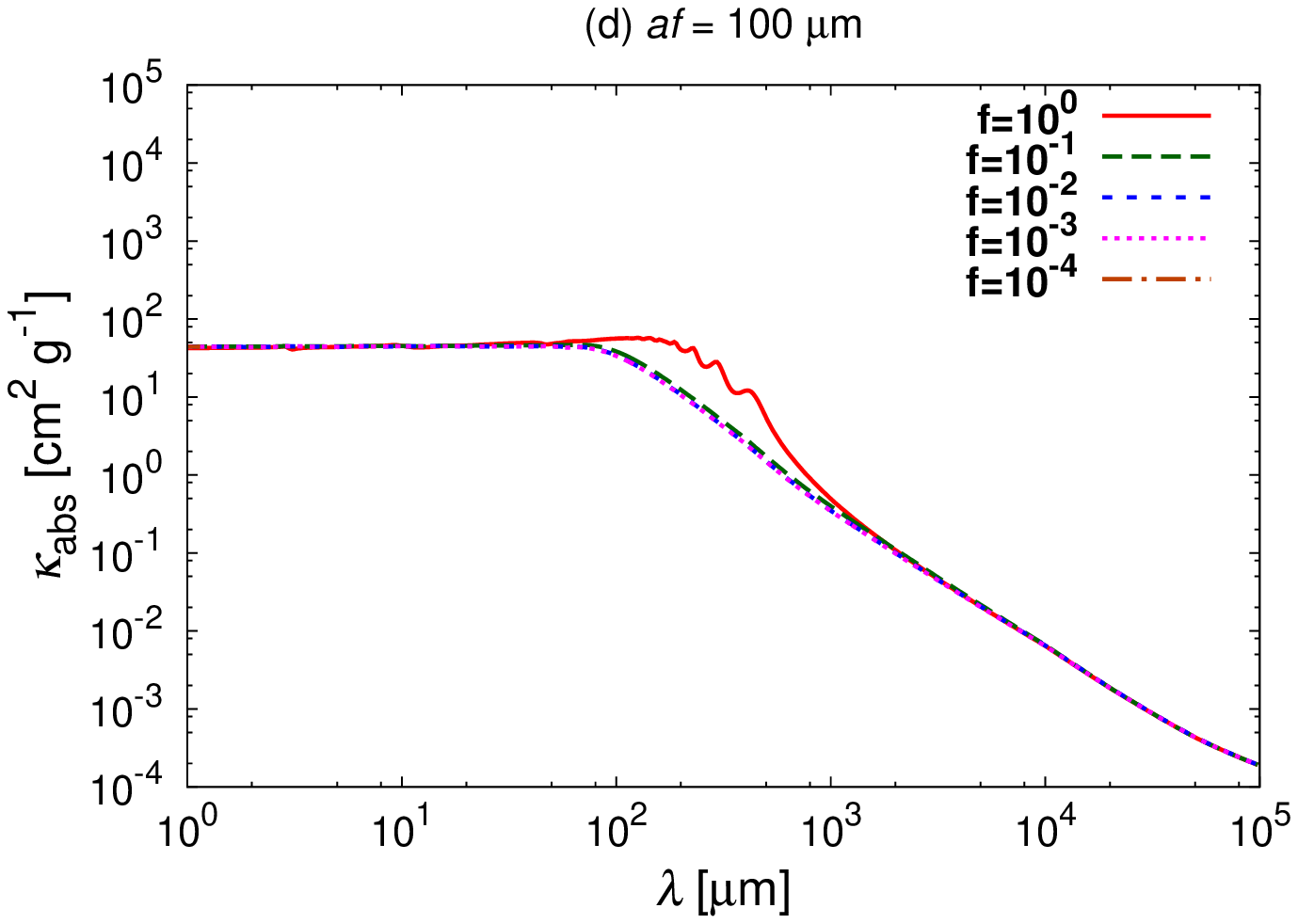}
	}\\
 	\subfigure{
 	 \includegraphics[width=80mm]{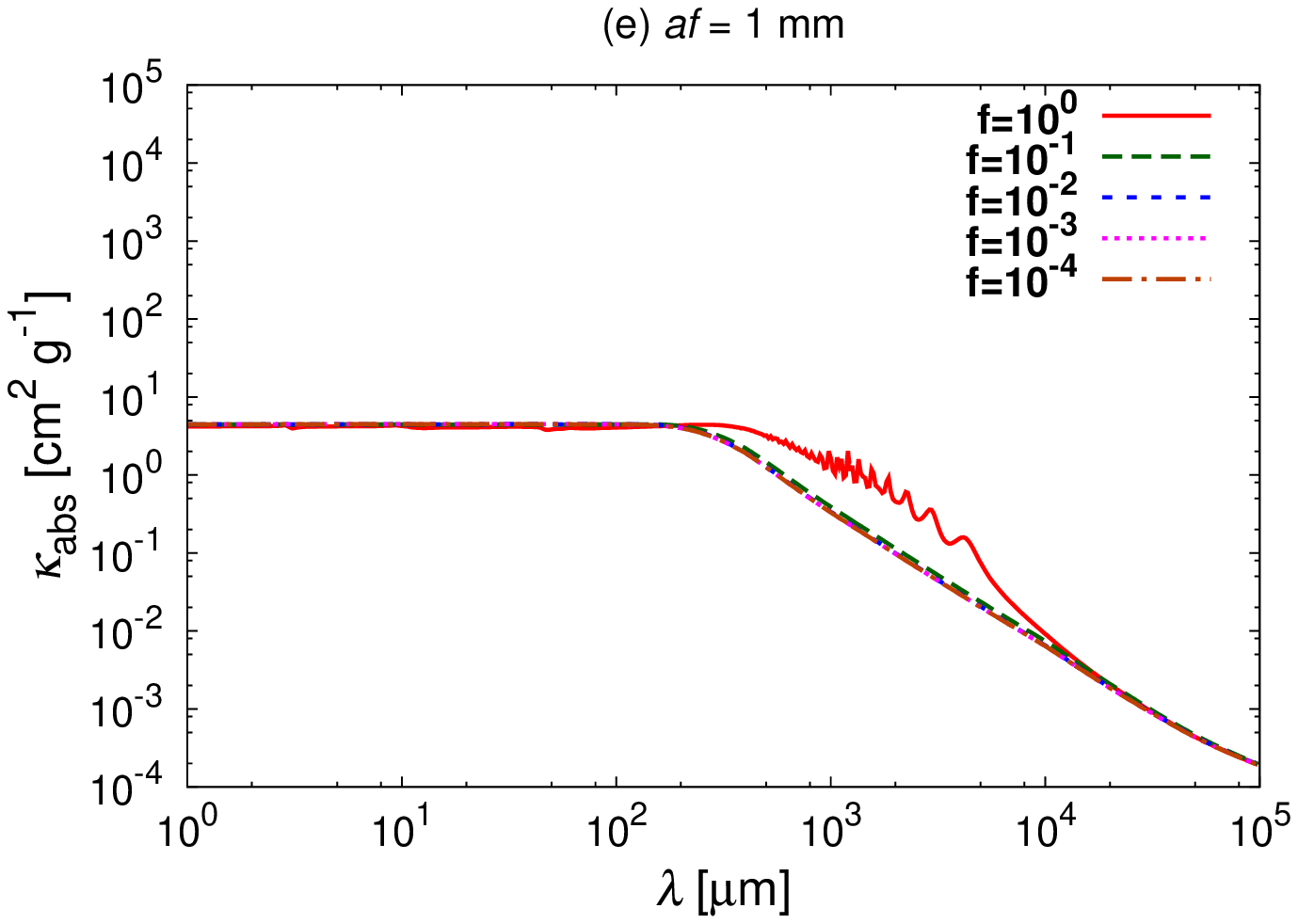}
	}
 	\subfigure{
 	 \includegraphics[width=80mm]{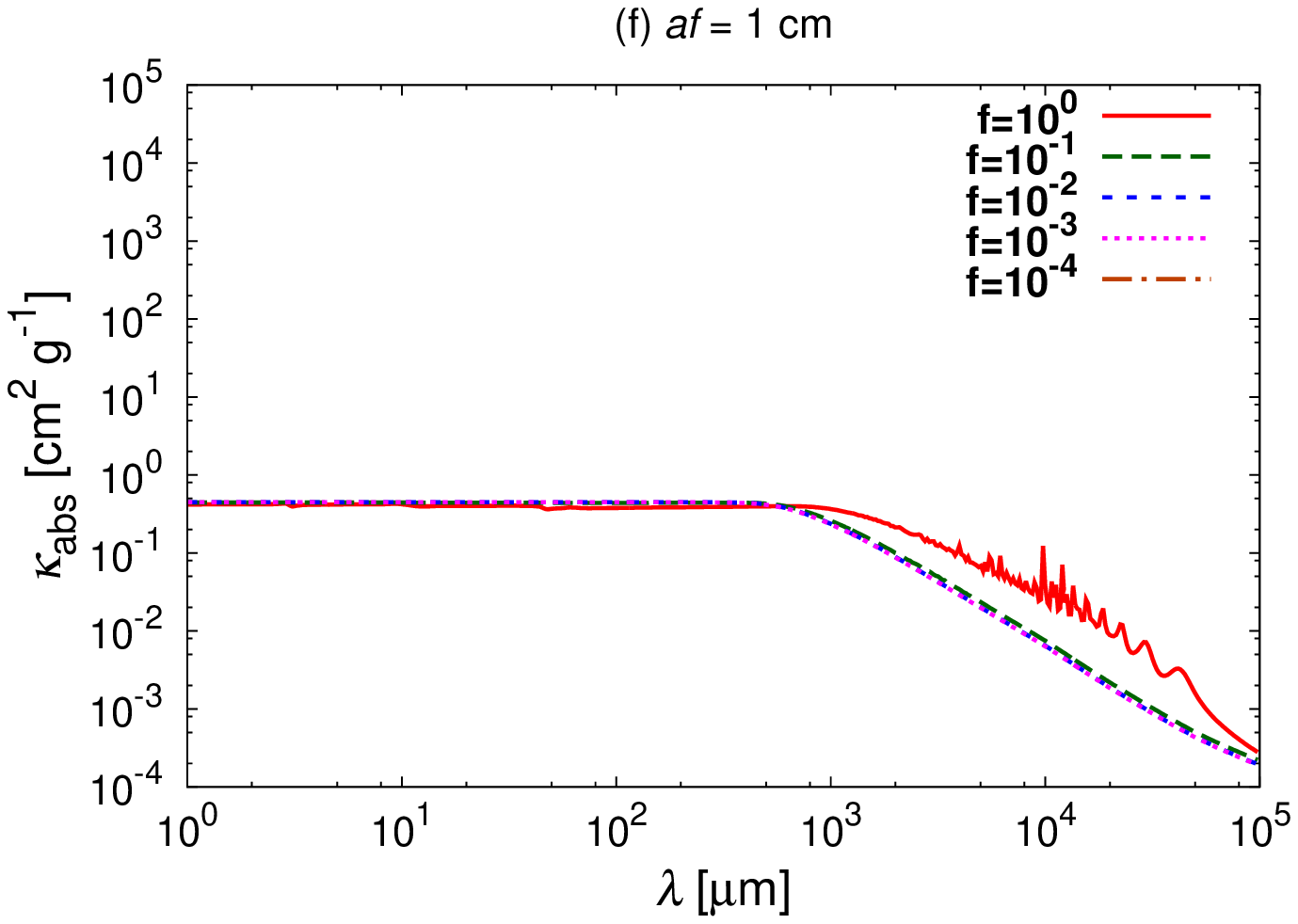}
	}
 \end{center}
 \caption{
 The absorption mass opacity in cases of different $af$.
 Panel (a) shows the parameter space of $f$ and $a$ in the cases of (b) to (f).
 The mass opacities are shown in the cases of (b) $af=1{~\rm \mu m}$, (c) $af=10{~\rm \mu m}$, (d) $af=100{~\rm \mu m}$, (e) $af=1{~\rm mm}$, and (f) $af=10{~\rm mm}$.
 }
 \label{fig:abs_dif}
\end{figure*}

Figure \ref{fig:abs_dif} shows the mass opacity where $af$ is fixed for each panel: $af$ has values of $1{~\rm \mu m}$, $10{~\rm \mu m}$, $100{~\rm \mu m}$, $1 {~\rm mm}$, and $1 {~\rm cm}$.
This figure clearly shows that the absorption mass opacity is almost the same in the cases of the same $af$.
For example, dust aggregates that have a size of 10 m and a filling factor of $10^{-4}$ are optically equivalent to 1 mm compact grains except for the interference structure.
We will show the reason why the absorption mass opacity is characterized by $af$ in Section \ref{sec:piecewise}.

That the absorption mass opacity is characterized by $af$ is a very naive result for observations: the filling factor cannot be measured because it is degenerated with the dust radius.
Therefore, to derive both the filling factor and the dust radius separately, we should find another clue rather than the general behavior of the absorption mass opacity.
We find two differences which might be ways to distinguish between $a$ and $f$: the interference structure of the absorption mass opacity and the scattering mass opacity at long wavelengths.
We discuss the interference first and will discuss the scattering mass opacity later in this section.

The only difference in the absorption mass opacity between the compact and fluffy cases if $af$ is the same is the interference structure, which appears when the size parameter $x$ is close to unity.
In the case of $af=$1 mm, for example, the absorption mass opacity in the compact case is one order of magnitude higher than the fluffy cases.
This is a way to distinguish between compact grains and fluffy aggregates in protoplanetary disks.
We will discuss the reason why the interference structure is unique only in the compact case in Section \ref{sec:piecewise}, and also discuss the feature as a way to distinguish between compact and fluffy aggregates by using the dust opacity index $\beta$ in Section \ref{sec:disk}.

\subsection{Scattering mass opacity}
\begin{figure*}
 \begin{center}
 	\subfigure{
 	 \includegraphics[width=80mm]{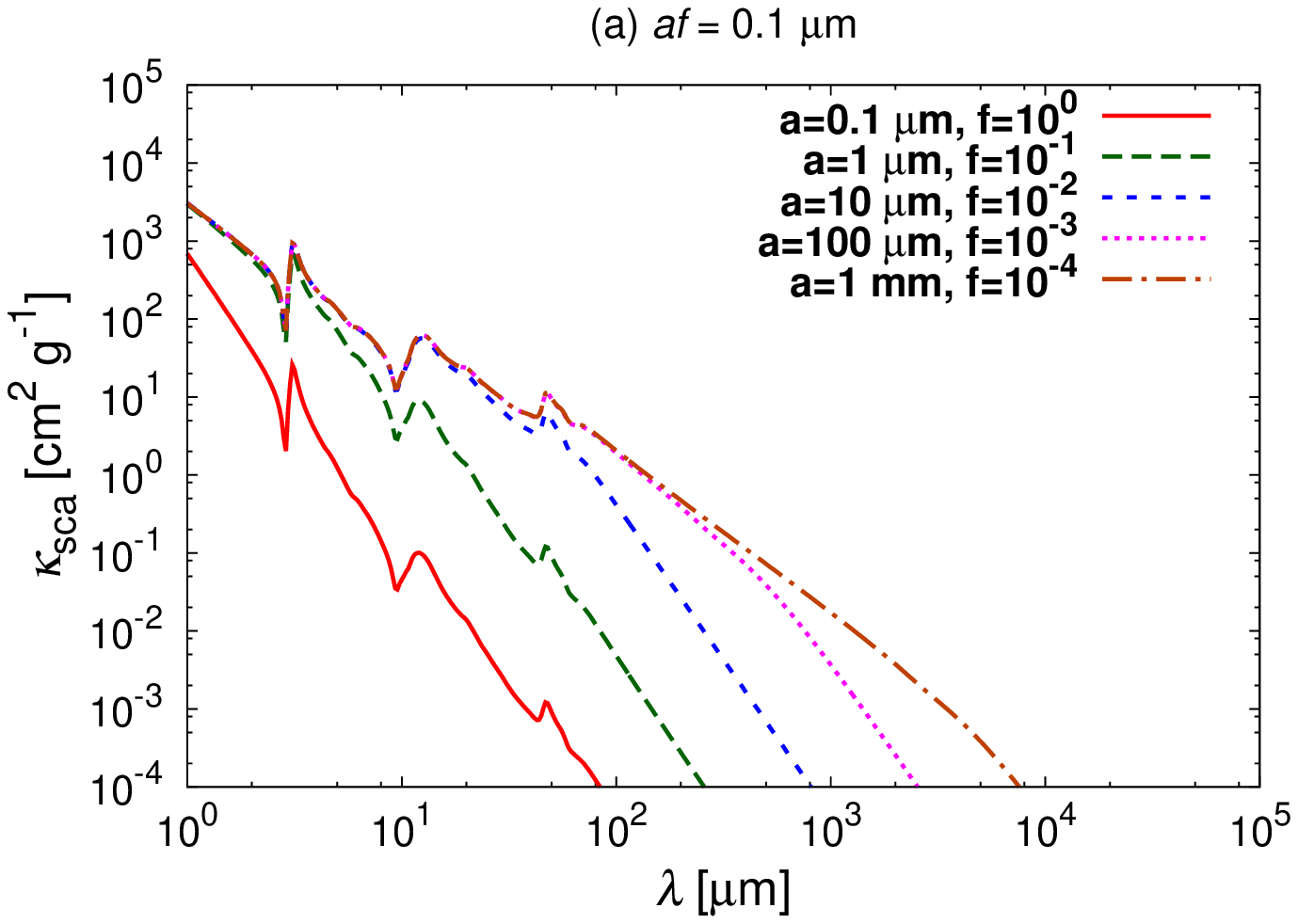}
	}
 	\subfigure{
 	 \includegraphics[width=80mm]{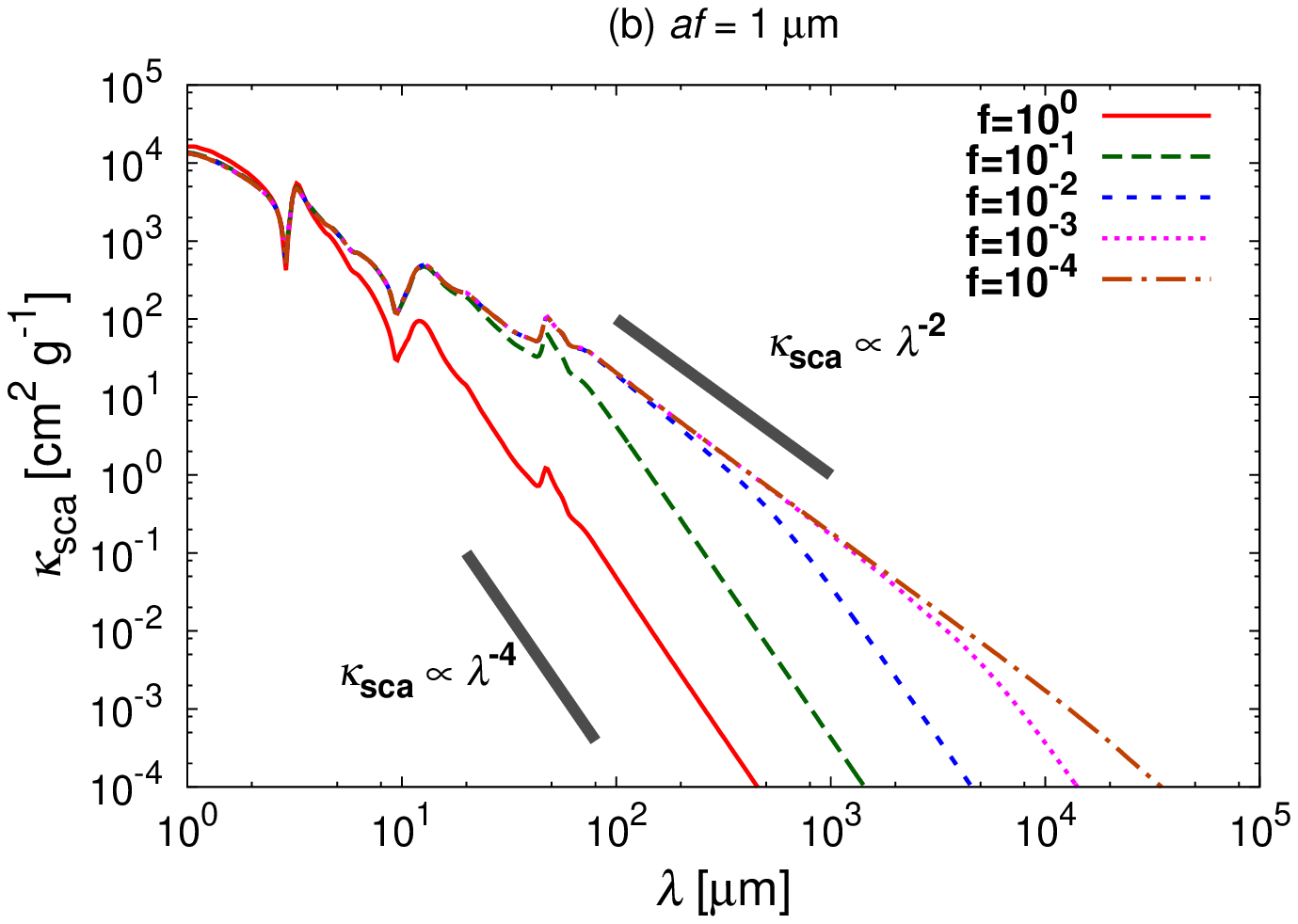}
	}\\
 	\subfigure{
 	 \includegraphics[width=80mm]{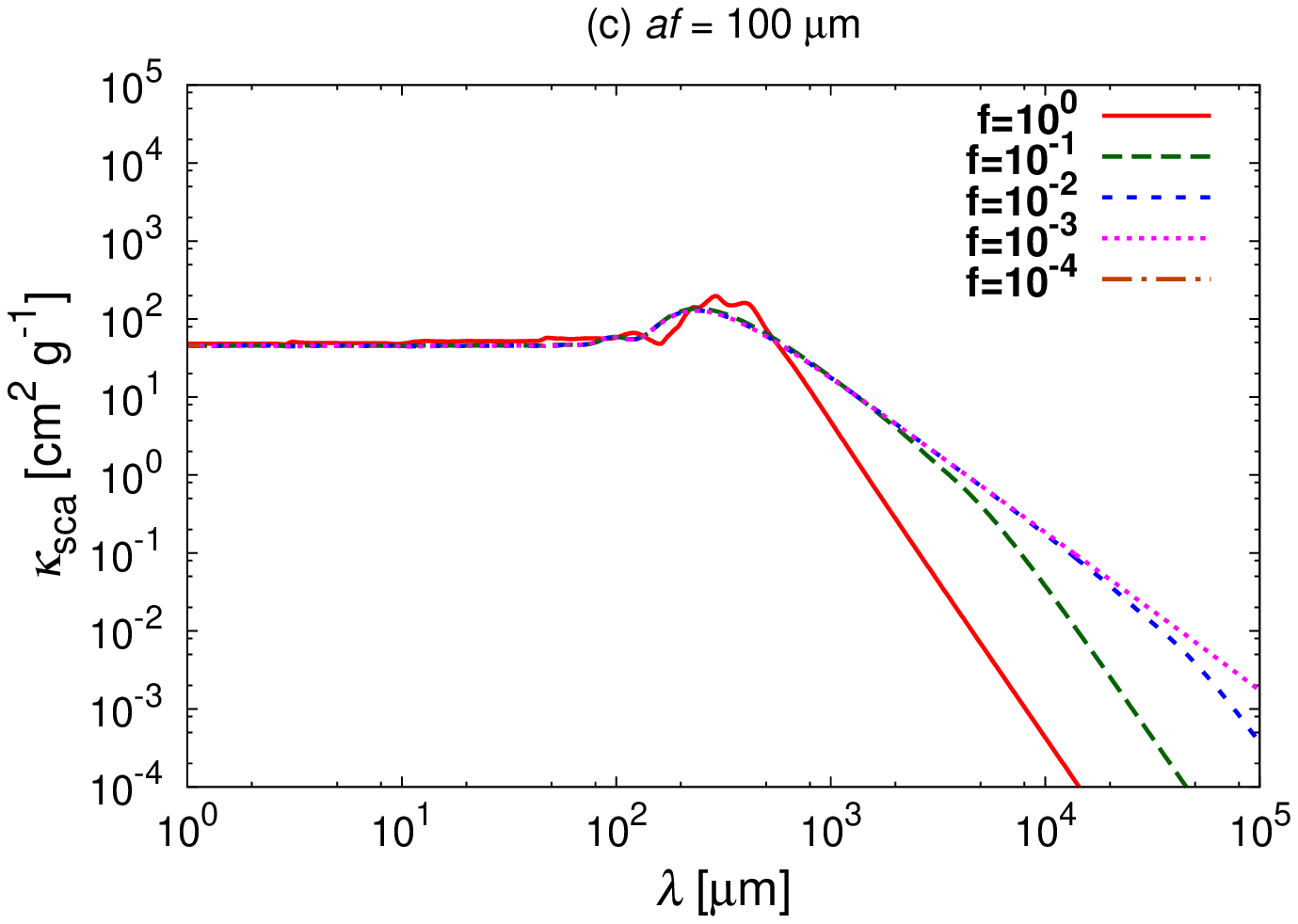}
	}
 	\subfigure{
 	 \includegraphics[width=80mm]{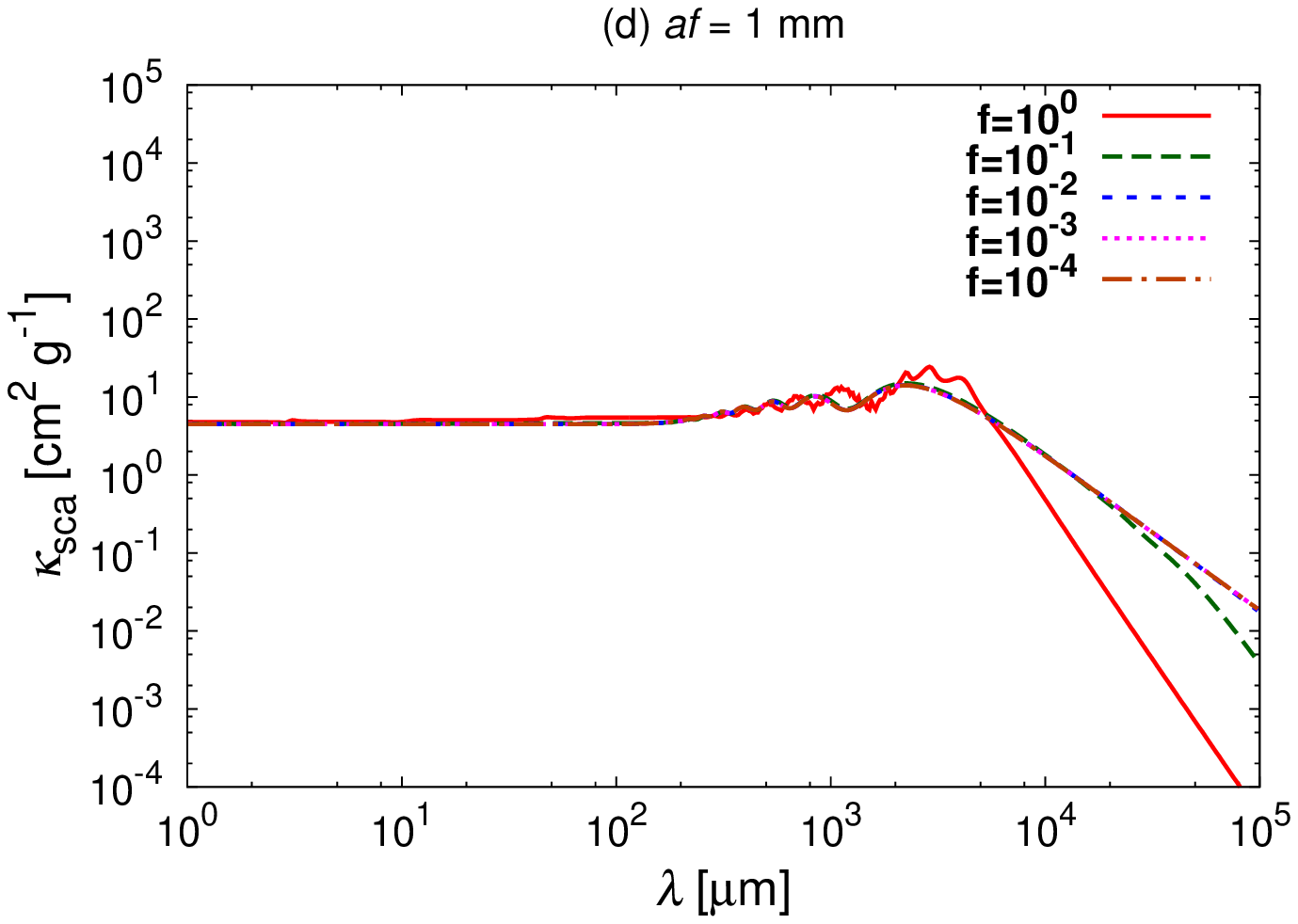}
	}
 \end{center}
 \caption{
 Same as Figure \ref{fig:abs_dif} but for scattering mass opacities.
 }
 \label{fig:case0sca}
\end{figure*}

We also calculate the scattering mass opacity by using Mie calculations.
Figure \ref{fig:case0sca} shows the scattering mass opacities in the case of $af=0.1{~\rm \mu m}$, $1{~\rm \mu m}$, $100{~\rm \mu m}$, and $1{~\rm mm}$.
This figure suggests that the scattering mass opacity is not characterized by $af$ at the longer wavelengths.
At the shorter wavelengths, the mass opacity corresponds to the geometric cross section.
In the compact case, the mass opacity scales as $\lambda^{-4}$ at the longer wavelengths.
On the other hand, in the fluffy case, the mass opacity scales as $\lambda^{-2}$ at the inter mediate wavelengths, then scales as $\lambda^{-4}$ at the longer wavelengths.
We will come back to this point with a physical explanation in Section \ref{sec:piecewise}.

\begin{figure*}
 \begin{center}
 	\subfigure{
 	 \includegraphics[width=80mm]{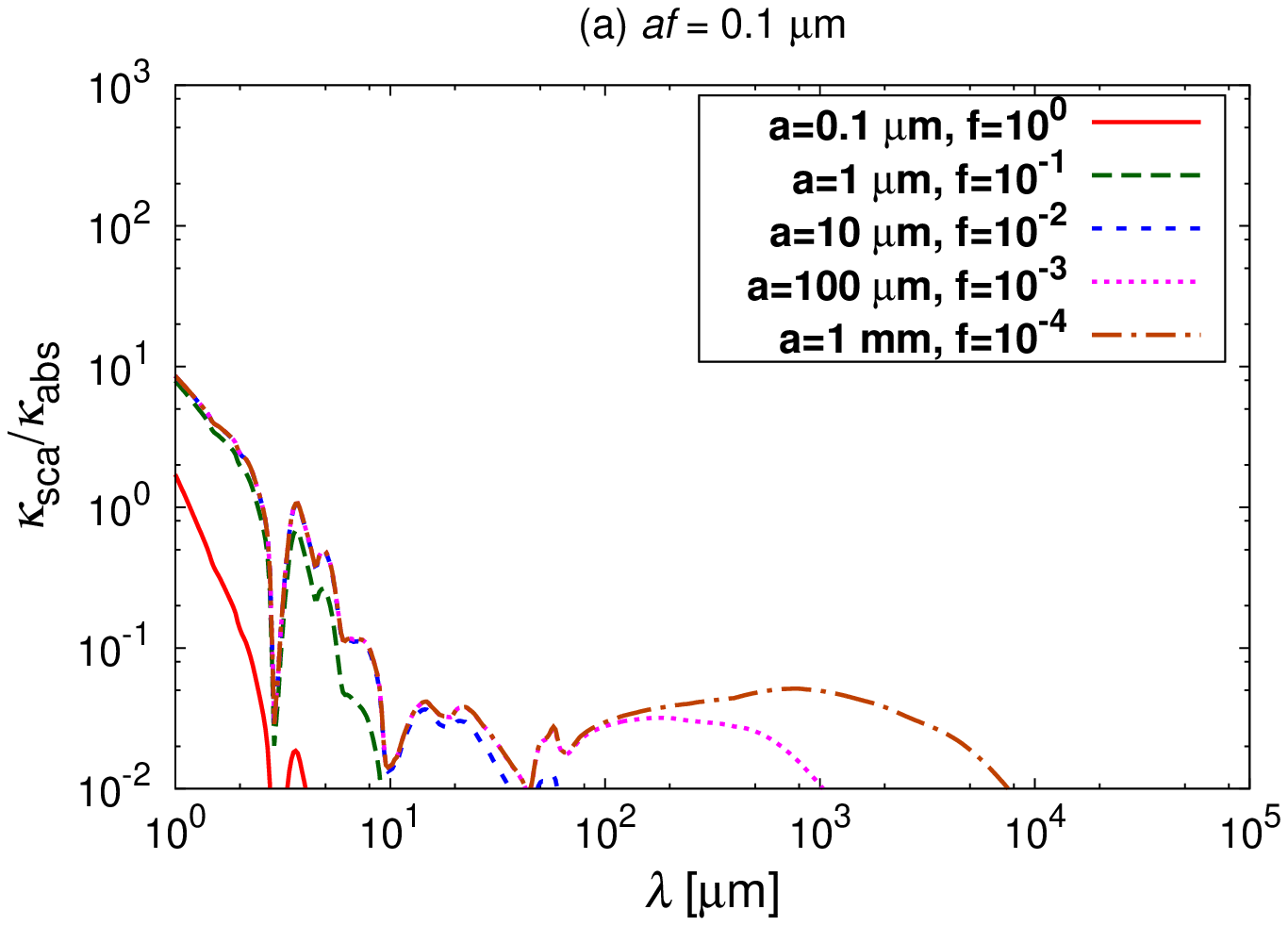}
	}
 	\subfigure{
 	 \includegraphics[width=80mm]{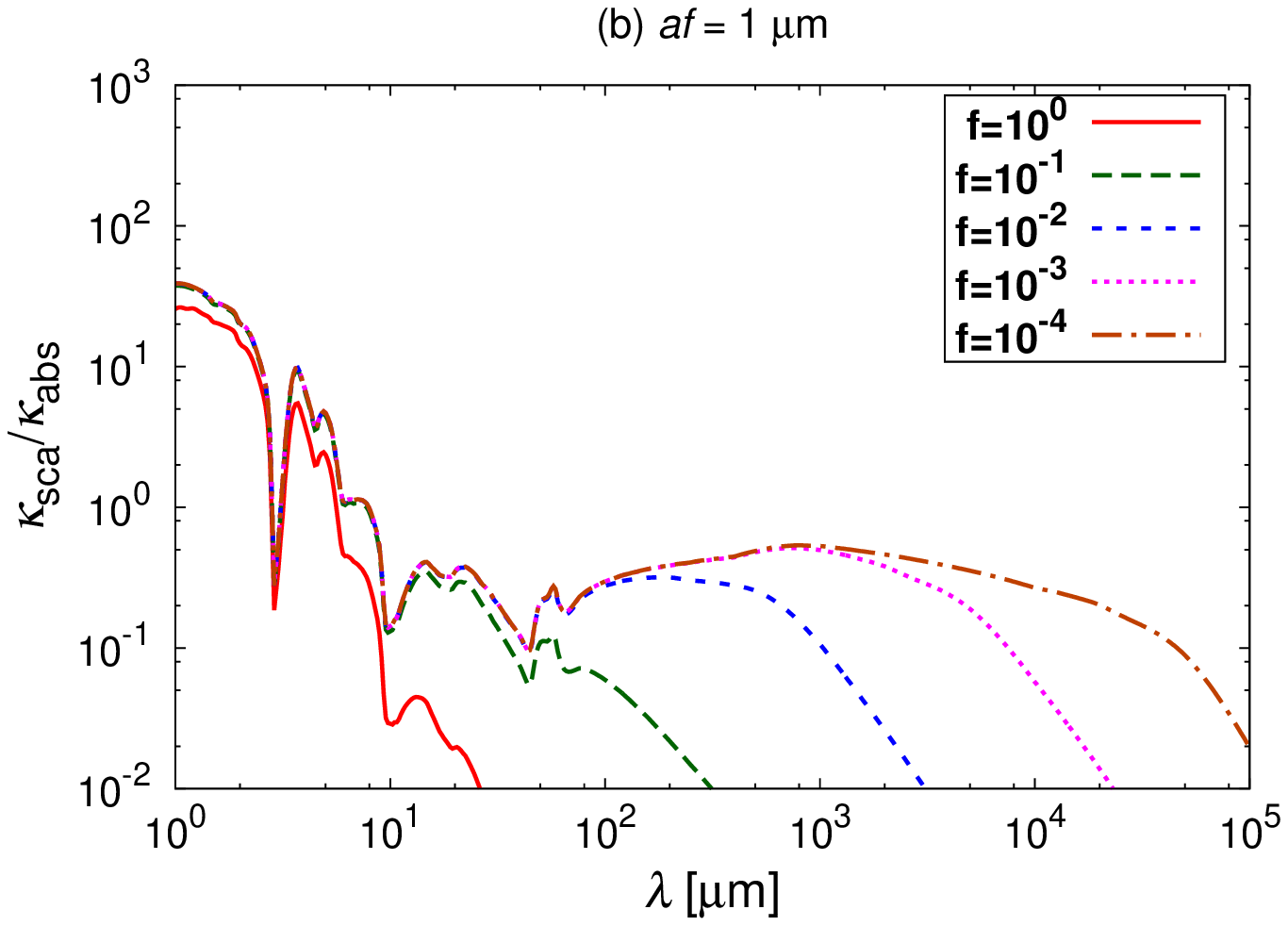}
	}\\
 	\subfigure{
 	 \includegraphics[width=80mm]{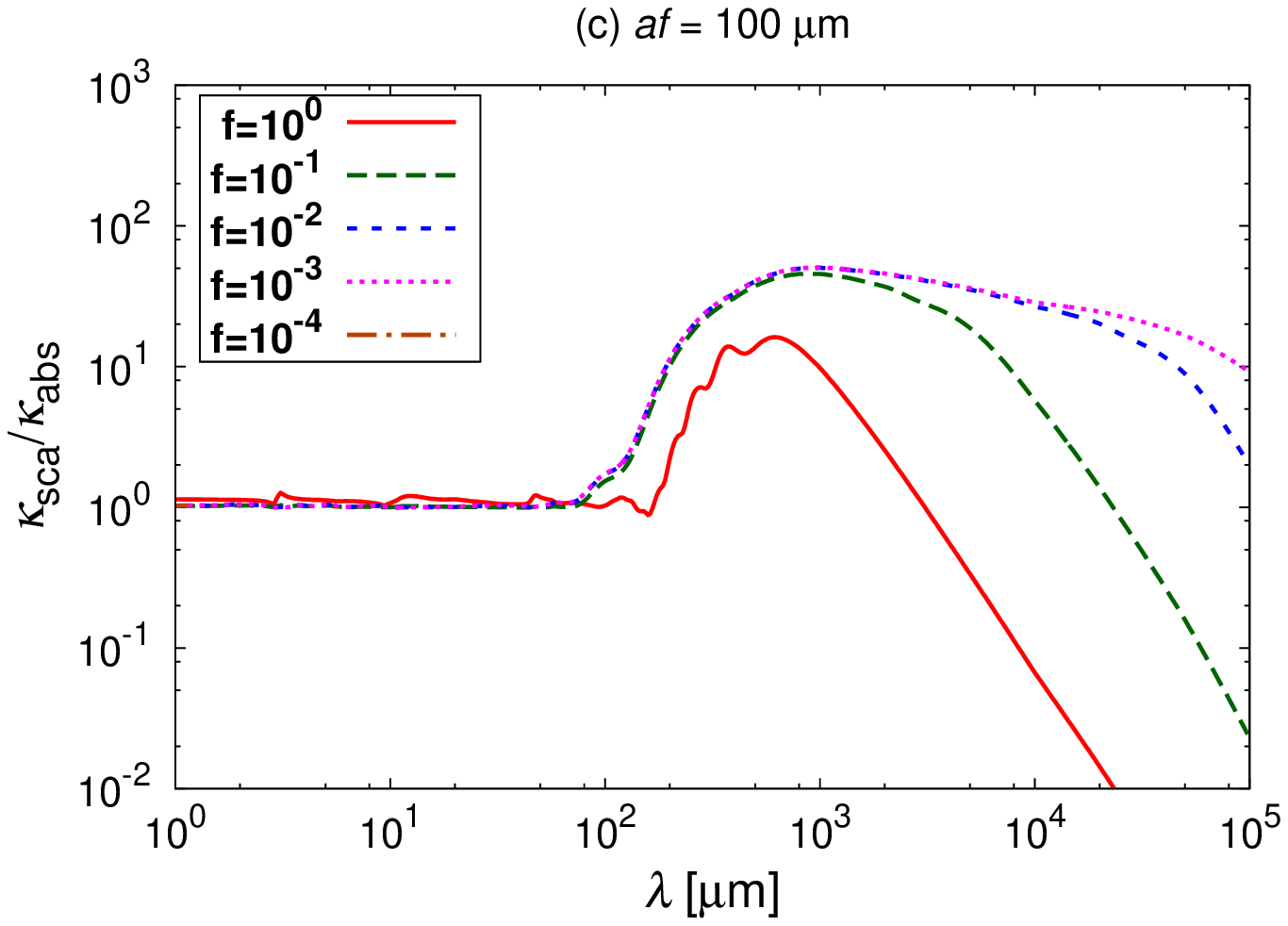}
	}
 	\subfigure{
 	 \includegraphics[width=80mm]{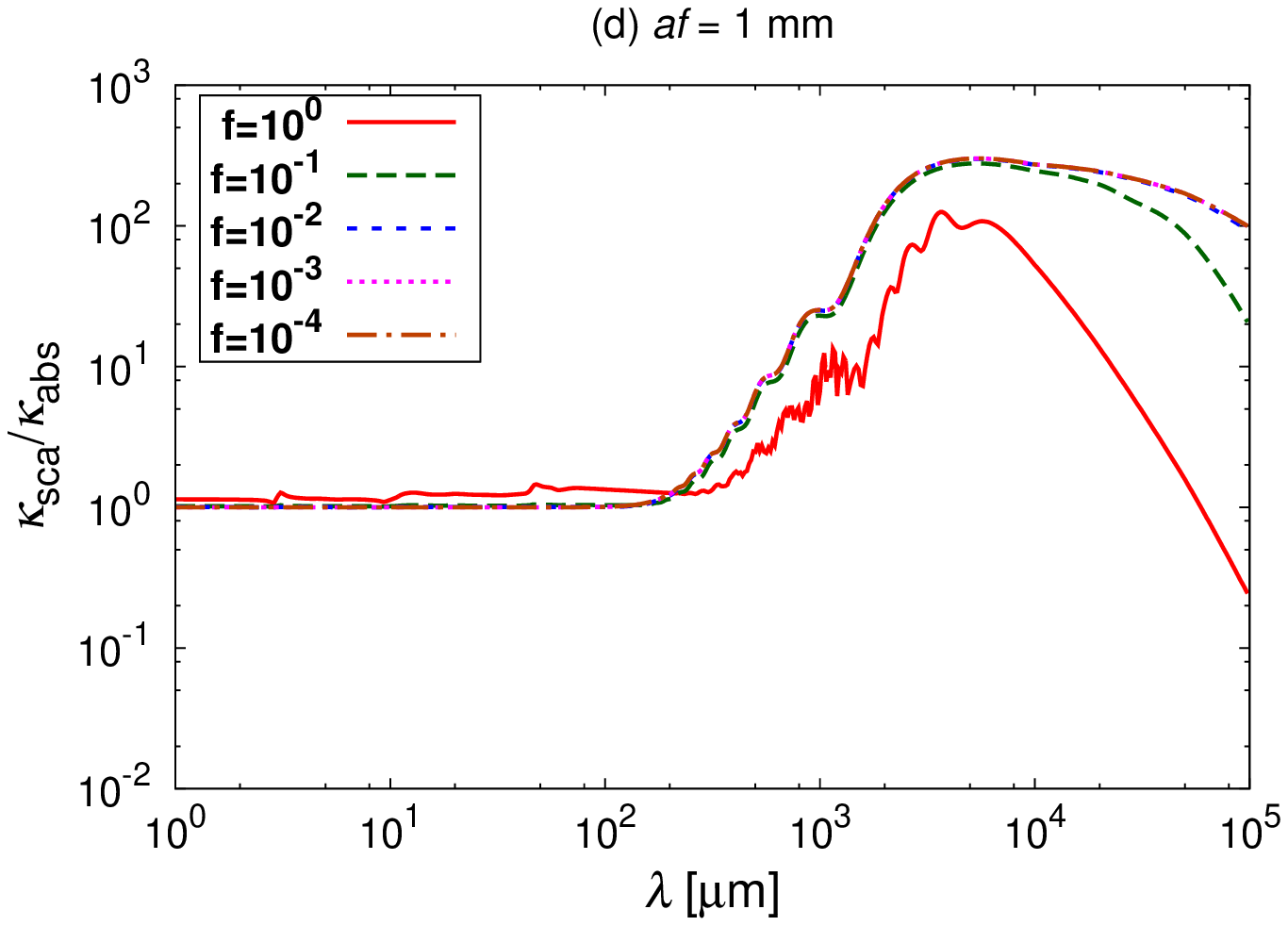}
	}
 \end{center}
 \caption{
 The ratio of scattering mass opacity over absorption mass opacity $\kappa_{\rm sca}/\kappa_{\rm abs}$ in the cases of $af=0.1{~\rm \mu m}$, $1{~\rm \mu m}$, $100{~\rm \mu m}$, and $1{~\rm mm}$.
 Each panel shows the cases with the same $af$, but the filling factor ranges are in $f=1, 10^{-1},10^{-2},10^{-3}$, and $10^{-4}$.
 }
 \label{fig:albedo}
\end{figure*}

As shown in Figure \ref{fig:case0sca}, the scattering mass opacity of the fluffy aggregates is expected to be higher than the compact case at the longer wavelengths even when the absorption mass opacity is almost the same.
Thus, we investigate the ratio of $\kappa_{\rm sca}$ against $\kappa_{\rm abs}$.
Figure \ref{fig:albedo} shows the ratio in each case corresponding to Fig. \ref{fig:case0sca}.
In the case of compact and $af=0.1{~\rm \mu m}$, the scattering mass opacity is less than absorption.
On the other hand, the scattering mass opacity dominates the absorption mass opacity in fluffy cases when $af=0.1{~\rm \mu m}$.
This greatly affects the infrared observations of dust grains.
For example, \citet{Pagani10} reported that dust grains in dense interstellar medium is composed of micron-sized grains (and not $0.1{~\rm \mu m}$) because of the high scattering efficiency observed by the Spitzer space telescope.
Thus, they infer that the monomer size must be micron.
However, figure \ref{fig:albedo} suggests that even if the monomers are $0.1{~\rm \mu m}$ in size, the aggregates of $0.1{~\rm \mu m}$ sized monomers represent the high albedo and thus might account for the observed high efficiency of scattering.

Figures \ref{fig:albedo} (c) and (d) show the scattering mass opacity in the case of $af=100{~\rm \mu m}$ and $1{~\rm mm}$.
The scattering mass opacity at the millimeter wavelengths is ten times larger than the absorption mass opacity in the compact case, and it is tens of times larger in the fluffy cases.
This suggests that the millimeter continuum emission is dominated not by direct thermal emission, but by scattered emission in transition disks if the dust aggregates are grown to have a millimeter size.
Moreover, determining the ratio of the scattering mass opacity over the absorption mass opacity at the millimeter wavelengths is a way to characterizing the porosity of the dust aggregates.

\section{Analytic formulae of the opacities}
\label{sec:piecewise}
In the previous section, we used the Mie calculations to obtain the mass opacity.
In this section, we derive the analytic formulae of the mass opacity and compare them to the results of Mie calculations.
By deriving analytic formulae, we explain why the mass opacity can be characterized by $af$.
In addition, the analytic formulae would be a computationally less expensive method to calculate the opacity of large aggregates.

\subsection{Approximation of refractive index}
When we consider fluffy aggregates, the filling factor satisfies $f\ll1$.
If $f\ll 1$, from Eq. (\ref{eq:eps_eff}) and $\epsilon=m^{2}$, we obtain
\begin{eqnarray}
n &\simeq &1+\frac{3}{2}f {\rm Re}(F),\\
k &\simeq &\frac{3}{2}f {\rm Im}(F),
\end{eqnarray}
to the first order of $f$.
From these equations, we obtain that $n-1 \propto f$ and $k \propto f$ in the case of fluffy aggregates where $f\ll1$.
We check the validity of the relations in Appendix \ref{sec:appendix1}.

We do not assume $f\ll1$ when deriving the analytic formulae.
After deriving the formulae, we assume $f\ll1$ and use the relations of $n-1 \propto f$ and $k \propto f$ to explain why the mass opacity is characterized by $af$.

\subsection{Absorption mass opacity}
\begin{figure}
 \begin{center}
 	 \includegraphics[width=80mm]{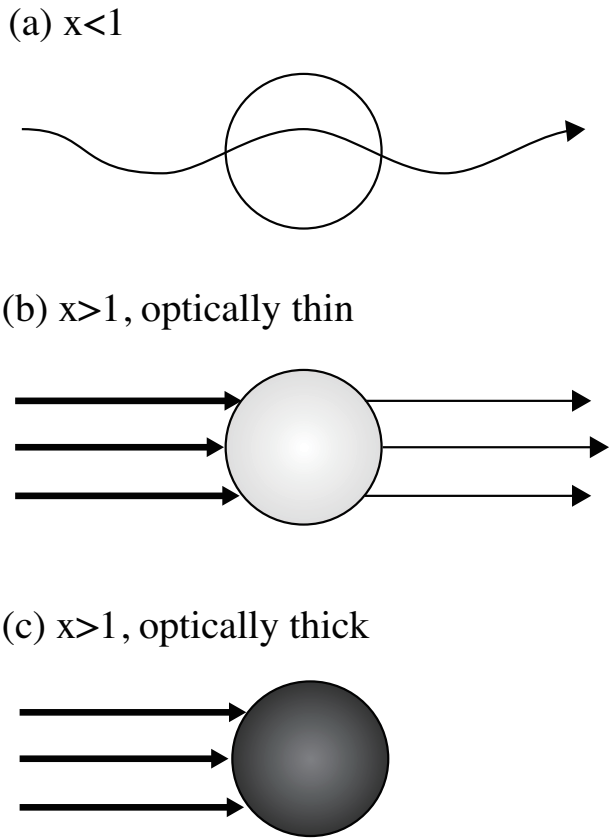}
 \end{center}
 \caption{
 The three limiting regime.
 (a) When $x \ll 1$, the opacity goes into the Rayleigh regimes.
 (b) When $x \gg 1$ and is optically thin ($kx \ll 3/8$), the opacity goes into the optically thin geometric regime.
 (c) When $x \gg 1$ and is optically thick ($kx \gg 3/8$), the opacity goes into the optically thick geometric regime. 
 }
 \label{fig:picture}
\end{figure}

We derive the approximated formulae of $Q_{\rm abs}$ in three limited cases, illustrated in Fig. \ref{fig:picture}: (1) $x \ll 1$, (2) $x \gg 1$ and optically thin ($kx \ll 3/8$) media, and (3) $x \gg 1$ and optically thick ($kx \gg 3/8$) media.
We note that the absorption mass opacity is $Q_{\rm abs}$ divided by the mass-to-area ratio of the aggregates.

\subsubsection{ $x \ll 1$}
When the dust radius $a$ is much smaller than the wavelength $\lambda$, the opacity goes into Reyleigh regime.
This corresponds to $x=2\pi a/\lambda \ll 1$.
In this case, $Q_{\rm abs}$ can be written as
\begin{equation}
Q_{\rm abs} \simeq 4x {\rm~ Im} \left(\frac{m^2-1}{m^2+2}\right) = \frac{24nkx}{(n^2-k^2+2)^2+(2nk)^2}.
\end{equation}
(see Eq. (5.11) in BH83).
The imaginary part $k$ of the refractive index is much smaller than the real part $n$ (see Fig. \ref{fig:check_ref} and Fig. \ref{fig:check_ref_0}).
So, we can approximate $Q_{\rm abs}$ as
\begin{equation}
Q_{\rm abs} \simeq Q_{\rm abs,1}\equiv \frac{24nkx}{(n^2+2)^2}.
\end{equation}
This equation explains the fact that the absorption mass opacity is characterized by mass-to-area ratio or $af$.
At the longer wavelengths, $n$ is almost unity while $k \propto f$.
Using $f\propto m/a^3$, we obtain that $kx$ is proportional to $m/a^2$, which is mass-to-area ratio.
Since $\kappa_{\rm abs}$ is $Q_{\rm abs}$ divided by mass-to-area ratio, $\kappa_{\rm abs}$ is independent of dust properties.

\subsubsection{$x \gg 1$ and optically thin}
When the dust radius $a$ is much larger than the wavelength $\lambda$, the opacity goes into geometric optics regime.
In this regime, the optical properties can be understood by tracing the ray inside the material.
The fraction of energy that transmits the material is $1-\exp(-\alpha \xi)$ where $\alpha=4\pi k/\lambda$, and $\xi$ is the path of the ray inside the material.
If $\alpha \xi < 1$, the incident light is weakly absorbed by the material because it is optically thin on the ray.
We set the length $\xi=2a$, the diameter of the sphere.
Thus, the condition $\alpha \xi < 1$ corresponds to $kx < 1$.

In the limit of $a\gg \lambda$ (or equivalently $x \gg 1$) and optically thin, we obtain
\begin{equation}
Q_{\rm abs} \simeq Q_{\rm abs,2} \equiv \frac{8kx}{3n}\left(n^3-(n^2-1)^{3/2}\right),
\end{equation}
(see Eq. (7.2) in BH83).

We note that if $n=1$, which is usually satisfied in the case of fluffy medium, $Q_{\rm abs}$ yields
\begin{equation}
Q_{\rm abs}=Q_{\rm abs,1}=Q_{\rm abs,2}=\frac{8kx}{3}.
\end{equation}
This equation is also characterized by the mass-to-area ratio or $af$ because $k\propto f$ and $x\propto a$.
We also note that the analytic formula between optically thick and thin regimes should be changed when $Q_{\rm abs}$ is unity.
Thus, we define optically thin as $kx \ll 3/8$.

\subsubsection{$x \gg 1$ and optically thick}
\label{sec:qabs3}
In the limit of $a\gg \lambda$ (equivalent to $x \gg 1$) and optically thick ($kx \gg 3/8$), on the other hand, the absorption coefficient is described as
\begin{equation}
Q_{\rm abs}\simeq Q_{\rm abs,3} \equiv \int^{\pi/2}_{0}(1-R(\theta_{i}))\sin 2\theta_{i} d\theta_{i},
\end{equation}
where the reflectance $R(\theta)$ is written as
\begin{equation}
R(\theta_{i})=\frac{1}{2}\left(\left|\frac{\cos \theta_{t} - m \cos \theta_{i}}{\cos \theta_{t} + m \cos \theta_{i}}\right|^2 + \left|\frac{\cos \theta_{i} - m \cos \theta_{t}}{\cos \theta_{i} + m \cos \theta_{t}}\right|^2\right),
\end{equation}
and
\begin{equation}
\sin \theta_{t}=\frac{\sin \theta_{i}}{m}
\end{equation}
(see Eq. (2.71),  Eq. (7.5), and Eq. (7.7) in BH83).
This regime is valid at shorter wavelengths (see Appendix \ref{sec:appendix3} for the optical depth of the aggregate).
As shown in Appendix \ref{sec:appendix2}, $Q_{\rm abs,3}\sim 1-0.1\times f$: $Q_{\rm abs,3}\sim 0.9$ for compact case and $Q_{\rm abs,3}\sim 0.99$ for $f=0.1$.
These values are regarded as unity in application to astronomical observations.
Therefore, $Q_{\rm abs,3}\sim 1$ for most cases: the absorption cross section yields the geometric cross section.
Because $Q_{\rm abs,3}$ has no dependency on $f$ and $a$, $\kappa_{\rm abs}$ is characterized by $af$.

\subsubsection{Analytic formula of absorption mass opacity}
\begin{figure*}
 \begin{center}
 	\subfigure{
 	 \includegraphics[width=80mm]{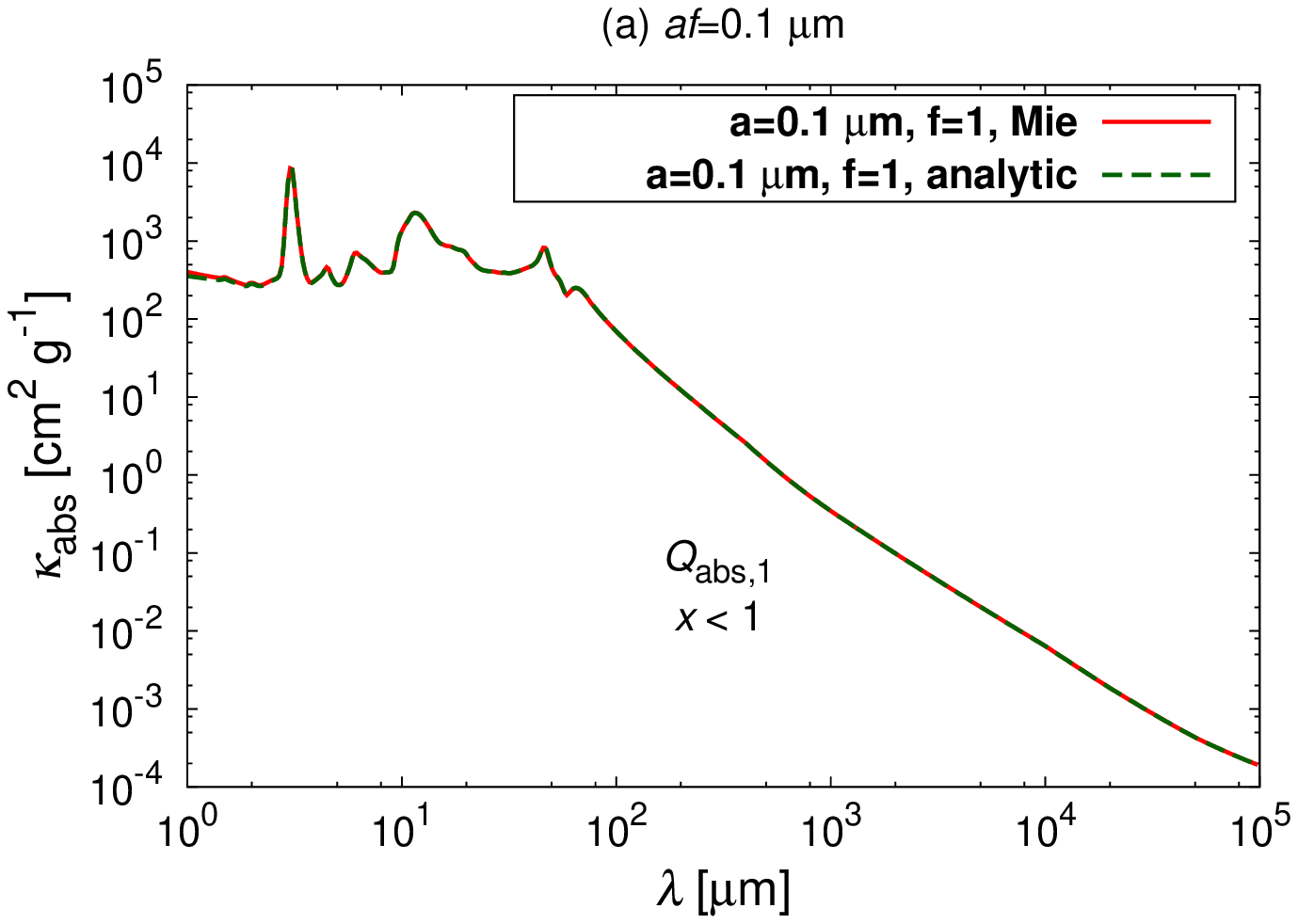}
	}
 	\subfigure{
 	 \includegraphics[width=80mm]{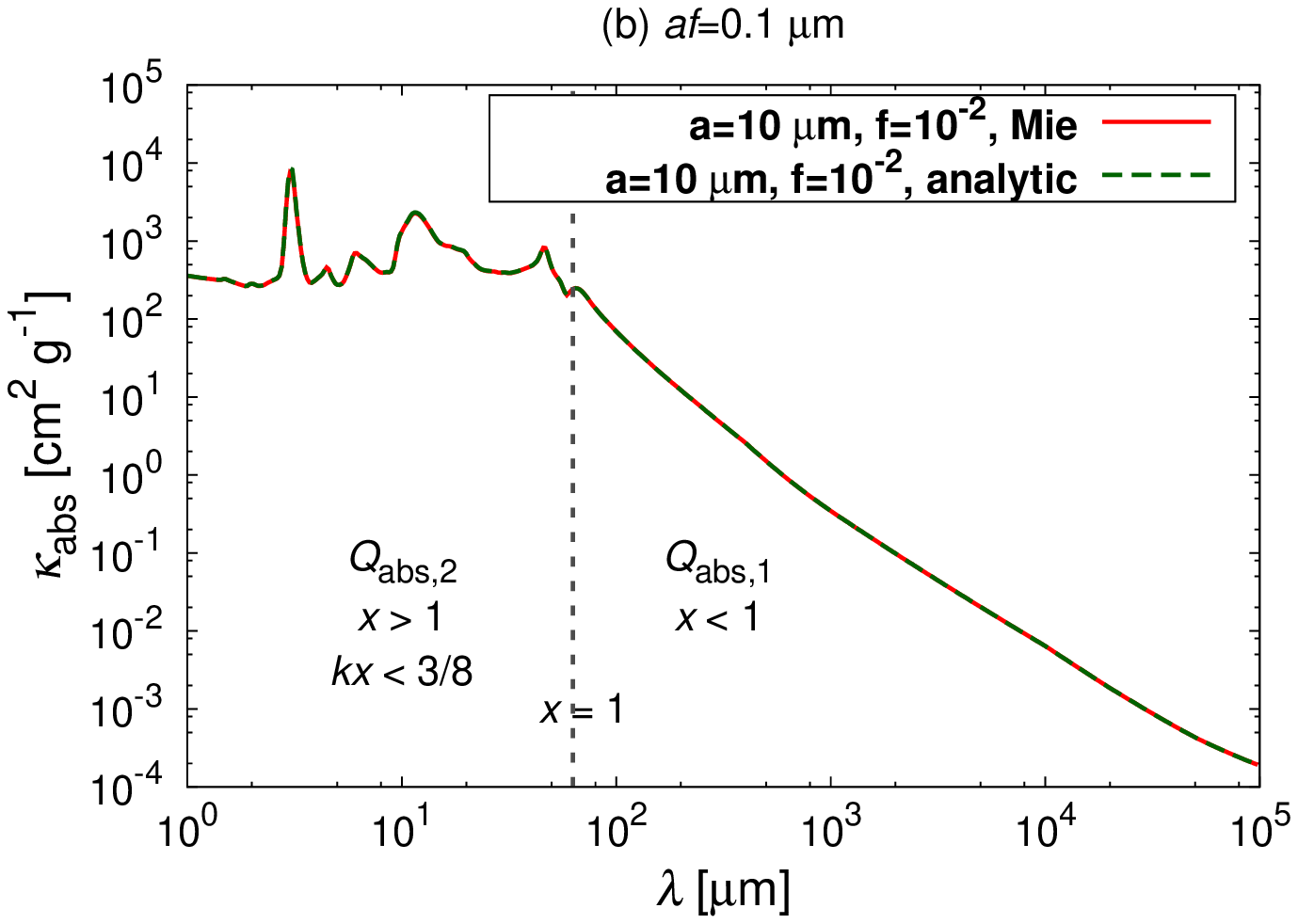}
	}\\
 	\subfigure{
 	 \includegraphics[width=80mm]{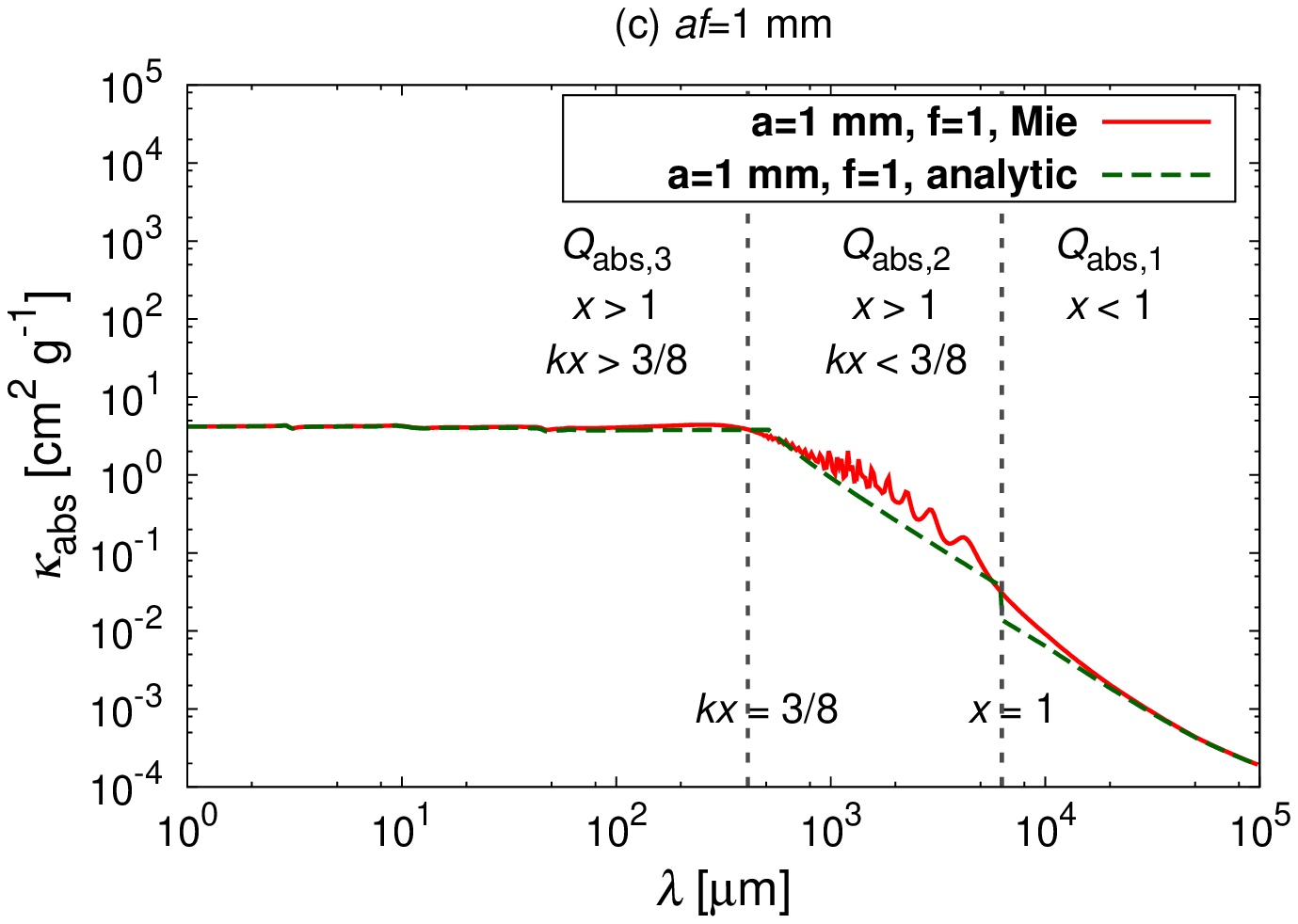}
	}
 	\subfigure{
 	 \includegraphics[width=80mm]{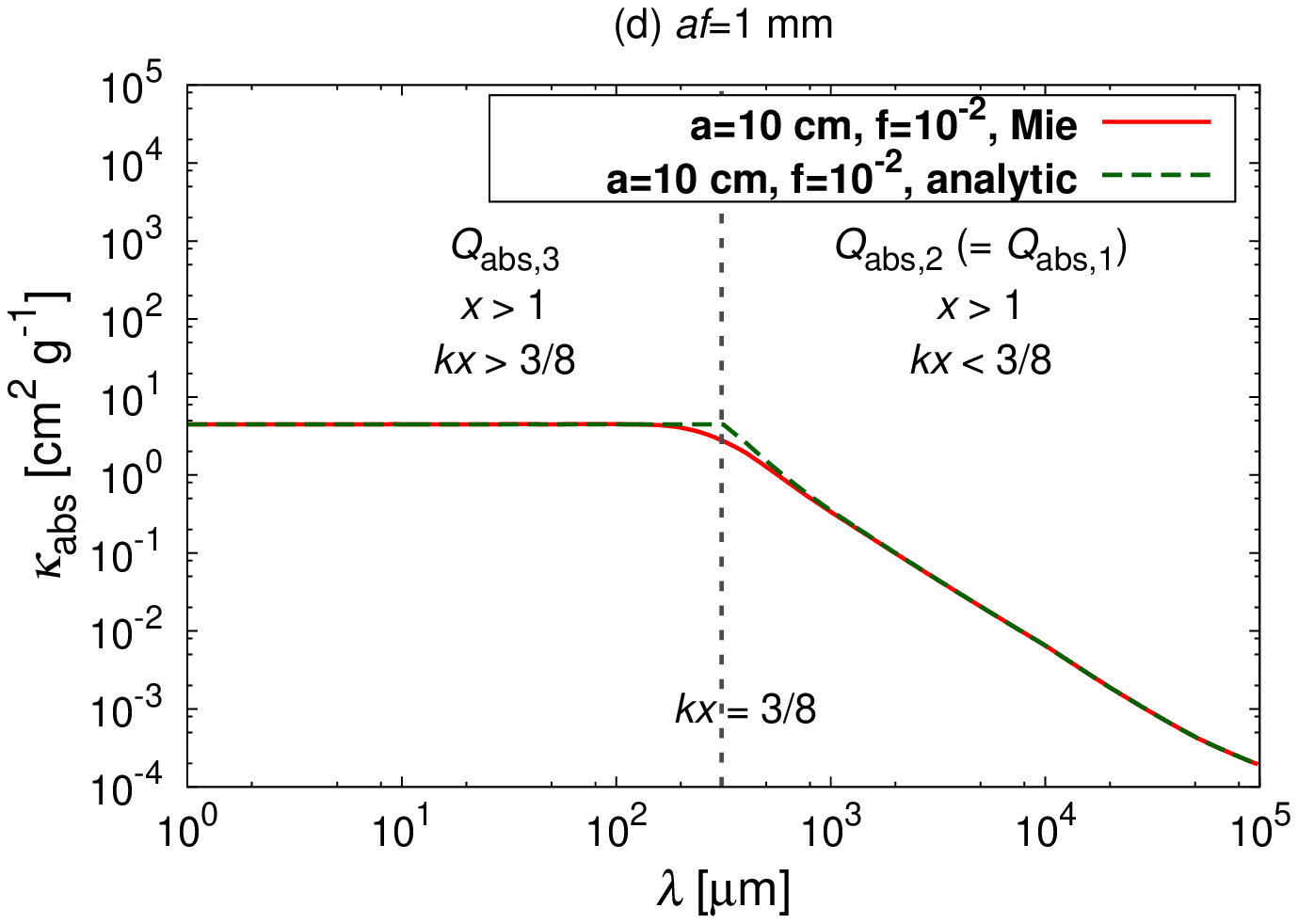}
	}
 \end{center}
 \caption{
 The comparison of Mie calculations and the analytic formulae.
 The dotted lines show where $x=1$ and $kx=3/8$.
 (a) The mass opacity in the case of $af=0.1{~\rm \mu m}$ and $f=1$.
 (b) $af=0.1{~\rm \mu m}$ and $f=10^{-2}$.
 (c) $af=1{~\rm mm}$ and $f=1$.
 (d) $af=1{~\rm mm}$ and $f=10^{-2}$.
 }
 \label{fig:piecewise}
\end{figure*}

Combining the three limiting regimes, we obtain the analytic formula of the absorption mass opacity of dust aggregates as
\begin{equation}
Q_{\rm abs}=
 \begin{cases}
   Q_{\rm abs,1}  & (x<1)\\
   \min(Q_{\rm abs,2}, Q_{\rm abs,3}) & (x>1)
  \end{cases}
  .
\end{equation}
Figure \ref{fig:piecewise} shows the absorption mass opacities calculated with both the Mie calculation and the analytic formula.
The absorption mass opacity of Figs. \ref{fig:piecewise} (a) and (b) are the same because $af=0.1{~\rm \mu m}$ but only the filling factor is different ($f=1$ and $f=0.01$), as shown in the previous sections (see Fig. \ref{fig:abs_dif}).
Figure \ref{fig:piecewise} (a) shows the case of $af=0.1{~\rm \mu m}$ and $f=1$ (compact).
The whole wavelengths in this panel satisfy $x<1$, and thus $Q_{\rm abs}\simeq Q_{\rm abs,1}$. 
The analytic formula greatly reproduce the Mie calculations.
Figure \ref{fig:piecewise} (b) shows the case of $af=0.1{~\rm \mu m}$ and $f=0.01$ (fluffy).
In this case, $a=10{~\rm \mu m}$, and thus $x=1$ at $\lambda=2\pi a \simeq 63{~\rm \mu m}$.
We use $Q_{\rm abs}=Q_{\rm abs,1}$ for $x>1$ and $Q_{\rm abs}=Q_{\rm abs,2}$ for $x<1$ and connect them at $x=1$.
This also reproduces the Mie calculation.

Figures \ref{fig:piecewise} (c) and (d) show the case of $af=1{~\rm mm}$, but the filling factor is 1 and 0.01, respectively.
The absorption mass opacity of Figs. \ref{fig:piecewise} (c) and (d) are almost the same except for the interference structure.
The interference structure corresponds to where $x>1$ and $kx<3/8$, the optically-thin geometric optics regime.
We note that the difference between the Mie calculation and the analytic formulae is the interference structure.
In Fig. \ref{fig:piecewise} (c), which is the compact case, the real part of the refractive index is greater than unity.
Thus, $Q_{\rm abs}$ has an enhancement because of the interference.
In Fig. \ref{fig:piecewise} (d), which is the fluffy case, on the other hand, the real part of the refractive index is almost unity and thus no enhancement appears and smoothly connects to $x<1$ region at the longer wavelengths.
From the analytic formula, we conclude that the interference structure only appears in the compact cases because $n$ is still larger than unity when $x>1$ and $kx<3/8$.

\subsection{Scattering mass opacity}
In the same manner of obtaining the analytic formula of the absorption mass opacity, we also derive the analytic formula of the scattering mass opacity.
In addition, by using the analysis, we explain why the mass opacity can and cannot be characterized by $af$.

\subsubsection{ $x \ll 1$}
When $x \ll 1$, in the Rayleigh regime, $Q_{\rm sca}$ can be written as
\begin{equation}
Q_{\rm sca} \simeq \frac{8}{3}x^{4} \left|\frac{m^2-1}{m^2+2}\right|^2,
\end{equation}
(see Eq. (5.8) in BH83).
At the longer wavelengths, $n-1\ll1$ and $k\ll1$.
Therefore, the equation can be approximated to
\begin{equation}
Q_{\rm sca} \simeq Q_{\rm sca,1} \equiv \frac{32}{27}x^{4} \left((n-1)^{2}+k^{2}\right).
\end{equation}
As shown in Appendix \ref{sec:appendix1}, $(n-1)>k$ at the longer wavelengths.
Therefore, $Q_{\rm sca,1}\propto x^{4}(n-1)^{2}$.
By using $x\propto a$ and $(n-1)\propto f$, we obtain $Q_{\rm sca,1}\propto a^{4}f^{2}$.
This is {\em not characterized} by $af$.
When we consider two aggregates whose $af$ is the same, the aggregate that has the larger radius has the larger scattering mass opacity at the longer wavelengths although $Q_{\rm abs}$ is same.
In other words, the scattering efficiency at the longer wavelengths is a way to determine the filling factor of fluffy aggregates.

\subsubsection{$x \gg 1$ and optically thin}
If $x\ll1$, the scattering mass opacity of an aggregate is regarded as the sum of the scattering mass opacity of each monomer because the scattered waves from all the constituent monomers have approximately the same phase.
If $x\gg1$, by contrast, scattered waves with scattering angle $\theta>\theta_{\rm max}\sim 1/x$ cancel out because of the phase difference.
Thus, the radiation within the solid angle of $\pi \theta_{\rm max}/4\pi$ is scattered.
With the condition that $\theta_{\rm max}\sim 1/x$ and that $Q_{\rm sca,2}$ is smoothly connected to $Q_{\rm sca,1}$ at $x=1$, we obtain
\begin{equation}
Q_{\rm sca} \simeq Q_{\rm sca,2} \equiv \frac{1}{x^2}Q_{\rm sca,1}.
\end{equation}
Using the same discussion in the previous section, we obtain $Q_{\rm sca,2}\propto x^{2}(n-1)^{2}\propto a^{2}f^{2}$.
This is again characterized by $af$.
We note that the optical depth of the aggregate is unity when $Q_{\rm sca}$ is unity.
Thus, the optical depth becomes unity where $x(n-1)\sim 1$ because $n-1>k$.

\subsubsection{$x \gg 1$ and optically thick}
When $x \gg 1$ and the medium is optically thick, $Q_{\rm sca}+Q_{\rm abs}=2$.
Therefore, in the same manner of the absorption mass opacity, we obtain
\begin{equation}
Q_{\rm sca}\simeq Q_{\rm sca,3} \equiv \int^{\pi/2}_{0}(1+R(\theta_{i}))\sin 2\theta_{i} d\theta_{i},
\end{equation}
(see Eq. (7.5) and Eq. (7.6) in BH83).

As discussed in Section \ref{sec:qabs3}, the integrated reflectance is $\sim 0.1\times f$.
Thus, $Q_{\rm sca,3}\sim 1+0.1\times f$.
This is regarded as $Q_{\rm sca,3}\sim 1$.
Thus, the scattering mass opacity also goes to the geometric cross section at shorter wavelengths and is characterized by $af$.

\subsubsection{Analytic formula of scattering mass opacity}
\begin{figure*}
 \begin{center}
 	\subfigure{
 	 \includegraphics[width=80mm]{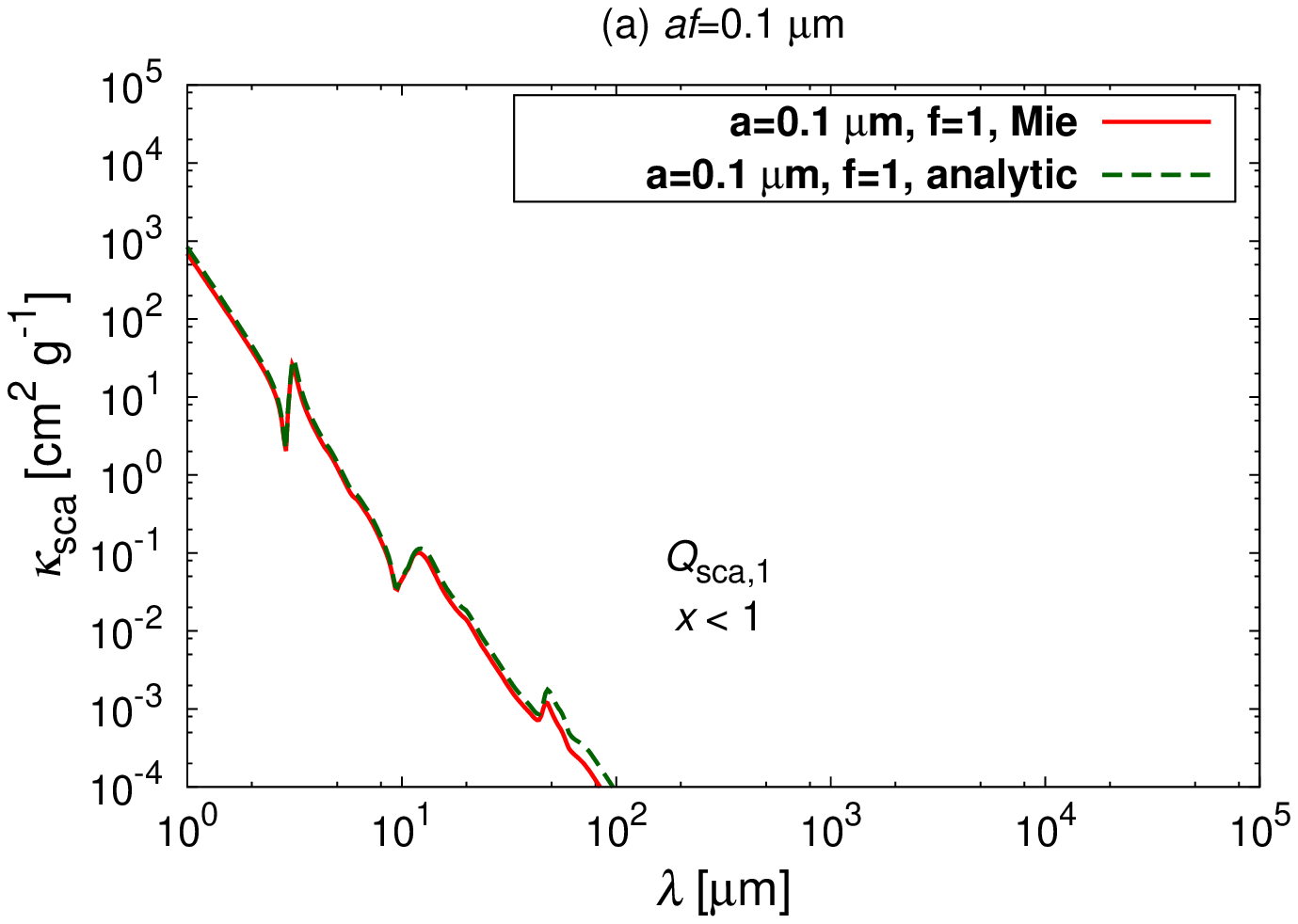}
	}
 	\subfigure{
 	 \includegraphics[width=80mm]{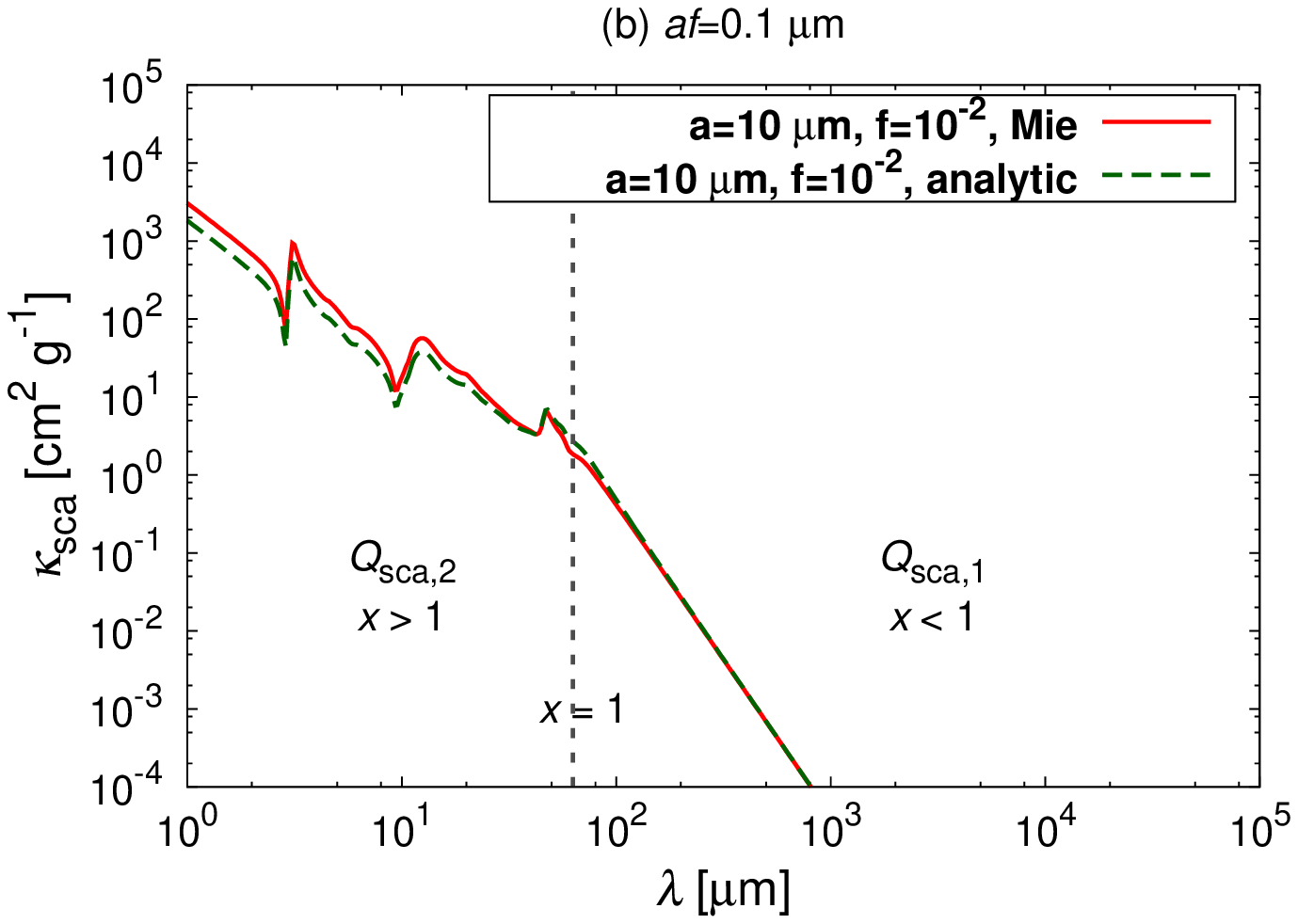}
	}\\
 	\subfigure{
 	 \includegraphics[width=80mm]{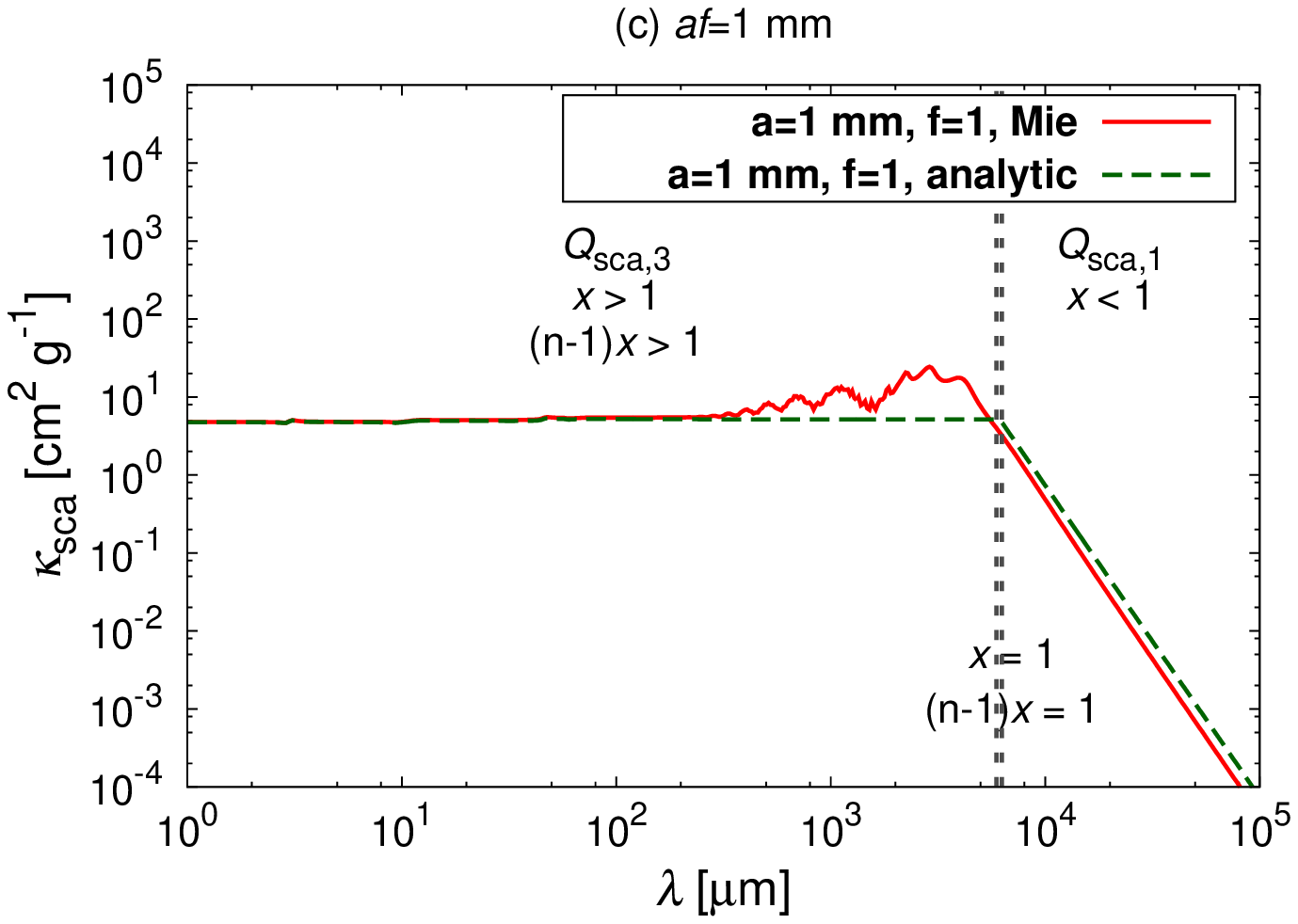}
	}
 	\subfigure{
 	 \includegraphics[width=80mm]{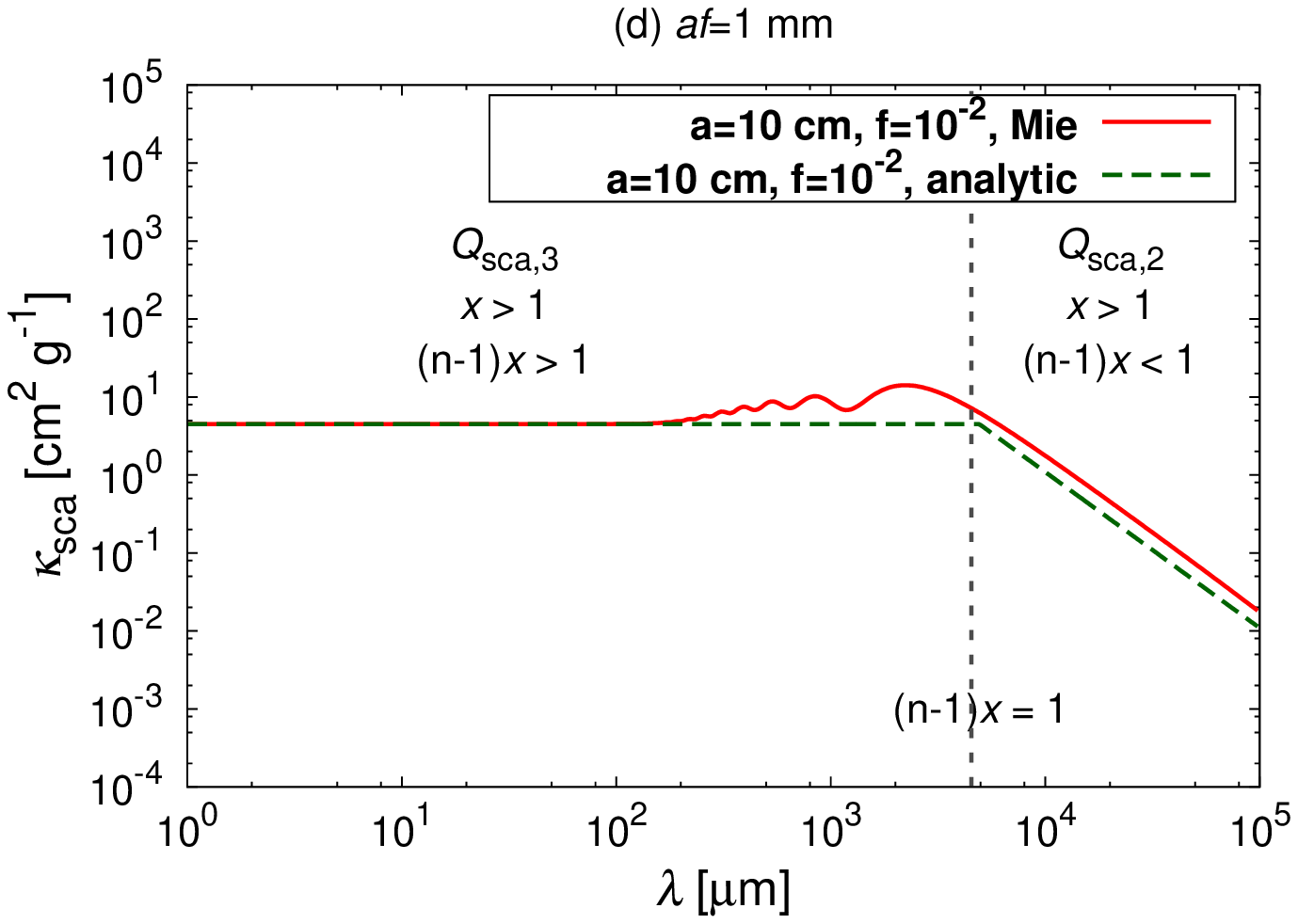}
	}
 \end{center}
 \caption{
 Same as Fig. \ref{fig:piecewise}, but for scattering mass opacity.
 }
 \label{fig:piecewise_sca}
\end{figure*}

Combining the three limiting regimes, we obtain the analytic formula of scattering mass opacity as
\begin{equation}
Q_{\rm sca}=
 \begin{cases}
   Q_{\rm sca,1}  & (x<1)\\
   \min(Q_{\rm sca,2}, Q_{\rm sca,3}) & (x>1)
  \end{cases}
  .
\end{equation}
Figure \ref{fig:piecewise_sca} shows the comparison of scattering mass opacity obtained with the Mie calculation and the analytic formula.
Figures \ref{fig:piecewise_sca} (a) and (b) show the case of $af=0.1{~\rm \mu m}$, but the filling factor is $f=1$ and $f=0.01$, respectively.
The analytic formula in both cases reproduces the Mie calculation.
Figures \ref{fig:piecewise_sca} (c) and (d) show the case of $af=1{\rm~ mm}$, but the filling factor is $f=1$ and $f=0.01$, respectively.
In this case, the analytic formula reproduces the Mie calculation except for the interference structure because we assume that all the amplification by interference is damped when $x>1$ and optically thin, which corresponds to $Q_{\rm sca}=Q_{\rm sca,2}$.
However, the difference of the mass opacity between the analytic and Mie calculations is less than one order.
Except for the interference structure, the analytic formula reproduces the Mie calculation even in the case of scattering mass opacity.

As already shown in Fig. \ref{fig:case0sca}, the scattering mass opacity is proportional to $\lambda^{-2}$ at the intermediate wavelengths and $\lambda^{-4}$ at the longer wavelengths in fluffy cases.
This can be explained by $Q_{\rm sca,2}$.
If $x=2\pi a/\lambda$ is less than unity, which occurs at the longer wavelengths, $Q_{\rm sca}=Q_{\rm sca,1}$ and thus scales as $\lambda^{-4}$ becasue $Q_{\rm sca,1}\propto x^{4}$.
However, in the case of fluffy aggregates, we find the region where $x>1$ and optically thin, and therefore $Q_{\rm sca}=Q_{\rm sca,2}=Q_{\rm sca,1} /x^{2}$.
This is the reason why the scattering mass opacity at the intermediate wavelengths in fluffy cases scales as $\lambda^{-2}$.

\section{Implications for opacity evolution in protoplanetary disks}
\label{sec:disk}
The index of the dust opacity $\beta$ has been widely used as an indicator of the dust growth.
In this section, we will show how $\beta$ changes as aggregates grow and drift both in compact and fluffy cases. Then, we propose a detection method of fluffy aggregates in protoplanetary disks by using the opacity index $\beta$.

\subsection{Fluffy dust growth and opacity evolution}
Before starting the discussion of $\beta$, we discuss the general mass opacity change as dust grains grow to fluffy aggregates in protoplanetary disks.
We adopt a fluffy dust growth model proposed by \citet{Kataoka13b}.
In this model, they reveal the overall porosity evolution from micron-sized grains to kilometer-sized planetesimals through direct sticking.
In the coagulation, icy particles are sticky and thus they are not disrupted or bounced, but grow to a larger size \citep{Wada09,Wada11, Wada13}.
Moreover, the large radius of fluffy aggregates enables them to grow rapidly to avoid the radial drift barrier \citep{Okuzumi12, Kataoka13b}.
Thus, the model is a complete scenario of growing path from dust grains to planetesimals by direct sticking.

\begin{figure}
 \begin{center}
 \includegraphics[width=80mm]{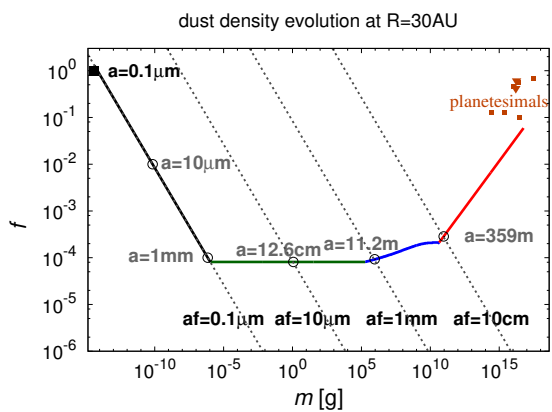}
 \end{center}
 \caption{
 The internal density evolution at $r=30$ AU in the minimum mass solar nebula model (see \citealt{Kataoka13b} for more details).
 The solid line represents the evolution.
 The black, green, blue, and red lines are in the coagulation phase of hit-and-stick, collisional compression, gas compression, and self-gravitational compression, respectively.
 The dotted lines show the $af=$(constant) lines, where $af=0.1{\rm~\mu m}$, $10{\rm~\mu m}$, $1{\rm ~mm}$, and $10{\rm ~cm}$, respectively.
 }
 \label{fig:fluffygrowth}
\end{figure}

Figure \ref{fig:fluffygrowth} shows the internal density evolution at 30 AU in orbital radius in a minimum mass solar nebula model, proposed by \citet{Kataoka13b}.
We note that the figure shows the local porosity evolution, but dust aggregates start to drift inward once they grow to be decoupled from the gas.
We discuss the radial drift later in this section.
The turbulent parameter $\alpha_{\rm D}$ is set to be $10^{-3}$ and the mean internal density is set to be 1.68 ${\rm g~cm^{-3}}$.
The picture of the overall porosity evolution is as follows.
As the dust grains first coagulate to form fluffy aggregates, the filling factor decreases to $f\sim 10^{-4}$.
Once the collisional compression becomes effective, the density keeps constant.
Then, the gas compression and the self-gravity compression make the dust aggregates compact.

The open circles represent the characteristic dust radius $a$, while the dotted lines show the lines of constant $af$.
From this figure, the dust aggregates in the initial growth stage is optically the same.
The initial growth is expected to be fractal.
The dust aggregates coagulate with aggregates of similar sizes, and thus the fractal dimension is expected to be 2 \citep{Okuzumi12}.
Thus, the mass-to-area ratio of the aggregates keeps constant.
In other words, $af$ keeps the same value.
Therefore, the initial growth is indistinguishable with no growth in the absorption mass opacity.
After the initial growth, when the compression mechanisms become effective, the opacity is expected to change because the mass-to-area ratio changes.

\begin{figure}
 \begin{center}
  \includegraphics[width=80mm]{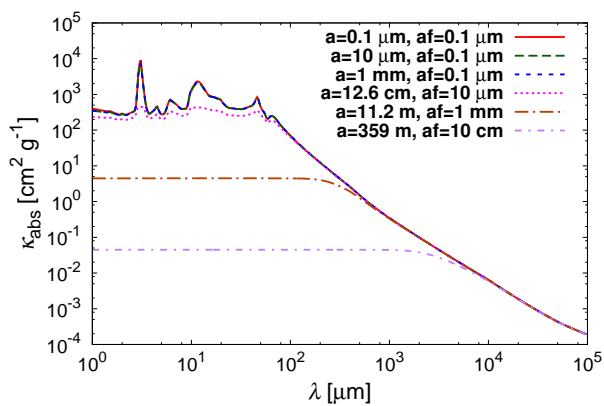}
 \end{center}
 \caption{
 The mass opacity change in the case of fluffy dust growth.
 Each line shows the mass opacity in the cases of different radius and filling factor, which corresponds to open circles in Fig. \ref{fig:fluffygrowth}.
 The dust properties of the lines are as follows: 
 red: $a=0.1{\rm~\mu m}$, and $af=0.1{\rm~\mu m}$;
 green: $a=10{\rm~\mu m}$, and $af=0.1{\rm~\mu m}$;
 blue: $a=1{\rm~ mm}$, and $af=0.1{\rm~\mu m}$;
 magenta: $a=12.6{\rm~ cm}$, and $af=10{\rm~\mu m}$;
 brown: $a=11.2{\rm~ m}$, and $af=1{\rm~ mm}$;
 purple: $a=359{\rm~ m}$, and $af=10{\rm~ cm}$. 
  }
 \label{fig:fluffyopacity}
\end{figure}

Figure \ref{fig:fluffyopacity} shows the mass opacity change, corresponding to open circles in Fig. \ref{fig:fluffygrowth}.
The first three cases are degenerated in mass opacity because $af$ is the same.
Once the compression becomes effective, the mass opacity changes as expected in the dust growth.
For example, when dust aggregates grow to have their radius of $a=12.6$ cm, they have almost the same opacity as $10{\rm~\mu m}$ compact grains.
We note that the interference structure does not appear as aggregates grow because the filling factor is typically $f\sim10^{-4}$ in this growth scenario.

\subsection{Dust opacity index beta}
\begin{figure*}
 \begin{center}
 	\subfigure{
 	 \includegraphics[width=80mm]{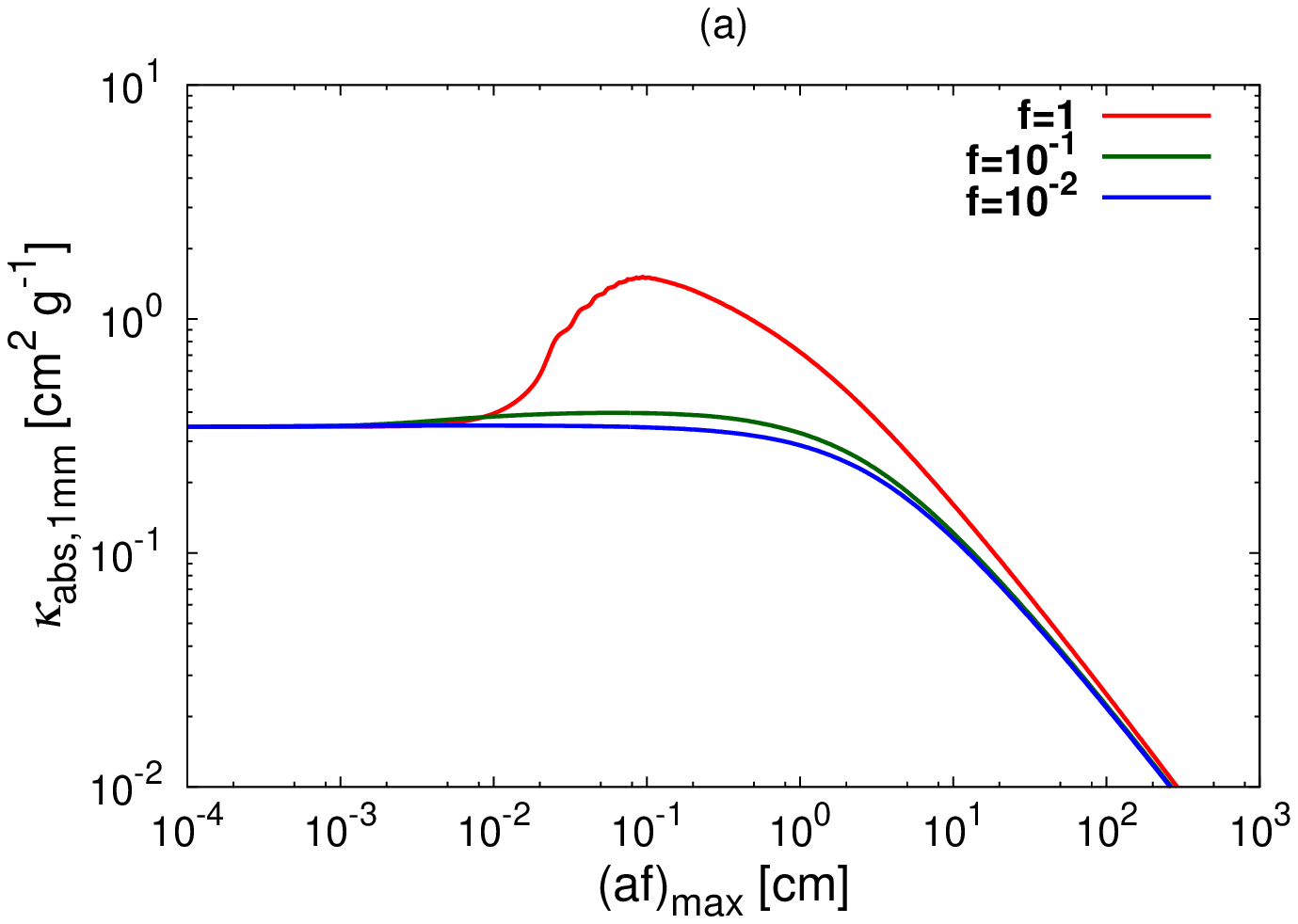}
	}
 	\subfigure{
 	 \includegraphics[width=80mm]{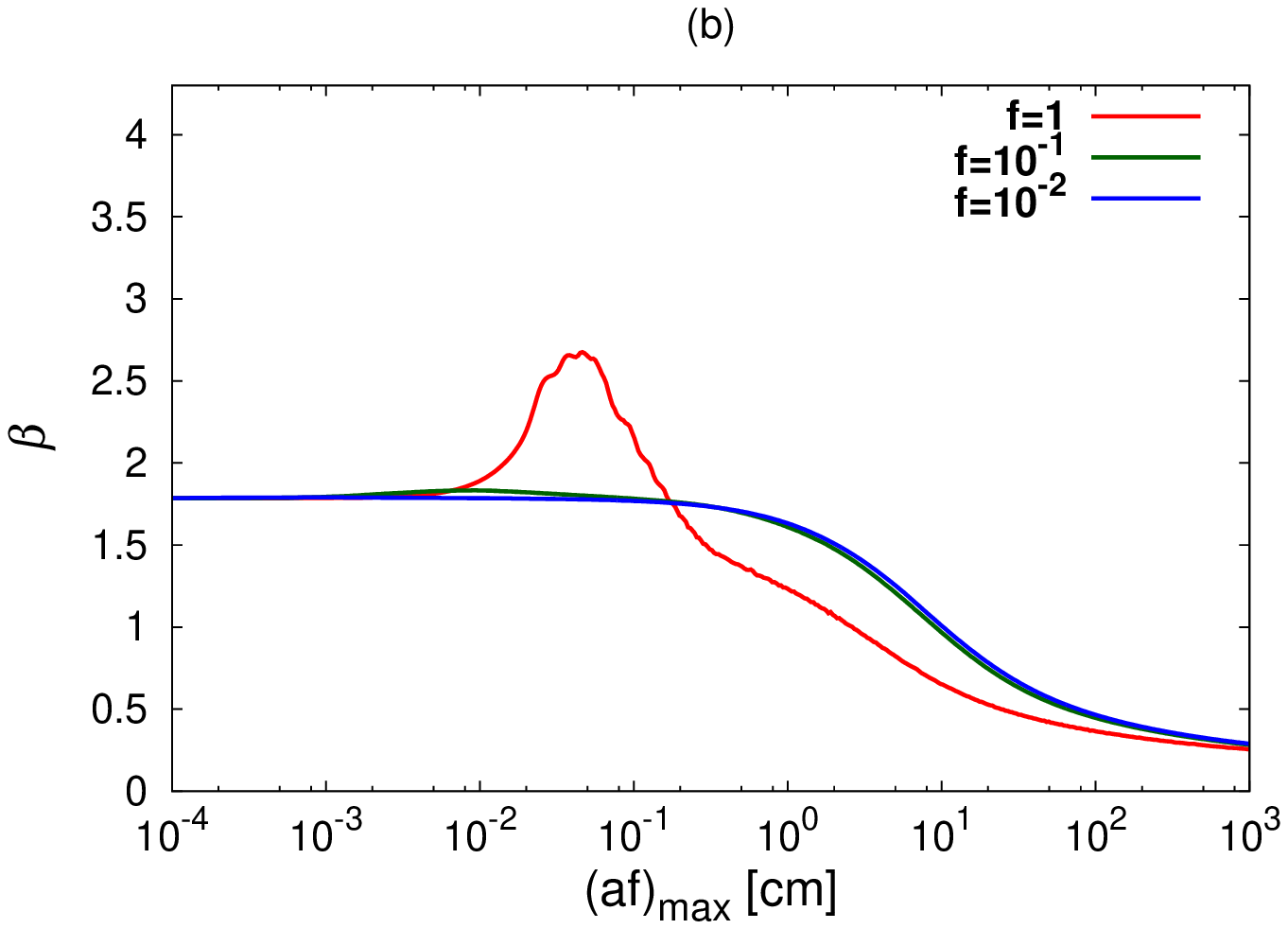}
	}
 \end{center}
 \caption{
 (a) The absorption mass opacity at $\lambda=1{\rm mm}$ against $(af)_{\rm max}$.
 The red, green, and blue lines show the cases of $f=1,10^{-1}$, and $10^{-2}$.
 The lines in the cases of $f<10^{-2}$ are indistinguishable from the line of $f=10^{-2}$.
 The aggregate size distribution is assumed to be $n\propto (af)^{-2}$ where $f$ is fixed in each case.
 (b) The $\beta$, the opacity slope between 1 mm and 3 mm, against $(af)_{\rm max}$.
 }
 \label{fig:beta}
\end{figure*}

We define $\beta$ as an opacity slope between 1 mm and 3 mm.
Here, we use $af$ again because optical properties are characterized by $af$.
We note that $af=a$ in the case of compact grains ($f=1$).
We consider several cases for calculating dust $\beta$ where the filling factor $f$ is fixed in each case.
Calculating $\beta$, we consider a grain size distribution with a power law as $n\propto (af)^{-2}$ between a minimum and a maximum size, $(af)_{\rm min}$ and $(af)_{\rm max}$, respectively; $(af)_{\rm min}$ is chosen to be 0.1 ${\rm \mu m}$.
Figure \ref{fig:beta} shows how the absorption mass opacity $\kappa_{\rm abs}$ at 1 mm and $\beta$ changes as the aggregate size increases.
The compact case in this figure corresponds to Fig. 3 in \citet{Ricci10a}.
The red line shows the compact case, while green and blue lines are fluffy cases.
The lines in the cases of $f<10^{-2}$ are indistinguishable from the line of $f=10^{-2}$, and thus we do not plot them.
The absorption mass opacity has a strong bump around $a_{\rm 0,max}\sim 1{\rm mm}$.
This bump corresponds to the interference structure where the size parameter $x\sim 1$.
On the other hand, the fluffy cases does not show such a bump because there is no interference.
The bump of the mass opacity results in a bump of $\beta$ in the compact case.
The dust $\beta$ increases up to $\sim 2.7$ in the compact case, but there is no bump in fluffy cases.
In other words, $\beta$ is always equal to or less than $\beta_{\rm ISM}=1.7$ in fluffy cases.

From this result, the difference in the absorption mass opacity between compact and fluffy dust appears in the intermediate size, which is between 0.1 mm and 1 mm.

\subsection{Radial profile of $\beta$}
The radial size distribution of dust aggregates is determined by both dust growth and drift.
In protoplanetary disks, the dust growth timescale strongly depends on orbital radius.
In the inner part of disks, dust grains coagulate faster than in the outer part because of the shorter Keplerian period.
Thus, dust aggregates have a larger size at the inner part and smaller at the outer part when considering only dust growth.
When dust aggregates grow to larger sizes, they start to drift inward.
Thus, the size where the aggregates start to drift is the maximum size of the aggregates at each orbital radius.
The maximum size also depends on an orbital radius: the maximum size is larger in the inner part and smaller in the outer part.
Combining both effects of dust growth and drift, the radial profile of the size of dust aggregates is expected to be smaller in the outer part and larger in the inner part.
Observationally, the radial profile of $\beta$ have the information of the radial size distribution.

To obtain the radial profile of $\beta$, we adopt the following simple dust growth and drift model.
We assume that there are initially $0.1 {~\rm \mu m}$ sized dust grains in the entire disk.
We trace the growth and drift motion of each set of dust grains initially located at each orbital radius.

To calculate the time evolution of dust mass $M=M(t)$ and the orbital radius $r=r(t)$ at each orbital radius, we assume that dust grains have a monodisperse distribution at each orbital radius.
Under this assumption, the dust growth and drift is described by \citep[e.g.,][]{Okuzumi12}
\begin{equation}
\frac{dM(t)}{dt}=\rho_{\rm d}\pi a^{2}\Delta v
\end{equation}
and
\begin{equation}
\frac{dr(t)}{dt}=-v_{\rm r}(M(t)),
\end{equation}
where $\rho_{\rm d}$ is the spatial dust density, $a$ the dust radius, $\Delta v$ the relative velocity of dust grains or aggregates, and $v_{\rm r}(M(t))$ the drift velocity.
We use the disk model of \citet{Kataoka13b} (see also \citealt{Okuzumi12} for the definitions of the dust velocity and the disk model).
Here, we briefly summarize the model.
At a radial distance $r(t)$ from the central star, the gas-surface density profile is $1700 (r/1{\rm AU})^{-p}  {\rm ~g~cm^{-2}} $ where we use $p=1$ in this paper.
We note that the gas-surface density does not change with time to clarify the effects of dust growth and drift \citep[e.g.,][]{Okuzumi12}.
The initial dust-to-gas mass ratio is 0.01.
The adopted temperature profile is $137 (r/1{\rm AU})^{-3/7} {\rm ~K} $, which corresponds to midplane temperature \citep{Chiang01}.
This is cooler than optically thin disk models to focus on the dust coagulation in the midplane.
The value of $\Delta v$ is assumed be the root mean square of Brownian motion and turbulent motion (see Eq. (32) in \citealt{Okuzumi12}).
The diffusion coefficient $\alpha_{\rm D}$ is taken to be $10^{-3}$.
For the velocity induced by turbulence, we denote the velocity difference of dust and gas as dust-dust velocity for simplicity.
We determine the dust scaleheight from the balance between sedimentation and turbulent diffusion \citep{Brauer08a}.
The filling factor is fixed to be unity in the compact case and changes as a function of the orbital radius in the fluffy case following \citet{Kataoka13b}.

\begin{figure*}
 \begin{center}
 	\subfigure{
 	 \includegraphics[width=80mm]{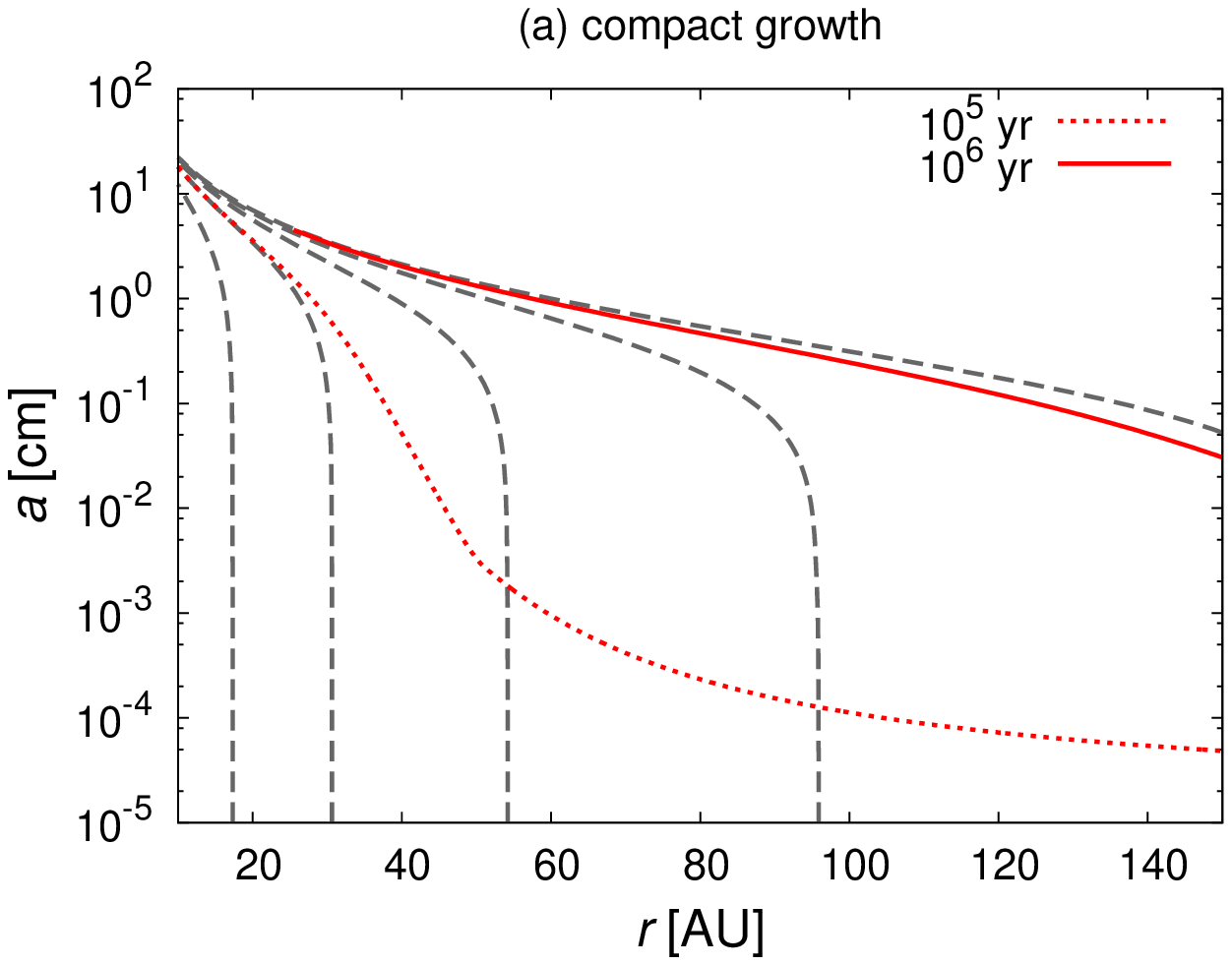}
	}
 	\subfigure{
 	 \includegraphics[width=80mm]{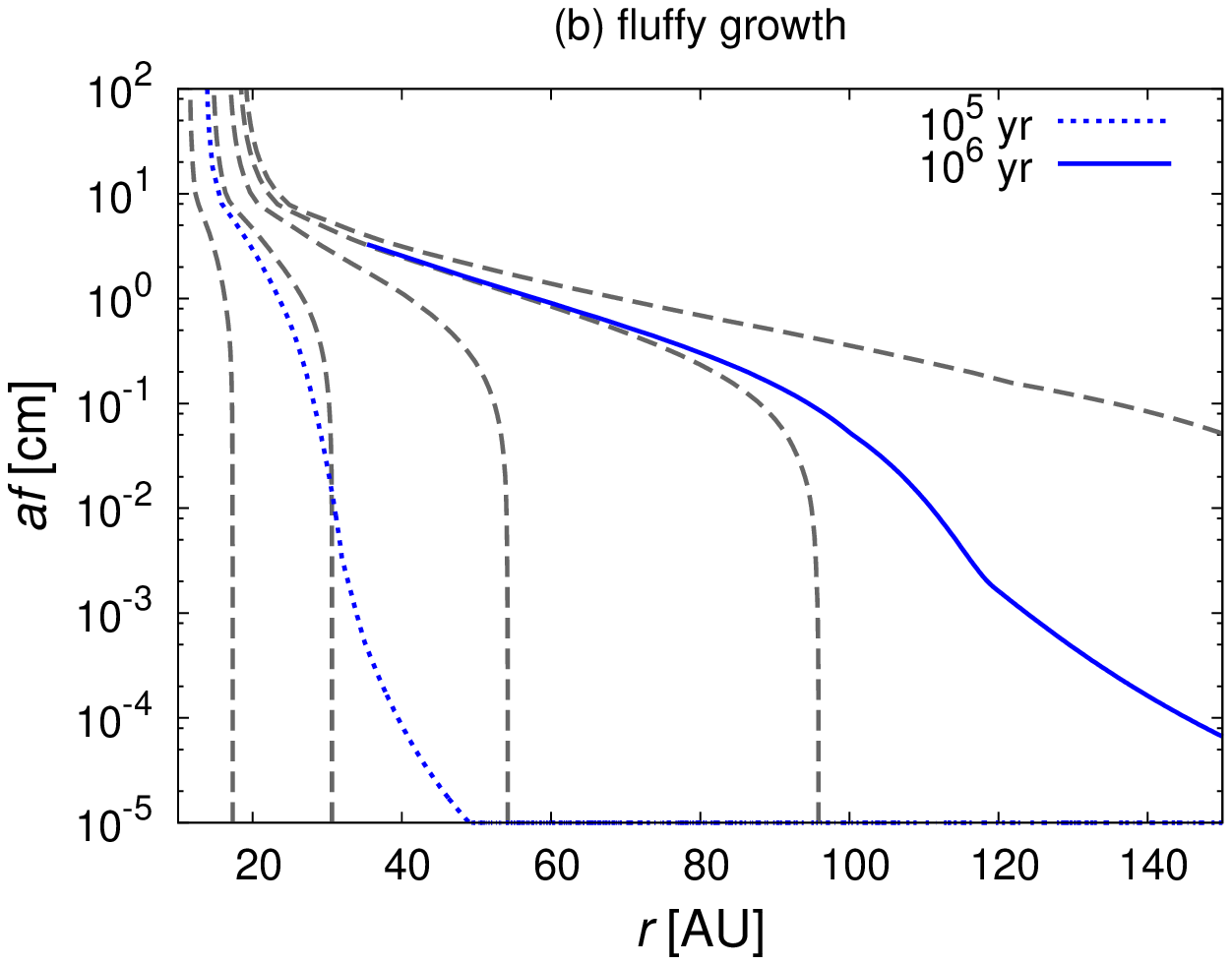}
	}
 \end{center}
 \caption{
 The gray dashed lines represent the paths of the growth and drift of dust aggregates at each initial orbital radius.
 The solid and dotted lines represent the radial size distribution of dust aggregates at the specific time.
 (a) The compact case, where the red dotted line shows the isochrone at $t=10^{5}$ years and the red solid line shows at $t=10^{6}$ years.
 (b) The fluffy growth case, where the blue dotted line shows the isochrone at $t=10^{5}$ years and the blue solid line at $t=10^{6}$ years.
 }
 \label{fig:dustgrowthdrift}
\end{figure*}

Figure \ref{fig:dustgrowthdrift} shows the dust growth and drift paths.
The dashed lines show the paths of growing dust aggregates and the colored dotted and solid lines represent isochrones at $t=10^{5}$ and $t=10^{6}$ years.
The isochrones represent the radial grain/aggregate-size distribution at the specific time.
The size of dust aggregates increases with a decreasing orbital radius.
This is caused by two effects: dust growth and drift.
For example, on the line of the isochrone at $10^5$ years in the compact case, shown in Fig. \ref{fig:dustgrowthdrift} (a), the dust growth determines the aggregate size beyond $\sim 30 {\rm~AU}$.
At 30 AU or larger in orbital radius, dust growth is faster at the inner part of the disk than the outer part because the growth timescale is proportional to the Keplerian period \citep[e.g.,][]{Okuzumi12}.
At the orbital radius less than 30 AU, the maximum size is determined by the drift motion.
The aggregates grow so large that they are decoupled from the gas, and drift inward.
Therefore, the aggregate size at orbital radius less than 30 AU corresponds to the maximum size determined by the radial drift.

\begin{figure}
 \begin{center}
  \includegraphics[width=80mm]{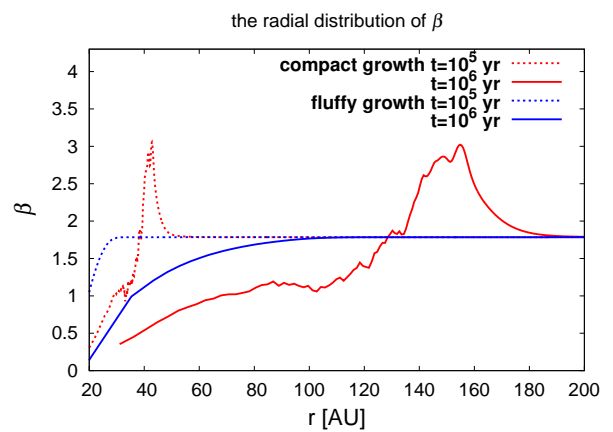}
 \end{center}
 \caption{
 The radial $\beta$ distribution for compact and fluffy cases.
 Each line corresponds to the same line in Fig. \ref{fig:dustgrowthdrift}.
 }
 \label{fig:dustgrowthdriftbeta}
\end{figure}

Figure \ref{fig:dustgrowthdriftbeta} shows the radial $\beta$ distribution for both compact and fluffy cases at the isochrones shown in Fig. \ref{fig:dustgrowthdrift}.
In the compact case, $\beta$ increases to around $\beta \sim 2.7$ at 40 AU in the range of 10 AU at $t=10^{5}$ years and at 150 AU in the range of 20 AU at $t=10^{6}$ years.
This means that protoplanetary disks have a specific radius where $\beta$ is greater than $\beta_{\rm ISM}=1.7$ in the compact case.
In the fluffy case, on the other hand, $\beta$ is always equal to or less than $\beta_{\rm ISM}=1.7$.
Therefore, if the radial $\beta$ distribution always has a value of $\beta_{\rm ISM}=1.7$ or less, it means that the millimeter emission comes from fluffy aggregates.

\subsection{Silicate feature}
In the inner part of the disk, ice particles are expected to be sublimated and there are dust aggregates whose constituent particles are made of silicate.
Micron-sized silicate grains show the broad feature at 10 ${\rm \mu m}$.
The feature is used as a signature of grain growth \citep[e.g.,][]{vanBoekel05}.
To show how the porosity affects the silicate feature, we also calculate the silicate feature of fluffy aggregates.
In this section, we change the material constant: we assume pure silicate monomers.
The material density is taken to be $\rho_{\rm mat}=3.5 {\rm ~g~cm^{-3}}$ and the refractive index is taken from \citet{WeingartnerDraine01}.

\begin{figure*}
 \begin{center}
 	\subfigure{
 	 \includegraphics[width=80mm]{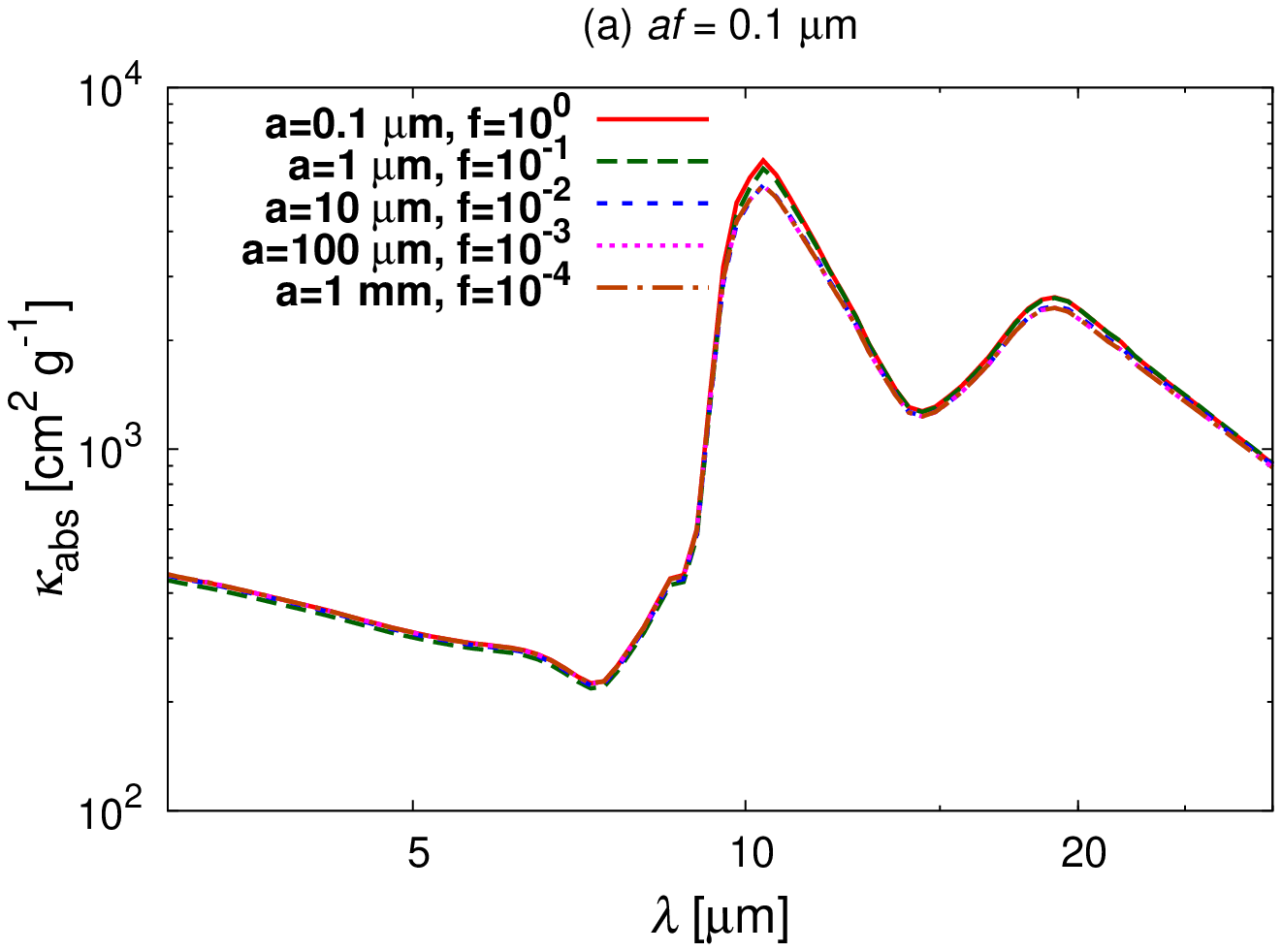}
	}
 	\subfigure{
 	 \includegraphics[width=80mm]{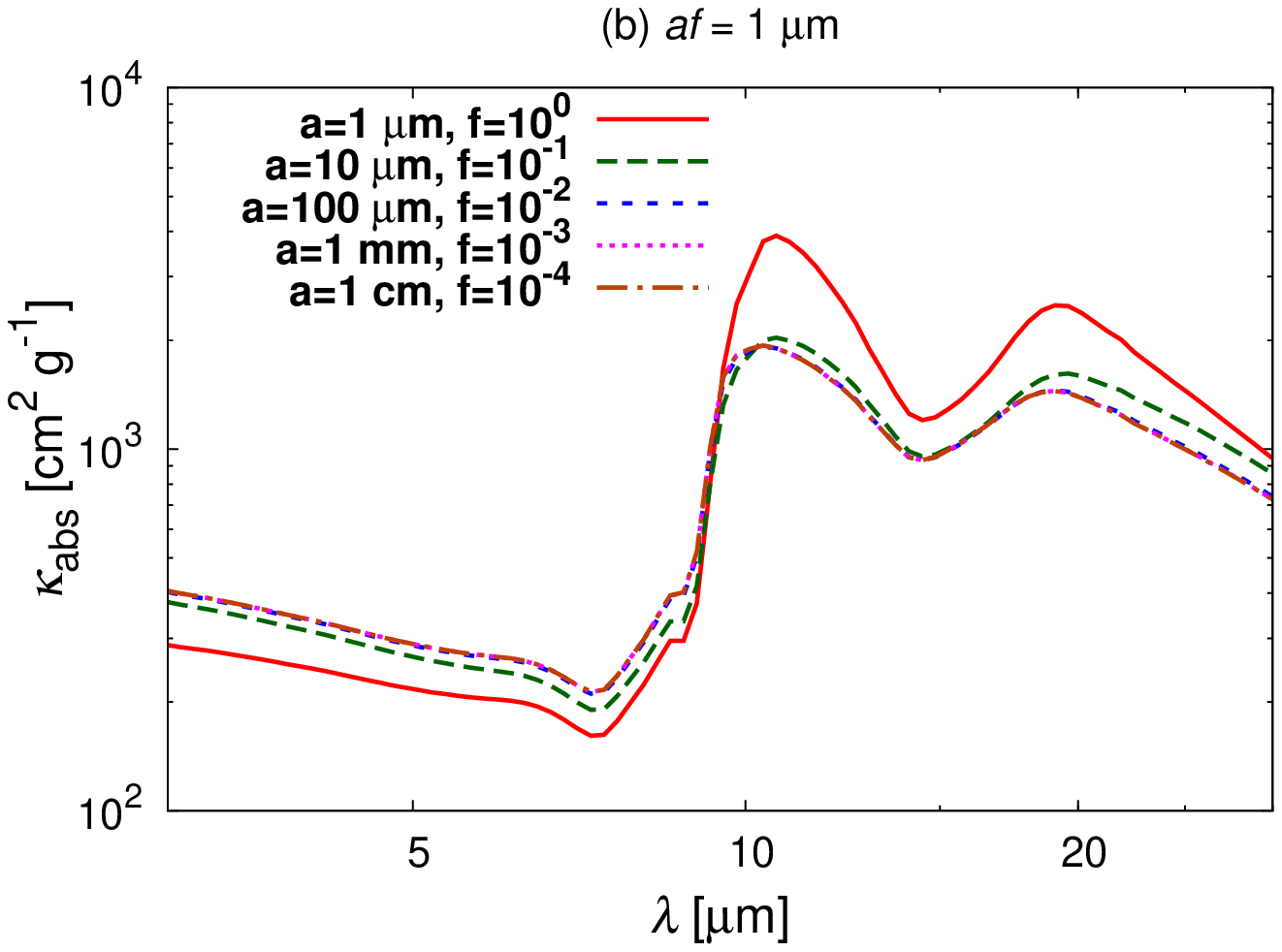}
	}\\
 	\subfigure{
 	 \includegraphics[width=80mm]{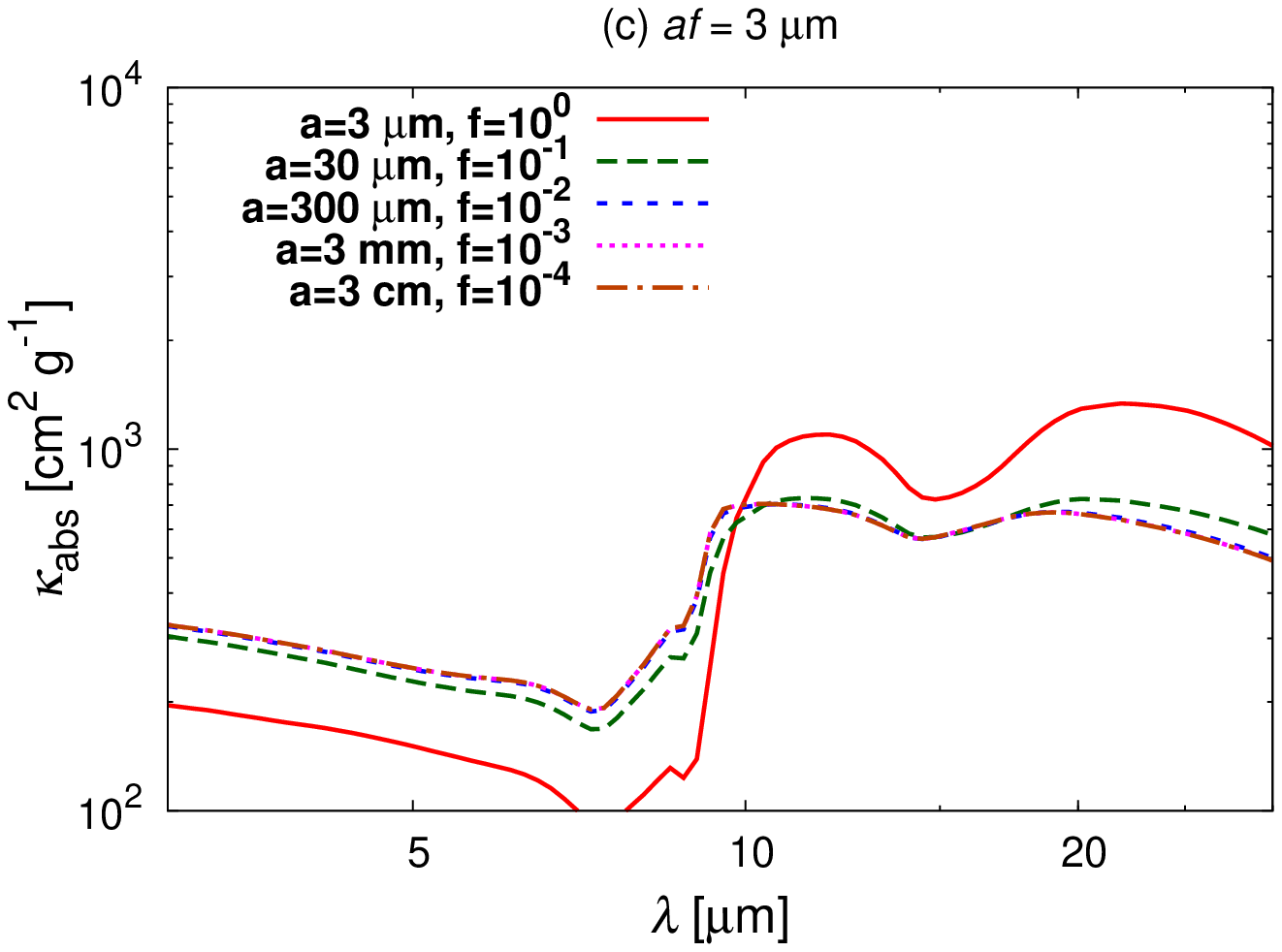}
	}
 	\subfigure{
 	 \includegraphics[width=80mm]{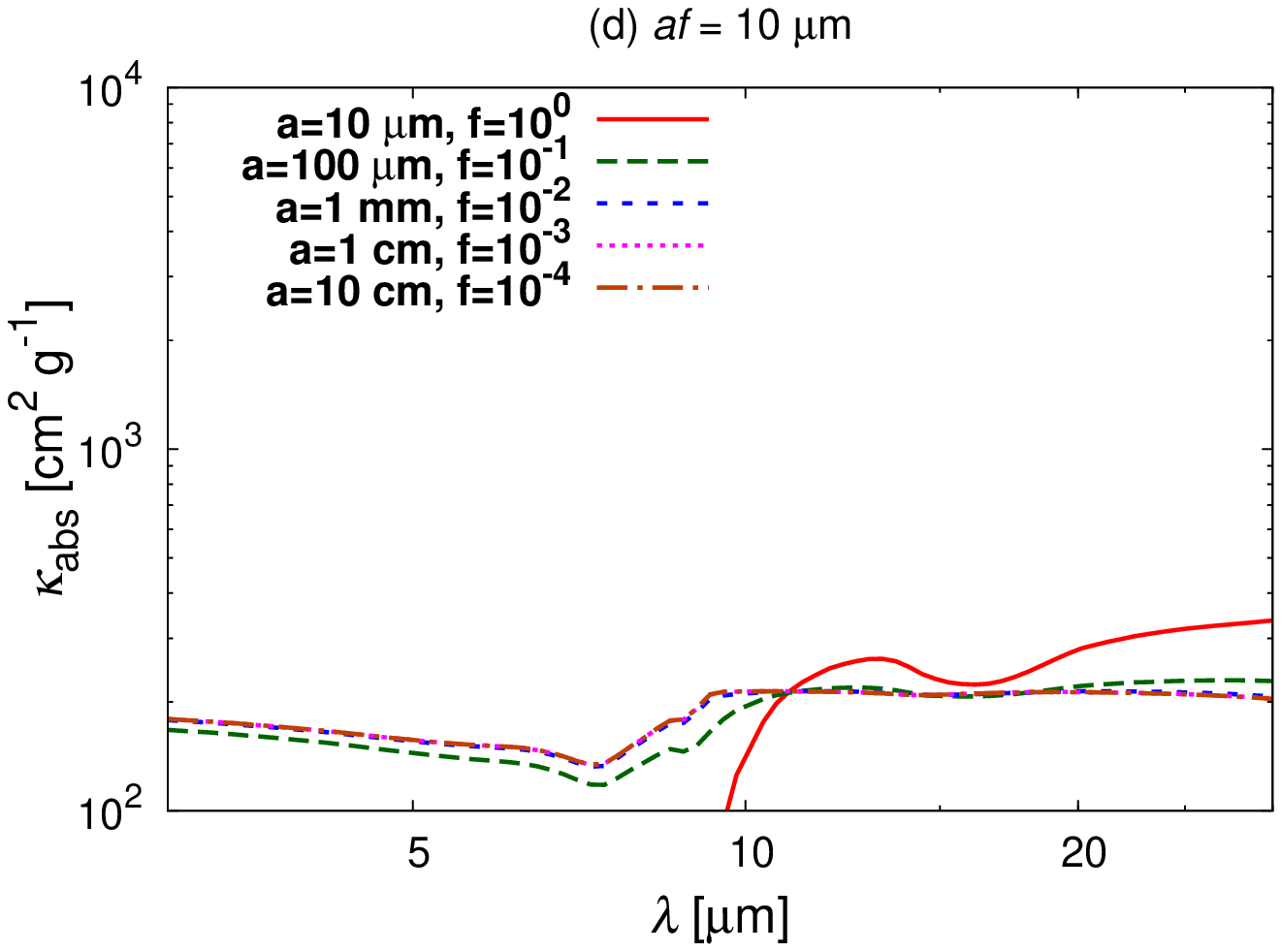}
	}
 \end{center}
 \caption{
 The absorption mass opacity for silicate aggregates in cases of different $af$.
 The mass opacities are shown in the case of (a) $af=0.1{~\rm \mu m}$, (b) $af=1{~\rm \mu m}$, (c) $af=3{~\rm \mu m}$, and (d) $af=30{~\rm \mu m}$.
 }
 \label{fig:silicate}
\end{figure*}

Figure \ref{fig:silicate} shows the absorption mass opacity in cases of silicate aggregates.
Each panel shows the mass absorption opacities where $af$ is constant.
When $af=0.1{~\rm \mu m}$, the broad silicate feature is seen at around $\lambda=10 {~\rm \mu m}$ and does not significantly change as the aggregate size increases from $0.1{~\rm \mu m}$ to $1{~\rm mm}$.
In the cases of $af=1$-$10{~\rm \mu m}$, the silicate feature becomes weaker as $af$ increases.
The absorption mass opacity of porous aggregates ($f \le 0.1$) is independent of the aggregate size as long as $af$ is constant, as is also shown in the case of icy aggregates.
The small differences between the cases of $f=1$ and $f\le0.1$ derive from the interference as well as the icy aggregates.
Thus, except for the interference, the silicate feature disappears as $af$ increases.

\section{Summary and discussion}
\label{sec:summary}
We calculated the mass opacity of fluffy dust aggregates expected to be in protoplanetary disks.
The wavelengths are in the range of $1{~\rm \mu m}~ <\lambda<~1{~\rm cm}$ and the filling factor in the range of $10^{-4} < f < 1$.
The assumed composition is the mixture of silicate, organics, and water ice \citep{Pollack94}.
We used the Mie calculation with the effective medium theory to calculate the mass opacity of fluffy aggregates.
Our main findings are as follows.

\begin{itemize}
\item
The absorption mass opacity of dust aggregates is characterized by $af$, where $a$ is the dust radius and $f$ is the filling factor.
The absorption mass opacity is almost independent of the aggregate size when $af$ is constant.
This makes it difficult to distinguish between fluffy aggregates and compact grains in observations.
The only difference of the absorption mass opacity between compact grains and fluffy aggregates where $af$ is the same appears as the interference structure in the compact case at the size parameter $x\sim 1$.

\item
The scattering mass opacity at short wavelengths is also characterized by $af$, but not at long wavelengths.
The scattering mass opacity at the long wavelengths is higher in more fluffy aggregates even if $af$ is the same.
The scattering mass opacity scales as $\lambda^{-2}$ at intermediate wavelengths and scales as $\lambda^{-4}$ at the longer wavelengths.

\item
We also derived the analytic formulae of the absorption and scattering mass opacities, connecting the three limiting cases, which are the Rayleigh regime, the optically-thin geometric regime, and the optically-thick geometric regime.
The analytic formulae reproduce the results of the Mie calculations.
The formulae are expected to greatly reduce the computational time to calculate the opacity of large fluffy aggregates.
By using the analytic formulae, we analytically showed that the absorption mass opacity is characterized by one parameter $af$ except for the interference structure.
We also showed that the scattering mass opacity at the shorter wavelengths is also characterized by $af$, but not at the longer wavelengths.
Thus, the fact that the mass opacity is characterized by $af$ is valid even out of the range investigated in the Mie calculation in this paper and is also applicable to other materials.

\item
The opacity index $\beta$ is a good way to distinguish between fluffy aggregates and compact grains at observations of protoplanetary disks.
If we assume the compact grain growth, with increasing grain size, the opacity index $\beta$ increases to $\sim 2.7$ and then decreases.
If we assume fluffy aggregate growth, the index $\beta$ monotonously decreases from its initial value $\beta_{\rm ISM}=1.7$ (see Fig. \ref{fig:beta}).
If dust grains are compact in protoplanetary disks, the radial distribution of the index $\beta$ should have a peak of $\sim 2.7$ (see Fig. \ref{fig:dustgrowthdriftbeta}).


\item
We also calculated the absorption mass opacity of silicate aggregates at around $\lambda=10 {\rm ~\mu m}$.
The opacity of fluffy aggregates has the $10 {\rm ~\mu m}$ feature as well as compact silicate grains.
The silicate feature is also appropriately characterized by $af$.
 
\end{itemize}

In this paper, as a first step, we use the effective medium theory.
\citet{Voshchinnikov05} have shown that EMT is a good approximation when considering small inclusions.
We considered $0.1{\rm ~\mu m}$ sized monomers and the wavelength ranges from $1{\rm ~\mu m}$ to 10 cm.
Because the monomer size is less than the wavelengths, EMT would be a good approximation.
However, the validity of EMT at infrared, especially at short wavelengths, is somewhat marginal because the wavelengths are close to the monomer size.
Thus, the validity of the effective medium theory should be further tested by future work.

We proposed that the radial profile of $\beta$ is a way to distinguish between compact grains and fluffy aggregates.
\citet{Perez12} have put a constraint on the radial $\beta$ distribution by observing a protoplanetary disk AS 209 with VLA, SMA, and CARMA.
They found that $\beta$ has a lower value inside the disk rather than the constant $\beta$ in the whole disk.
It is consistent with the model of Fig. \ref{fig:dustgrowthdrift}, where the grain size is distributed because of the difference of growth time and the maximum grain size is limited by radial drift.
The results of \citet{Perez12} also prefer the fluffy growth scenario to the compact because there is no signature that $\beta$ is large as $\beta \sim 3$.
However, the observation has little information about $\beta$ in the outer part of the disk because of the sensitivity limitation at the longer wavelengths.
Thus, to clearly determine whether the emission comes from compact grains or fluffy aggregates, we need higher sensitivity at the longer wavelengths.
Moreover, to reject the possibility of $\beta>\beta_{\rm ISM}$, we need a high spatial resolution to resolve 20 AU bump in the compact case at $t=10^{6}$ years.
This observation is challenging, but would be a good target of ALMA.

The fact that the scattering mass opacity at the longer wavelengths cannot characterized by $af$ is another way to distinguish between compact grains and fluffy aggregates.
Here, we propose the polarization observation of the millimeter continuum emission.
The scattered light is expected to be linearly polarized, and thus by comparing the intensity and the polarized intensity, the ratio of the scattering and absorption mass opacity can be directly observed.
Although the polarized emission depends on the disk geometry and has many uncertainties, it would be a good target in the next phase of ALMA.

In addition, \citet{Mulders13} proposed that to interpret the low effective albedo of protoplanetary disks, there are large particles at the outer disk surface and they should be porous structures to be stirred up to the surface.
However, we showed that the infrared scattering opacity is determined by $af$
In addition, the coupling efficiency of aggregates to the disk gas is also determined by $af$.
Thus, the optical and kinematical properties are degenerated.
Therefore, the porous aggregates would not help to interpret the observations of the low effective albedo.

\begin{appendix}
\section{Refractive index of fluffy aggregates}
In this section, we confirm the validity of the assumptions used to derive the analytic formula in Section \ref{sec:piecewise}.

\subsection{$(n-1)>k$ at the longer wavelengths}
\label{sec:appendix1}
\begin{figure}[htbp]
 \begin{center}
  \includegraphics[width=80mm]{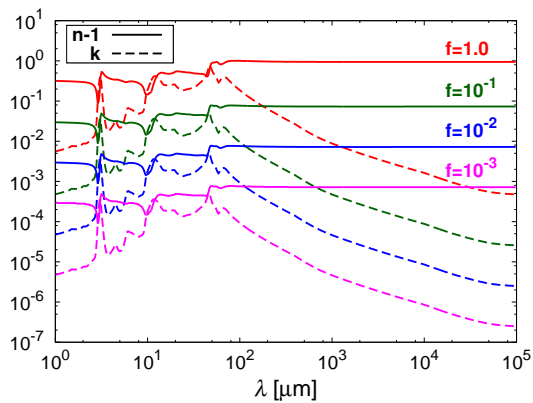}
 \end{center}
 \caption{
 $n-1$ and $k$ against the wavelengths when $f=1, 10^{-1}, 10^{-2}$, and $10^{-3}$.
 }
 \label{fig:check_ref_0}
\end{figure}

Figure \ref{fig:check_ref_0} shows the comparison of $n-1$ and $k$ when $f=1, 10^{-1}, 10^{-2}, 10^{-3}$, and $10^{-4}$.
We confirm that the refractive index always satisfies $n-1> k$ at the longer wavelengths in any value of $f$.
Moreover, we also confirm that $(n-1) \propto f$ and  $k \propto f$ when $f<1$.

\subsection{Reflectance}
\label{sec:appendix2}
\begin{figure}[htbp]
 \begin{center}
  \includegraphics[width=80mm]{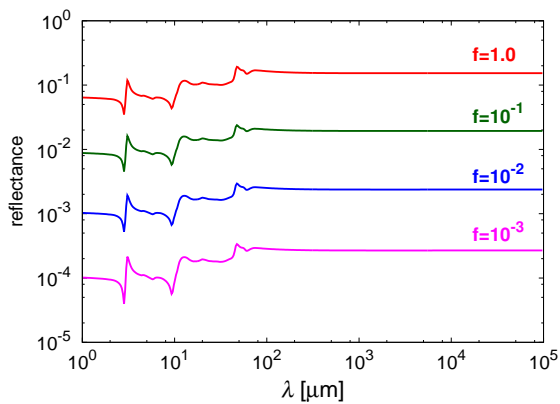}
 \end{center}
 \caption{
The integrated reflectance when $f=1,10^{-1},10^{-2}$, and $10^{-3}$.
 }
 \label{fig:reflectance}
\end{figure}

We define the integrated reflectance as
\begin{equation}
R \equiv \int^{\pi/2}_{0}R(\theta_{i})\sin 2\theta_{i} d\theta_{i}.
\end{equation}
When the medium satisfies $x\gg1$ and is optically thick, $Q_{\rm abs}$ and $Q_{\rm sca}$ are written as $Q_{\rm abs}=Q_{\rm abs,3}=1-R$ and $Q_{\rm sca}=Q_{\rm sca,3}=1+R$.
We assume that $R\ll 1$, and therefore $Q_{\rm abs}$ and $Q_{\rm sca}$ are unity in Section \ref{sec:piecewise}.
Figure \ref{fig:reflectance} shows the integrated reflectance $R$.
Roughly speaking, $R \sim 0.1\times f$ and thus we can assume that $Q_{\rm abs}$ and $Q_{\rm sca}$ are unity in the regime.

\subsection{Optical thickness inside the material}
\label{sec:appendix3}
\begin{figure}[htbp]
 \begin{center}
  \includegraphics[width=80mm]{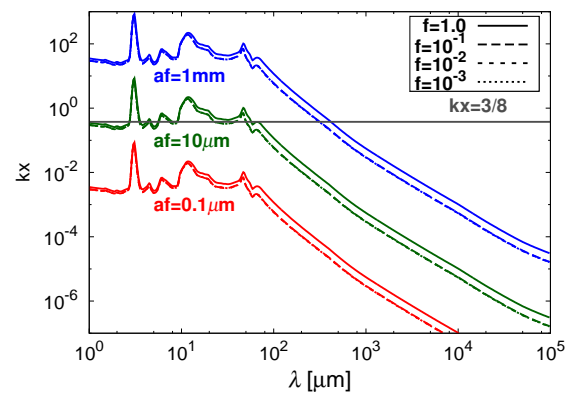}
 \end{center}
 \caption{
$kx$ against wavelengths when $af=0.1 {\rm ~\mu m}, 10 {\rm ~\mu m}$, and $1 {\rm ~mm}$.
The solid lines represent the compact cases while the dashed lines fluffy cases.
 }
 \label{fig:kx}
\end{figure}

We discuss the optical thickness inside the medium by considering whether $kx$ is greater than $3/8$.
Figure \ref{fig:kx} shows $kx$ in the case of $af=0.1 {\rm ~\mu m}, 10 {\rm ~\mu m}$, and $1 {\rm ~mm}$.
We confirm that if $af$ is the same, $kx$ is almost the same at all wavelengths.
In the case of $af=0.1 {\rm ~\mu m}$, the medium is totally optically thin.
In the case of $af=10 {\rm ~\mu m}$, the medium is optically thin at the longer wavelengths and marginally optically thick when $\lambda\lesssim10^{-2}{\rm ~cm}$.
In the case of $af=1 {\rm ~mm}$, the medium is optically thin at the longer wavelengths and optically thick when $\lambda\lesssim6\times10^{-2}{\rm ~cm}$.

\end{appendix}

%
%

\begin{acknowledgements}
We acknowledge Masaomi Tanaka for the fruitful discussion, Takaya Nozawa, Andrea Isella for their kind cooperation about opacity tables, Munetake Momose, Takashi Tsukagoshi for great discussion from observational point of view.
A.K. acknowledges Kohji Tomisaka for his support for PhD life.
A.K. is supported by the Research Fellowship from JSPS for Young Scientists (24$\cdot$2120).
\end{acknowledgements}

\bibliography{cite}

\newcommand{\SortNoop}[1]{}
\begin{thebibliography}{56}
\expandafter\ifx\csname natexlab\endcsname\relax\def\natexlab#1{#1}\fi

\bibitem[{{Andrews} \& {Williams}(2005)}]{AndrewsWilliams05}
{Andrews}, S.~M. \& {Williams}, J.~P. 2005, \apj, 631, 1134

\bibitem[{{Beckwith} \& {Sargent}(1991)}]{BeckwithSargent91}
{Beckwith}, S.~V.~W. \& {Sargent}, A.~I. 1991, \apj, 381, 250

\bibitem[{{Beckwith} {et~al.}(1990){Beckwith}, {Sargent}, {Chini}, \&
  {Guesten}}]{Beckwith90}
{Beckwith}, S.~V.~W., {Sargent}, A.~I., {Chini}, R.~S., \& {Guesten}, R. 1990,
  \aj, 99, 924

\bibitem[{{Birnstiel} {et~al.}(2010){Birnstiel}, {Ricci}, {Trotta},
  {Dullemond}, {Natta}, {Testi}, {Dominik}, {Henning}, {Ormel}, \&
  {Zsom}}]{Birnstiel10b}
{Birnstiel}, T., {Ricci}, L., {Trotta}, F., {et~al.} 2010, \aap, 516, L14

\bibitem[{{Blum} \& {Wurm}(2008)}]{BlumWurm08}
{Blum}, J. \& {Wurm}, G. 2008, \araa, 46, 21

\bibitem[{{Bohren} \& {Huffman}(1983)}]{BohrenHuffman83}
{Bohren}, C.~F. \& {Huffman}, D.~R. 1983, {Absorption and scattering of light
  by small particles}

\bibitem[{{Brauer} {et~al.}(2008){Brauer}, {Dullemond}, \&
  {Henning}}]{Brauer08a}
{Brauer}, F., {Dullemond}, C.~P., \& {Henning}, T. 2008, \aap, 480, 859

\bibitem[{{Chiang} {et~al.}(2001){Chiang}, {Joung}, {Creech-Eakman}, {Qi},
  {Kessler}, {Blake}, \& {van Dishoeck}}]{Chiang01}
{Chiang}, E.~I., {Joung}, M.~K., {Creech-Eakman}, M.~J., {et~al.} 2001, \apj,
  547, 1077

\bibitem[{{D'Alessio} {et~al.}(2001){D'Alessio}, {Calvet}, \&
  {Hartmann}}]{DAlessio01}
{D'Alessio}, P., {Calvet}, N., \& {Hartmann}, L. 2001, \apj, 553, 321

\bibitem[{{Draine} \& {Flatau}(1994)}]{DraineFlatau94}
{Draine}, B.~T. \& {Flatau}, P.~J. 1994, Journal of the Optical Society of
  America A, 11, 1491

\bibitem[{{Dr{\c a}{\.z}kowska} {et~al.}(2013){Dr{\c a}{\.z}kowska},
  {Windmark}, \& {Dullemond}}]{Drazkowska13}
{Dr{\c a}{\.z}kowska}, J., {Windmark}, F., \& {Dullemond}, C.~P. 2013, \aap,
  556, A37

\bibitem[{{Guilloteau} {et~al.}(2011){Guilloteau}, {Dutrey}, {Pi{\'e}tu}, \&
  {Boehler}}]{Guilloteau11}
{Guilloteau}, S., {Dutrey}, A., {Pi{\'e}tu}, V., \& {Boehler}, Y. 2011, \aap,
  529, A105

\bibitem[{{Isella} {et~al.}(2009){Isella}, {Carpenter}, \&
  {Sargent}}]{Isella09}
{Isella}, A., {Carpenter}, J.~M., \& {Sargent}, A.~I. 2009, \apj, 701, 260

\bibitem[{{Kataoka} {et~al.}(2013{\natexlab{a}}){Kataoka}, {Tanaka}, {Okuzumi},
  \& {Wada}}]{Kataoka13a}
{Kataoka}, A., {Tanaka}, H., {Okuzumi}, S., \& {Wada}, K. 2013{\natexlab{a}},
  \aap, 554, A4

\bibitem[{{Kataoka} {et~al.}(2013{\natexlab{b}}){Kataoka}, {Tanaka}, {Okuzumi},
  \& {Wada}}]{Kataoka13b}
{Kataoka}, A., {Tanaka}, H., {Okuzumi}, S., \& {Wada}, K. 2013{\natexlab{b}},
  \aap, 557, L4

\bibitem[{{Kimura} {et~al.}(2003){Kimura}, {Kolokolova}, \& {Mann}}]{Kimura03}
{Kimura}, H., {Kolokolova}, L., \& {Mann}, I. 2003, \aap, 407, L5

\bibitem[{{Kimura} {et~al.}(2006){Kimura}, {Kolokolova}, \& {Mann}}]{Kimura06}
{Kimura}, H., {Kolokolova}, L., \& {Mann}, I. 2006, \aap, 449, 1243

\bibitem[{{Kolokolova} {et~al.}(2007){Kolokolova}, {Kimura}, {Kiselev}, \&
  {Rosenbush}}]{Kolokolova07}
{Kolokolova}, L., {Kimura}, H., {Kiselev}, N., \& {Rosenbush}, V. 2007, \aap,
  463, 1189

\bibitem[{{Kothe} {et~al.}(2013){Kothe}, {Blum}, {Weidling}, \&
  {G{\"u}ttler}}]{Kothe13}
{Kothe}, S., {Blum}, J., {Weidling}, R., \& {G{\"u}ttler}, C. 2013, \icarus,
  225, 75

\bibitem[{{Kozasa} {et~al.}(1992){Kozasa}, {Blum}, \& {Mukai}}]{Kozasa92}
{Kozasa}, T., {Blum}, J., \& {Mukai}, T. 1992, \aap, 263, 423

\bibitem[{{Kudo} {et~al.}(2002){Kudo}, {Kouchi}, {Arakawa}, \&
  {Nakano}}]{Kudo02}
{Kudo}, T., {Kouchi}, A., {Arakawa}, M., \& {Nakano}, H. 2002, Meteoritics and
  Planetary Science, 37, 1975

\bibitem[{{Li} \& {Greenberg}(1997)}]{LiGreenberg97}
{Li}, A. \& {Greenberg}, J.~M. 1997, \aap, 323, 566

\bibitem[{{Lommen} {et~al.}(2010){Lommen}, {van Dishoeck}, {Wright},
  {Maddison}, {Min}, {Wilner}, {Salter}, {van Langevelde}, {Bourke}, {van der
  Burg}, \& {Blake}}]{Lommen10}
{Lommen}, D.~J.~P., {van Dishoeck}, E.~F., {Wright}, C.~M., {et~al.} 2010,
  \aap, 515, A77

\bibitem[{{Min} {et~al.}(2006){Min}, {Dominik}, {Hovenier}, {de Koter}, \&
  {Waters}}]{Min06}
{Min}, M., {Dominik}, C., {Hovenier}, J.~W., {de Koter}, A., \& {Waters},
  L.~B.~F.~M. 2006, \aap, 445, 1005

\bibitem[{{Min} {et~al.}(2003){Min}, {Hovenier}, \& {de Koter}}]{Min03}
{Min}, M., {Hovenier}, J.~W., \& {de Koter}, A. 2003, \aap, 404, 35

\bibitem[{{Min} {et~al.}(2005){Min}, {Hovenier}, \& {de Koter}}]{Min05}
{Min}, M., {Hovenier}, J.~W., \& {de Koter}, A. 2005, \aap, 432, 909

\bibitem[{{Min} {et~al.}(2007){Min}, {Waters}, {de Koter}, {Hovenier},
  {Keller}, \& {Markwick-Kemper}}]{Min07}
{Min}, M., {Waters}, L.~B.~F.~M., {de Koter}, A., {et~al.} 2007, \aap, 462, 667

\bibitem[{{Miyake} \& {Nakagawa}(1993)}]{MiyakeNakagawa93}
{Miyake}, K. \& {Nakagawa}, Y. 1993, \icarus, 106, 20

\bibitem[{{Mukai} {et~al.}(1992){Mukai}, {Ishimoto}, {Kozasa}, {Blum}, \&
  {Greenberg}}]{Mukai92}
{Mukai}, T., {Ishimoto}, H., {Kozasa}, T., {Blum}, J., \& {Greenberg}, J.~M.
  1992, \aap, 262, 315

\bibitem[{{Mulders} {et~al.}(2013){Mulders}, {Min}, {Dominik}, {Debes}, \&
  {Schneider}}]{Mulders13}
{Mulders}, G.~D., {Min}, M., {Dominik}, C., {Debes}, J.~H., \& {Schneider}, G.
  2013, \aap, 549, A112

\bibitem[{{Okuzumi} {et~al.}(2012){Okuzumi}, {Tanaka}, {Kobayashi}, \&
  {Wada}}]{Okuzumi12}
{Okuzumi}, S., {Tanaka}, H., {Kobayashi}, H., \& {Wada}, K. 2012, \apj, 752,
  106

\bibitem[{{Okuzumi} {et~al.}(2009){Okuzumi}, {Tanaka}, \&
  {Sakagami}}]{Okuzumi+09}
{Okuzumi}, S., {Tanaka}, H., \& {Sakagami}, M.-a. 2009, \apj, 707, 1247

\bibitem[{{Ormel} {et~al.}(2007){Ormel}, {Spaans}, \& {Tielens}}]{Ormel07}
{Ormel}, C.~W., {Spaans}, M., \& {Tielens}, A.~G.~G.~M. 2007, \aap, 461, 215

\bibitem[{{Pagani} {et~al.}(2010){Pagani}, {Steinacker}, {Bacmann}, {Stutz}, \&
  {Henning}}]{Pagani10}
{Pagani}, L., {Steinacker}, J., {Bacmann}, A., {Stutz}, A., \& {Henning}, T.
  2010, Science, 329, 1622

\bibitem[{{P{\'e}rez} {et~al.}(2012){P{\'e}rez}, {Carpenter}, {Chandler},
  {Isella}, {Andrews}, {Ricci}, {Calvet}, {Corder}, {Deller}, {Dullemond},
  {Greaves}, {Harris}, {Henning}, {Kwon}, {Lazio}, {Linz}, {Mundy}, {Sargent},
  {Storm}, {Testi}, \& {Wilner}}]{Perez12}
{P{\'e}rez}, L.~M., {Carpenter}, J.~M., {Chandler}, C.~J., {et~al.} 2012,
  \apjl, 760, L17

\bibitem[{{Pollack} {et~al.}(1994){Pollack}, {Hollenbach}, {Beckwith},
  {Simonelli}, {Roush}, \& {Fong}}]{Pollack94}
{Pollack}, J.~B., {Hollenbach}, D., {Beckwith}, S., {et~al.} 1994, \apj, 421,
  615

\bibitem[{{Ricci} {et~al.}(2010{\natexlab{a}}){Ricci}, {Testi}, {Natta}, \&
  {Brooks}}]{Ricci10b}
{Ricci}, L., {Testi}, L., {Natta}, A., \& {Brooks}, K.~J. 2010{\natexlab{a}},
  \aap, 521, A66

\bibitem[{{Ricci} {et~al.}(2010{\natexlab{b}}){Ricci}, {Testi}, {Natta},
  {Neri}, {Cabrit}, \& {Herczeg}}]{Ricci10a}
{Ricci}, L., {Testi}, L., {Natta}, A., {et~al.} 2010{\natexlab{b}}, \aap, 512,
  A15

\bibitem[{{Shen} {et~al.}(2008){Shen}, {Draine}, \& {Johnson}}]{Shen08}
{Shen}, Y., {Draine}, B.~T., \& {Johnson}, E.~T. 2008, \apj, 689, 260

\bibitem[{{Shen} {et~al.}(2009){Shen}, {Draine}, \& {Johnson}}]{Shen09}
{Shen}, Y., {Draine}, B.~T., \& {Johnson}, E.~T. 2009, \apj, 696, 2126

\bibitem[{{Suyama} {et~al.}(2008){Suyama}, {Wada}, \& {Tanaka}}]{Suyama08}
{Suyama}, T., {Wada}, K., \& {Tanaka}, H. 2008, \apj, 684, 1310

\bibitem[{{Suyama} {et~al.}(2012){Suyama}, {Wada}, {Tanaka}, \&
  {Okuzumi}}]{Suyama12}
{Suyama}, T., {Wada}, K., {Tanaka}, H., \& {Okuzumi}, S. 2012, \apj, 753, 115

\bibitem[{{Tanaka} {et~al.}(2005){Tanaka}, {Himeno}, \& {Ida}}]{Tanaka05}
{Tanaka}, H., {Himeno}, Y., \& {Ida}, S. 2005, \apj, 625, 414

\bibitem[{{van Boekel} {et~al.}(2005){van Boekel}, {Min}, {Waters}, {de Koter},
  {Dominik}, {van den Ancker}, \& {Bouwman}}]{vanBoekel05}
{van Boekel}, R., {Min}, M., {Waters}, L.~B.~F.~M., {et~al.} 2005, \aap, 437,
  189

\bibitem[{{van der Marel} {et~al.}(2013){van der Marel}, {van Dishoeck},
  {Bruderer}, {Birnstiel}, {Pinilla}, {Dullemond}, {van Kempen}, {Schmalzl},
  {Brown}, {Herczeg}, {Mathews}, \& {Geers}}]{vanderMarel13}
{van der Marel}, N., {van Dishoeck}, E.~F., {Bruderer}, S., {et~al.} 2013,
  Science, 340, 1199

\bibitem[{{Voshchinnikov} {et~al.}(2005){Voshchinnikov}, {Il'in}, \&
  {Henning}}]{Voshchinnikov05}
{Voshchinnikov}, N.~V., {Il'in}, V.~B., \& {Henning}, T. 2005, \aap, 429, 371

\bibitem[{{Wada} {et~al.}(2013){Wada}, {Tanaka}, {Okuzumi}, {Kobayashi},
  {Suyama}, {Kimura}, \& {Yamamoto}}]{Wada13}
{Wada}, K., {Tanaka}, H., {Okuzumi}, S., {et~al.} 2013, \aap

\bibitem[{{Wada} {et~al.}(2008){Wada}, {Tanaka}, {Suyama}, {Kimura}, \&
  {Yamamoto}}]{Wada08}
{Wada}, K., {Tanaka}, H., {Suyama}, T., {Kimura}, H., \& {Yamamoto}, T. 2008,
  \apj, 677, 1296

\bibitem[{{Wada} {et~al.}(2009){Wada}, {Tanaka}, {Suyama}, {Kimura}, \&
  {Yamamoto}}]{Wada09}
{Wada}, K., {Tanaka}, H., {Suyama}, T., {Kimura}, H., \& {Yamamoto}, T. 2009,
  \apj, 702, 1490

\bibitem[{{Wada} {et~al.}(2011){Wada}, {Tanaka}, {Suyama}, {Kimura}, \&
  {Yamamoto}}]{Wada11}
{Wada}, K., {Tanaka}, H., {Suyama}, T., {Kimura}, H., \& {Yamamoto}, T. 2011,
  \apj, 737, 36

\bibitem[{{Warren}(1984)}]{Warren84}
{Warren}, S.~G. 1984, \ao, 23, 1206

\bibitem[{{Weingartner} \& {Draine}(2001)}]{WeingartnerDraine01}
{Weingartner}, J.~C. \& {Draine}, B.~T. 2001, \apj, 548, 296

\bibitem[{{Windmark} {et~al.}(2012){Windmark}, {Birnstiel}, {G{\"u}ttler},
  {Blum}, {Dullemond}, \& {Henning}}]{Windmark12}
{Windmark}, F., {Birnstiel}, T., {G{\"u}ttler}, C., {et~al.} 2012, \aap, 540,
  A73

\bibitem[{{Zsom} {et~al.}(2011){Zsom}, {Ormel}, {Dullemond}, \&
  {Henning}}]{Zsom11b}
{Zsom}, A., {Ormel}, C.~W., {Dullemond}, C.~P., \& {Henning}, T. 2011, \aap,
  534, A73

\bibitem[{{Zsom} {et~al.}(2010){Zsom}, {Ormel}, {G{\"u}ttler}, {Blum}, \&
  {Dullemond}}]{Zsom10}
{Zsom}, A., {Ormel}, C.~W., {G{\"u}ttler}, C., {Blum}, J., \& {Dullemond},
  C.~P. 2010, \aap, 513, A57

\bibitem[{{Zubko} {et~al.}(1996){Zubko}, {Mennella}, {Colangeli}, \&
  {Bussoletti}}]{Zubko96}
{Zubko}, V.~G., {Mennella}, V., {Colangeli}, L., \& {Bussoletti}, E. 1996,
  \mnras, 282, 1321

\end{thebibliography}
\end{document}